\documentclass[journal]{IEEEtran}

\usepackage{cite}
\usepackage{setspace}
\usepackage{amsmath,amssymb,amsthm}
\usepackage{graphicx}
\usepackage{caption}
\usepackage{subcaption}
\usepackage{float}
\usepackage{bm}

\usepackage{tikz}
\usetikzlibrary{spy,decorations.fractals}
\usetikzlibrary{lindenmayersystems}
\usetikzlibrary{shadings}
\usepackage{pgf}
\usetikzlibrary{patterns,backgrounds,calc,arrows.meta}
\pgfdeclarelayer{foreground}
\pgfsetlayers{background,main,foreground}

\usepackage{url}
\usepackage{algorithm}
\usepackage{algorithmicx}

\usepackage{algpseudocode}

\usepackage{indentfirst}
\usepackage{color}
\usepackage{booktabs}

\usepackage{epstopdf}
\usepackage{grffile}
\usepackage{multirow}
\usepackage{diagbox}
\usepackage{grffile}

\usepackage{cancel}
\DeclareMathOperator*{\argmin}{argmin} 

\algdef{SE}[FUNCTION]{Function}{EndFunction}
   [2]{\algorithmicfunction\ \textproc{#1}\ifthenelse{\equal{#2}{}}{}{(#2)}}
   {\algorithmicend\ \algorithmicfunction}

\newcommand{\x}		{\mathbf{x}}	
\newcommand{\X}		{\mathbf{X}}	
\newcommand{\y}		{\mathbf{y}}
\newcommand{\A}		{\mathbf{A}}
\newcommand{\z}	 {\mathbf{z}}
\newcommand{\W}	 {\mathbf{W}}
\renewcommand{\P}	 {\mathbf{P}}
\newcommand{\omg}	 {\mathbf{\Omega}}

\newcommand{\Z}		{\mathbf{Z}}

\newcommand{\R}	     {\mathsf{R}}

\newcommand{\BLUE}{\textcolor{black}}

\newcommand*{\MG}[1]{{\color{black}#1}}
\newcommand*{\BLK}[1]{{\color{black}#1}}

\newcommand{\subparagraph}{}
\usepackage{titlesec}
\titlespacing{\section}{0pt}{1.8ex plus .2ex minus .25ex}{\dimexpr1.0ex-3pt plus .1ex}
\titlespacing{\subsection}{0pt}{0.8ex plus .2ex minus .1ex}{\dimexpr0.6ex-2pt plus .08ex}
\titlespacing{\subsubsection}{0pt}{\dimexpr2.0ex-6pt plus .5ex minus .1ex}{0pt}

\title{Unified Supervised-Unsupervised (SUPER) Learning for X-ray CT Image Reconstruction}

\author{Siqi Ye$^{\dagger}$, Zhipeng Li$^{\dagger}$, Michael T. McCann, Yong Long$^*$, Saiprasad Ravishankar

	\thanks{ 
				This work was supported by NSFC (61501292). \textit{$\dagger$ indicates equal contribution to this work. Asterisk indicates the corresponding author.} 
		
		S. Ye, Z. Li and Y. Long are with the University of Michigan - Shanghai Jiao Tong University Joint Institute, Shanghai Jiao Tong University, Shanghai 200240, China (email: yesiqi@sjtu.edu.cn, zhipengli@sjtu.edu.cn, yong.long@sjtu.edu.cn).
		
M. McCann is with the Department of Computational Mathematics, Science and Engineering, Michigan State University, East Lansing, MI, 48824 USA (email: {mccann13@msu.edu}).

S. Ravishankar is with the Department of Computational Mathematics, Science and Engineering, and the Department of Biomedical Engineering, Michigan State University, East Lansing, MI, 48824 USA (email: {ravisha3@msu.edu}).
}}

\begin{document}

	\maketitle
	\begin{abstract}
	Traditional model-based image reconstruction (MBIR) methods combine forward and noise models with simple object priors. Recent machine learning methods for image reconstruction typically involve supervised learning or unsupervised learning, both of which have their \BLK{advantages and disadvantages}. 
	In this work, we propose a unified supervised-unsupervised (SUPER) learning framework for X-ray computed tomography (CT) image reconstruction.
	The proposed \BLK{learning} formulation combines both unsupervised learning-based priors \BLK{(or even simple analytical priors)} together with (supervised) deep network-based priors in a unified MBIR framework \BLK{based on a fixed point iteration analysis. 
	The proposed training algorithm is also an approximate scheme for a bilevel supervised training optimization problem,
	wherein the network-based regularizer in the lower-level MBIR problem is optimized using an upper-level reconstruction loss.}
The training problem is optimized 
by alternating between updating the network weights and iteratively updating the reconstructions based on those weights. 
We demonstrate the learned SUPER models' efficacy for low-dose CT image reconstruction, for which we use the NIH AAPM Mayo Clinic Low Dose CT Grand Challenge dataset for training and testing. In our experiments, we studied different combinations of supervised deep network priors and unsupervised learning-based or analytical priors.
Both numerical and visual results show the \BLK{superiority of the proposed unified SUPER methods over standalone supervised learning-based methods, iterative MBIR methods, and variations of SUPER obtained via ablation studies.}
We also show that the proposed algorithm converges rapidly in practice. 

	\end{abstract}

	\begin{IEEEkeywords}
	Low-dose X-ray CT, image reconstruction, deep learning, transform learning, iterative reconstruction, \BLK{mixture of priors, fixed point iteration}, bilevel optimization.
	\end{IEEEkeywords}
\section{Introduction} 
X-ray computed tomography (CT) image reconstruction is a fundamental process in medical imaging. It generates latent anatomical images from measurements (i.e., sinograms), that do not directly reflect anatomical features. Similar to other medical imaging modalities, X-ray CT image reconstruction is often formulated as an inverse problem, which can be solved by analytical methods, or iterative optimization algorithms for model-based image reconstruction problems. More recently, deep learning methods have also been used for CT reconstruction \cite{book19wang,ongie_deep_2020}. In this section, we first review existing image reconstruction methods, and then propose a unified framework combining model-based image reconstruction and supervised and unsupervised machine learning priors.
Although we focus on X-ray CT image reconstruction in this paper, our proposed method can be easily adapted for other imaging modalities.

\subsection{Background}
Analytical methods for X-ray CT image reconstruction such as the filtered backprojection (FBP) method \cite{fbp1974, feldkamp1984practical}, often involve short reconstruction times but have
poor noise-resolution trade-offs when dealing with incomplete or degraded measurement data, e.g., sparse-view or low-dose sinogram data.

More sophisticated iterative algorithms have also been developed for image reconstruction. 
They are often referred to as model-based image reconstruction (MBIR) methods, as they iteratively optimize a cost function that incorporates imaging physics, statistical model of measurements, and prior knowledge of the unknown object. Classical MBIR methods in CT solve a penalized weighted-least squares (PWLS) problem, where a weighted quadratic data-fidelity term captures the imaging forward model and measurement statistics, and a penalty term (a.k.a. regularizer) models prior information about the object \cite{pwls1994jf,thibault:07:atd,beister2012iterative}. 
\BLUE{Many effective regularizers have been designed to capture sparsity priors. Examples include hand-crafted priors such as edge-preserving regularizers \cite{EP1998} as well as data-driven or learning-based priors such as prior image constrained compressive sensing priors \cite{PICCS2008,ncpiccs2011,piccs-app2012}, dictionary learning-based priors \cite{xu:12:ldx}, sparsifying transform learning-based priors \cite{pwls-ultra2018,ye2019TMI}, etc. 
In particular, sparsifying transform learning adopts computationally efficient thresholding operations to sparsify signals in a learned transform domain~\cite{STlearning13}. 
Various structures are used for the transforms during learning, which prove useful during reconstruction, such as 
doubly-sparse transforms \cite{doublyST13}, unions of transforms~\cite{ravishankar:16:tci}, rotation invariance~\cite{wen2017frist}, and filterbank models~\cite{pfister2018learning}.}

In the past few years, deep learning methods have also been gaining popularity for medical image reconstruction.
Depending on whether the training relies on paired data (low-dose and corresponding regular-dose CT data) or not, deep learning methods can be roughly categorized into supervised learning and unsupervised learning-based methods. Supervised learning-based image reconstruction methods use the paired data to learn deep neural network mappings that regress low-quality inputs to high-quality outputs. 

A typical class of supervised learning methods work in the image domain, with both inputs and outputs of the network being images. For example, a residual encoder-decoder convolutional neural network (RED-CNN) framework that combines the autoencoder, deconvolution network, and shortcut connections was proposed for low-dose CT imaging \cite{red-cnn17}. Another \BLK{U-Net based framework FBPConvNet} \cite{jin:17:dcn} learns a CNN that maps FBP reconstructed X-ray CT images to suitable high-quality images. The WavResNet framework \cite{WavResNet18} learns a CNN-based image mapping after transforming images into the wavelet domain, where image features may be better preserved. Image-domain learning methods do not directly need raw measurement data, so they can be conveniently deployed with existing imaging systems. However, not directly exploiting the measurement data and imaging physics \BLK{may limit} the ability of image-domain learning methods for recovering missing or corrupted information in the measurement domain. 

In order to better exploit the measurement data, several attempts have been made to exploit deep learning in the measurement domain. For example, \cite{wurfl2016deep} proposes a neural network to learn projection-domain weights in the FBP method. However, the designed network does not include other components in the FBP, such as ramp filtering and back-projection operations. This idea was recently improved in \cite{iRadonMap20}, where the designed neural network incorporates all fundamental steps in FBP: filtering, back-projection, and image post-processing.  
This method achieved competitive results with total variation based PWLS methods and some \BLK{image-domain deep learning-based} denoising methods. A drawback of the approach is that network trained with data acquired with a specific imaging geometry may not be suitable for reconstructing images with other imaging geometries. 

The third type of supervised learning methods exploit both imaging physics and iterative image reconstruction methods in the neural network architecture \BLK{and learn the parameters of simple unrolled model-based iterative algorithms. Examples include unrolling the} 
alternating direction method of multipliers (ADMM) algorithm \cite{admm-net2016}, 
primal-dual algorithms \cite{adler2018primaldual}, `fields of experts' (FoE)-based iterative algorithms \cite{LEARN18}, the block coordinate descent algorithm~\cite{ravchfess17,chfess18,Chun&etal:19MICCAI}, 
\BLUE{the gradient descent algorithm~\cite{adler2017solving}, etc. 
The unrolled methods can be combined with the plug-and-play strategy to replace the explicit regularizers in the MBIR cost with a denoising step using some off-the-shelf neural networks~\cite{PnP-Stanley-17,3pADMM,schlemper2017deep,mardani2017recurrent,aggarwal2018modl}. 
}

\BLK{In contrast to supervised methods,} unsupervised learning-based methods do not need paired training data. 
Typical examples include iterative methods that use dictionary learning or sparsifying transform learning based priors, where the dictionaries or transforms can be learned from unpaired clean images.
The generative adversarial network (GAN) based methods form another important class of unsupervised learning-based schemes. GAN-based approaches attempt to generate target data by somehow minimizing the difference between the probability distributions of the generator output and the target data. While several recent works \cite{Wolterink17,cycleGAN19,OT-cycleGAN-19}\BLUE{\cite{mardani2017deep}} have applied GAN-based methods to low-dose CT \BLUE{or other imaging modalities}, it is still challenging to avoid artificial features created by generators in such GAN frameworks.
\BLUE{A recent unsupervised learning technique named deep image prior (DIP) does not need external prior data to train the network. Instead, DIP is task-specific and takes a fixed input such as random noise to train the network~\cite{DIP2018}. DIP implicitly regularizes the MBIR problem by replacing the unknown image in the MBIR cost with the network, and training is performed via minimizing the network involved MBIR cost with respect to the network parameters. DIP has been used for medical image reconstructions~\cite{DIP-PET:18:qzli,DIP-MRI:19:Unser,DIP-CT:20}, but selecting a proper network architecture to capture informative priors and deciding when to stop iterations to prevent overfitting is often done in an ad hoc manner.
}

Both supervised learning and unsupervised learning have their \BLK{advantages}. Often, supervised learning provides superior results to unsupervised learning when there is a high similarity between training and testing samples. However, supervised learning methods also usually need large amounts of paired data (to train complex networks), which is not always feasible in medical imaging.
Unsupervised learning methods involving dictionaries or sparsifying transforms typically require relatively small training sets, and may have better generalization properties than supervised learning methods \cite{ye2019TMI}.

\subsection{Contributions}
In this paper, we \BLK{present} a unified reconstruction framework combining supervised and unsupervised learning, and physics and statistical models. We significantly extend our recent preliminary conference work~\cite{super-19} in several aspects.
First, we develop a systematic and unified mathematical framework for supervised-unsupervised (SUPER) training and reconstruction. We use an MBIR formulation consisting of a data-fidelity term incorporating forward models and statistical models, and regularizer terms incorporating unsupervised learning-based priors or simple analytical priors together with supervised deep network priors. 
The deep network in the MBIR formulation is trained in a supervised way with an alternating scheme to approximate solutions to the corresponding challenging fixed point iteration problem or a bilevel optimization problem.
At testing or reconstruction time, a similar MBIR optimization is used with learned unified priors. 
\BLUE{We allow using various types of explicit unsupervised regularizers for SUPER, including nonsmooth regularizers or integer-valued variable-based regularizers, which are usually not compatible with many (plug-and-play) unrolled methods.}
\BLUE{Second,}
we selected the recent FBPConvNet \cite{jin:17:dcn} and WavResNet \cite{WavResNet18} as example networks in our formulation that are learned in a supervised manner. We then incorporate different analytical and unsupervised learning-based priors in the proposed unified framework including the nonadaptive edge-preserving regularizer and a regularizer using a union of sparsifying transforms learned from (unpaired) regular-dose images.
The experimental results show that the proposed (unified) SUPER learning approaches achieve much better image reconstruction quality in low-dose CT than standalone deep learning methods and iterative reconstruction schemes, \BLUE{and they can outperform a plug-and-play unrolled method~\cite{PnP-Stanley-17} and a state-of-the-art post-processing method~\cite{shan2019competitive} as well}. In particular, combining both supervised and unsupervised learning in our framework leads to the best reconstruction performance. 
Our results also show the practical rapid convergence of the MBIR-based SUPER reconstruction. 
Finally, we consider several special cases of the SUPER model (akin to an ablation study) and demonstrate the superior reconstruction performance of the general unified approach compared to the special schemes.

\subsection{Organization}
We organize the rest of this paper as follows. In Section \ref{sec:formulation}, we describe the SUPER training and reconstruction formulations, and interpret this \BLK{model} in detail. In Section \ref{sec:algorithm}, we develop the algorithms for the proposed problems. In Section \ref{sec:experiments}, we show experimental results with the proposed method, compare results among various image reconstruction methods, and study the proposed methods' properties and behavior in detail. Finally, we conclude in Section \ref{sec:conclusion}.


\section{Proposed Model and Problem Formulations}
\label{sec:formulation}
This section presents the general SUPER model along with
formulations for training the model and using it at reconstruction time. The proposed approach could be useful for a variety of imaging modalities.
We provide interpretations of our formulations, and give specific examples of SUPER models for the low-dose CT application that is a focus of this paper.

In the low-dose CT image reconstruction problem, the goal is to reconstruct an image $\x\in \mathbb{R}^{N_p}$ from its observed noisy sinogram data $\y \in \mathbb{R}^{N_d}$, and we assume a given measurement matrix or forward operator $\A \in \mathbb{R}^{N_d \times N_p}$.


\subsection{Proposed Model}
The main idea of the SUPER framework is to combine supervised deep learning-based approaches with unsupervised or iterative model-based reconstruction approaches.
First, we can state reconstruction with an image-domain deep network learned in a supervised manner as
\begin{equation} \label{eq:intro-sup}
\hat{\x}_{\bm{\theta}}(\y) =
G_{\bm{\theta}}(\BLK{\hat{\x}}(\y)),
\vspace{-0.06in}
\end{equation}
where \BLK{$\hat{\x}(\y)$ is a reconstructed image using a specific reconstruction method.}
While the most common choice of \BLK{method} for $\hat{\x}$ \BLK{in the X-ray CT application} is the filtered back projection (FBP) \cite{fbp1974}, \BLK{it would be reasonable to consider}
an iterative method as well. 
Here, $G_{\bm{\theta}}(\cdot)$ denotes the (supervised) deep network operator with parameters $\bm{\theta}$.
Note that \BLK{$\hat{\x}$} 
and therefore $\hat{\x}_{\bm{\theta}}$ 
depend on the measurements, $\y$.

On the other hand, a typical iterative reconstruction method can be made to depend on the results of a trained \BLK{deep model} via adding a penalty term \BLK{to a usual MBIR cost as}
\begin{equation} \label{eq:intro-unsup}
    \hat{\x}(\y) = \argmin_\x L(\A \x, \y) + \beta \R(\x) +\mu \|\x - \hat{\x}_{\bm{\theta}}(\y)\|_2^2,
    \vspace{-0.06in}
\end{equation}
\BLK{where $\hat{\x}_{\bm{\theta}}(\y)$ is a fixed image (obtained with a pre-trained network) when solving~\eqref{eq:intro-unsup},  
$L(\A \x, \y)$ and $\R(\x)$ comprise the data-fidelity term and an analytical or unsupervised learning-based regularizer, 
and \BLK{$\beta$ and $\mu$ are non-negative weights (scalars) that trade off between the data-fidelity term and the regularizers.}}

Equations \eqref{eq:intro-sup} and \eqref{eq:intro-unsup} show that the supervised network's reconstruction can depend on an iterative reconstruction and vice-versa;
our proposed approach is to complete the cycle,
i.e. substitute \BLK{$\hat{\x}_{\bm{\theta}}$} 
in \eqref{eq:intro-sup} into \eqref{eq:intro-unsup}.
Doing so leads to an expression where the (unknown) reconstruction, \BLK{$\hat{\x}(\y)$},
appears on both the left and right hand side,
\begin{equation} \label{eq:intro-SUPER}
\BLK{\hat{\x}(\y) = 
     \argmin_\x J(\x,\y) +
     \mu \|\x - G_{\bm{\theta}}(\hat{\x}(\y))\|_2^2,}
     \vspace{-0.06in}
\end{equation}
\BLK{where ${J(\x,\y)\triangleq L(\A\x,\y) + \beta \R(\x)}$.}
Roughly speaking, \eqref{eq:intro-SUPER} 
seeks an image that is the solution to 
a regularized
reconstruction problem,
but where a deep neural \BLK{network} applied to the same image (solution)
acts as a regularizer.
In this way, regularization effects from the iterative reconstruction (involving $\R(\x)$)
and from the deep network are combined.
\BLK{We assume a unique global minimizer on the right hand side of~\eqref{eq:intro-SUPER}, else, we can replace `$=$' with `$\in$' therein.}

While one could attempt to directly use~\eqref{eq:intro-SUPER} as a reconstruction method,
computing $\hat{\x}(\y)$ (to say nothing of training the deep network)
turns out to be very challenging.
Instead, we consider solving~\eqref{eq:intro-SUPER} via the fixed point iteration
\begin{equation} \label{eq:fixed-point}
\hat{\x}_{\bm{\theta}}^{(l)}\BLK{(\y)} = 
    \argmin_\x \BLK{J(\x,\y)}
    + \mu \| \x -  G_{\bm{\theta}} \left(\hat{\x}_{\bm{\theta}}^{(l-1)}\BLK{(\y)} \right) \|_2^2,
    \vspace{-0.06in}
\end{equation}
\BLK{where $\hat{\x}_{\bm{\theta}}^{(l)}\BLK{(\y)}$ represents the reconstruction of the $l$th iteration ($l=1,2,\cdots, L$) based on the deep network weights $\bm{\theta}$, loss $J(\x,\y)$, and measurements $\y$. The initial reconstruction $\hat{\x}^{(0)}_{\bm{\theta}}(\y)$ is}
set to some fixed function of the measurements, 
e.g., FBP.
The opposite substitution,
i.e., substituting \eqref{eq:intro-unsup} into \eqref{eq:intro-sup},
leads to a similar fixed point,
but with the opposite alternation between the 
\BLK{deep and iterative reconstructions.}

We also found that sharing weights between these steps decreased performance,
so we learn a different set of supervised weights at each step.
Thus, the 
simplified
SUPER reconstruction framework is
\begin{equation} \label{eq:SUPER-reconstruction}
\hat{\x}_{ \bm{\theta}^{(l)} }^{(l)}(\y) = 
     \argmin_\x J(\x, \y) 
    + 
    \mu \| \x - 
    G_{\bm{\theta}^{(l)}} \left(\hat{\x}_{\bm{\theta}^{(l-1)}}^{(l-1)}(\y) \right) \|_2^2,
 \tag{P0}
\end{equation}
where the final reconstruction is $\hat{\x}_{\bm{\theta}^{(L)}}^{(L)}(\y)$,
and
$\hat{\x}^{(0)}_{\bm{\theta}^{(0)}}(\y)  = \hat{\x}^{(0)}(\y)$ is an initial reconstruction that does not depend on 
a network.
We can view the iterations in (P0) as layers in a larger 
neural network;
we call these ``SUPER layers".

\subsection{SUPER Learning Formulation} 
The SUPER model includes a regularizer $\R(\x)$ that can be an analytical prior or based on unsupervised learning from unpaired data (e.g., regular-dose images).
On the other hand, the network parameters in the deep network-based regularizer are learned in a supervised manner from paired training data.
For this training process (referred to as SUPER learning), we denote the paired low-dose and regular-dose (reference) training images as
\BLK{$\{(\hat{\x}^{(0)}(\y_n), \x_n^*)\}_{n=1}^{N}$,}
and the corresponding low-dose sinograms (measurements) as $\{\y_n\}_{n=1}^N$.
To learn the supervised weights 
in \eqref{eq:SUPER-reconstruction}, we 
take a greedy approach by learning each $\bm{\theta}^{(l)}$ in sequence according to
\begin{equation}\label{eq:super-all-sim}
\bm{\theta}^{(l)}=\argmin_{\bm{\theta}} 
\sum_{n=1}^{N}
\|
G_{\BLUE{\bm{\theta}}} \left(\hat{\x}^{(l-1)}_{\bm{\theta}^{(l-1)}}(\y_n)\right) - \x_n^*
\|_2^2, 
\tag{P1}
\vspace{-0.06in}
\end{equation}
for $l = 1, 2, \dots, L$.
We solve the sequence of problems~\eqref{eq:super-all-sim} by
alternating between learning weights for the supervised method
and solving the iterative reconstruction problem (to compute $\hat{\x}^{(l)}_{\bm{\theta}^{(l)}}(\y_n)$); 
we describe this approach in detail in Section~\ref{sec:super-alg}.

Another perspective on this training approach is that it is a heuristic for solving the following bilevel problem: 
\vspace{-0.05in}
\begin{equation} \label{eq:bilevel}
\begin{split}
   \bm{\theta} &= \argmin_{\bm{\theta}}
    \sum_{n=1}^{N} \| G_{\bm{\theta}}( \hat{\x}_{\bm{\theta}}(\y_n)) - \x^*_n \|_{2}^{2} 
  \\
\text{s.t.} \;\; \hat{\x}_{\bm{\theta}}(\y_n) &= \argmin_\x J(\x, \y_n) +
     \mu \|\x - G_{\bm{\theta}}(\x)\|_2^2.
     \end{split}
\vspace{-0.06in}
\end{equation}
The problem is called bilevel because the network input $\hat{\x}_{\bm{\theta}}(\y_n)$ in the main cost arises as the minimizer of another (i.e., the lower-level) optimization problem.
\BLUE{The bilevel optimization problem is very challenging to solve in general.
Some of the authors have proposed an algorithm for a simple form of~\eqref{eq:bilevel} in~\cite{MikeandSai:bilevel:20} (without deep networks). Here, the proposed alternating training approach could be viewed as a plausible heuristic for the challenging bilevel problem~\eqref{eq:bilevel}.
}
We can interpret~\eqref{eq:bilevel} as learning part of the regularizer of an MBIR problem (involving the network $G_{\bm{\theta}}$) in a supervised manner.

\subsection{SUPER Reconstruction Formulation}
\BLK{With the trained supervised network parameters $\{\bm{\theta}^{(l)}\}$ for $l = 1,2, \cdots, L$, the reconstruction (or testing) step becomes optimizing the MBIR formulation constructed in \eqref{eq:SUPER-reconstruction} in every SUPER layer to obtain the final layer reconstruction $\hat{\x}^{(L)}_{\bm{\theta}^{(L)}}(\y)$.}


\subsection{Examples of SUPER Modeling}\label{sec:examples_super}
The SUPER framework is flexible, and allows incorporating
various deep \BLK{networks} $G_{\bm{\theta}}(\cdot)$ in the supervised network-based regularizer and various unsupervised regularizers $\R(\cdot)$. 
In this work, we focus on studying some examples of the SUPER model. For the supervised component, we choose the recent FBPConvNet (FCN) ~\cite{jin:17:dcn} and the (feed-forward version of) WavResNet (WRN)~\cite{WavResNet18}.
For the regularizer $\R(\cdot)$, we study both a non-adaptive and an unsupervised learning-based one, namely the non-adaptive edge-preserving (EP) regularizer,
and a state-of-the-art union of learned sparsifying transforms (ULTRA) \cite{ravishankar:16:tci} regularizer. The union of transforms is learned in an unsupervised manner from a set of (unpaired) regular-dose images. We refer to the resulting SUPER models as SUPER-FCN-EP, SUPER-FCN-ULTRA, SUPER-WRN-EP, and SUPER-WRN-ULTRA, respectively. For simplicity, we refer to any regularizer $\R(\cdot)$ that is not learned in a supervised manner as an unsupervised regularizer. In the following, we further describe the models chosen above.

\subsubsection{Supervised Networks}
We work with FBPConvNet and WavResNet, both of which are CNN-based \BLK{image-domain} denoising architectures.
FBPConvNet was originally designed for sparse-view CT, \BLK{while we applied it to the low-dose CT case}; it is a U-Net like CNN \BLK{and we took} low-dose FBP images as input.
The neural network is trained so that the denoised versions of the input images closely match the high-quality reference images.
Traditional U-Net uses a multilevel decomposition, and a dyadic scale decomposition based on max pooling.
Similar to U-Net, FBPConvNet adopts multichannel filters to increase the capacity of the network.  

WavResNet is an interpretable framelet-based denoising neural network that employs contourlet transforms, \BLK{a concatenation layer, and a skip connection}. Contourlet transforms increase the input data size according to the number of transform levels, which can create memory bottlenecks during training. Hence, a patch-based training strategy is adopted.
WavResNet can be applied either with a feed-forward scheme or a recursive scheme \cite{WavResNet18}. We chose the feed-forward scheme in this paper.

\subsubsection{Unsupervised \BLK{MBIR Components}}
We adopt the weighted-least squares (WLS) data-fidelity term ${L(\A\x,\y) = \|\y-\A \x\|_{\W}^2}$,
where $\W\in\mathbb{R}^{N_d\times N_d}$ is a diagonal weighting matrix whose diagonal elements are \BLK{the estimated inverse variance of $y_i$ \cite{thibault:07:atd}}.
For the regularizer $\R(\x)$, we adopt 
a traditional EP regularizer $\R_{\textup{EP}}$ and a state-of-the-art ULTRA regularizer $\R_{\textup{ULTRA}}$. For the EP regularizer,
${\R_{\textup{EP}}(\x) = \sum_{j  =1}^{N_p} \sum_{k\in N_{j}}\kappa_{j} \kappa_{k} \varphi(x_j - x_k)}$, where $N_j$ is the size of the neighborhood, $x_j$ is the $j$th pixel of $\x$, $\kappa_j$ and $\kappa_k$ are analytically determined weights that encourage resolution uniformity \cite{cho:15:rdf}, and 
the potential function ${\varphi (t) \triangleq \delta^2(|t/\delta| - \log(1+|t/\delta|))}$ with $\delta>0$ being the EP parameter. 

PWLS-ULTRA pre-learns a union of sparsifying transforms from image patches. 
With the pre-learned transforms $\{\omg_k\}$, the regularizer $\R_{\textup{ULTRA}}(\x)$ for image reconstruction is
\begin{equation} \min_{\{\z_j, C_k\}} \sum_{k=1}^{K} \sum_{j\in{C_k}} \tau_j \left\{\|\omg_k\P_j\x-\z_j\|^2_2+\gamma^2\|\z_j\|_0\right\},
\label{eq:ultra_reg}
\vspace{-0.06in}
\end{equation}
where the operator $\P_j\in \mathbb{R}^{m\times N_p}$ extracts the $j$th patch of size $\sqrt{m}\times \sqrt{m}$ from $\x$, 
vector $\z_j\in \mathbb{R}^m$ denotes the sparse coefficients for the $j$th image patch, 
\BLUE{$C_k$ denotes the indices of all patches matched to the $k$th transform,}
$\{\tau_j\}$ are patch-based weights to encourage uniform spatial resolution or uniform noise in the reconstructed images \cite{pwls-ultra2018}, and $\gamma$ is a parameter controlling sparsity in the model. 
\subsection{Discussion of the SUPER Framework} \label{sec:interp}
Having described the SUPER Framework,
we now describe a few of its conceptual advantages.
SUPER unifies several distinct models---%
a supervised learning-based model, an unsupervised (iterative) model,
a physics-based forward model, and a statistical model of measurements and noise---%
in a common MBIR-type framework.
This combination affords \BLK{training} algorithms based on SUPER extra flexibility:
the supervised part can benefit from paired training data (e.g., low-dose and corresponding regular-dose images/measurements),
while the unsupervised part, e.g., based on ULTRA, can use \BLK{a few regular-dose} training images without corresponding measurements.
And, because of the 
physics-based forward models and statistical models,
the algorithm can perform well even when training data of any kind is scarce.
The relative importance of each of these models can be tuned simply by adjusting the corresponding scalar parameters
\BLUE{(see Sections~\ref{subsec:mu} and \ref{subsec:beta} in this manuscript)}.

Our training approach~\eqref{eq:super-all-sim} can also be viewed as optimization that alternates between two different \emph{modules}, \BLK{i.e., the supervised module and the unsupervised (iterative) module.
While the supervised module 
involves
layer-wise neural network weights to effectively remove noise and artifacts,
the unsupervised module could 
substantially optimize each image 
by incorporating various physical and image properties.}
\BLUE{The SUPER framework is closely related to the idea of plug-and-play~\cite{VBW_prior13}, with which various unsupervised and supervised regularizers can be plugged. Different from many plug-and-play unrolled methods that exclude explicit regularizers in the MBIR step (corresponding to the iterative module in SUPER) while implicitly regularizing the reconstruction with an additional denoising step using some neural networks (corresponding to the supervised module in SUPER)~\cite{admm-net2016,LEARN18,PnP-Stanley-17,3pADMM,schlemper2017deep,aggarwal2018modl,mardani2017recurrent}, SUPER allows to use explicit regularizers in the iterative module, including nonsmooth regularizers (e.g., based on the $\ell_0$ ``norm") and integer-valued variables (e.g., clusters with the ULTRA regularizer), and allows optimizing the MBIR cost with such explicit regularizers using many iterations 
during training and reconstruction. Moreover, rather than learning a common network for all layers (or iterations in terms of unrolled methods), our approach is akin to greedy layer-wise learning to circumvent the need to propagate gradients through a large number of iterations of MBIR (which is well-known to lead to vanishing gradients), and also to more readily handle regularizers incomptabile with gradient backpropagation. The layer-wise training also encourages the supervised networks in each SUPER layer to optimally denoise the images with certain noise levels (in that layer), which shares 
similarities with
layer-wise training in~\cite{Chun&etal:19MICCAI} and MBIR with varying regularization strength (and transforms/dictionaries) for varying noise or artifacts over iterations (sometimes called a continuation strategy)~\cite{ravishankar:16:tci,jacob:blindCS:13}.
}

\section{Algorithms}\label{sec:algorithm}
This section describes the algorithms to solve the training and reconstruction optimization problems in Section~\ref{sec:formulation}.
We first briefly introduce the unsupervised learning of a union of transforms~\cite{pwls-ultra2018} and then describe the proposed methods.
\subsection{Learning a Union of Sparsifying Transforms}
We pre-learn a union of transforms $\{\omg_k\}_{k=1}^K$ to effectively group and sparsify a training set of image patches by solving
\begin{equation}
\begin{aligned}
\min_{\{\omg_k,\,\Z_i,\,C_k \}} \sum_{k=1}^{K}& \sum_{i\in C_k} \left\{ \|\omg_k \X_i-\Z_i \|_2^2 +\eta^2\|\Z_i\|_0 \right\} \\
&+\sum_{k=1}^{K} \lambda_k Q(\omg_k),\quad \textup{s.t.} \quad \{C_k \} \in \mathcal{\bm{G}}.
\end{aligned}
\vspace{-0.07in}
\label{eq:train_ultra}
\end{equation}
where $\X_i\in\mathbb{R}^m$ denotes the $i$th vectorized (overlapping) image patch extracted from training images, $\Z_i\in\mathbb{R}^m$ is the corresponding transform-domain sparse approximation (with sparsity measured using the $\ell_0$ ``norm" that counts the number of nonzeros in a vector), parameter $K$ denotes the number of clusters, $C_k$ denotes the indices of all the patches matched to the $k$th transform, and the set $\mathcal{G}$ is the set of all possible partitions of $\{1,2,\dots,N'\}$ into $K$ disjoint subsets, with $N'$ denoting the total number of patches.
We use $K$ regularizers $Q(\omg_k)=\|\omg_k\|_F^2-\log|\det\omg_k|, 1\le k \le K$, which control the properties of each transform $\omg_k$, and prevent trivial solutions (e.g., matrices with zero or repeated rows).
We set the weights $\lambda_k=\lambda_0 \sum_{i \in C_k} \|\X_{i}\|^2_2$, where $\lambda_0$ is a constant \cite{pwls-ultra2018}.
We adopt an alternating algorithm for (\ref{eq:train_ultra}) that alternates between a \textit{transform update step} (solving for $\{\omg_k\}$) and  a \textit{sparse coding and clustering step } (solving for $\{\Z_i,\,C_k\}$), with closed-form solutions in each step~\cite{ravishankar:16:tci}.
As a patch-based unsupervised learning method, ULTRA typically only needs a few regular-dose training images to learn rich features. 

\subsection{SUPER Training \BLK{and Reconstruction} Algorithms} \label{sec:super-alg}
As stated in Section~\ref{sec:formulation}, we \BLK{train a sequence of supervised network parameters $\{\bm{\theta}^{(l)}\}_{l=1}^L$ by alternating between optimizing $\eqref{eq:super-all-sim}$ in a supervised manner to get $\bm{\theta}^{(l)}$ and optimizing \eqref{eq:SUPER-reconstruction} with iterative algorithms to obtain $\hat{\x}^{(l)}_{\bm{\theta}^{(l)}}(\y)$. When updating the network parameters, the network inputs are fixed to the most recent iterative reconstructions.
Specifically, training $\bm{\theta}^{(l)}$ in a single ($l$th) SUPER layer coincides with a conventional network training problem which can be solved by stochastic gradient descent (SGD) algorithms or Adam \cite{Adam15}. }
\noindent
\begin{minipage}{0.5\textwidth} 
\centering
\begin{algorithm}[H]  
	\caption{SUPER Training Algorithm}
	\label{alg: super-alg}
	\begin{algorithmic}[1]
		\Require~~\\
		 $N$ pairs of low-dose FBP images and corresponding regular-dose reference images \BLK{$\{(\hat{\x}^{(0)}_{\bm{\theta}^{(0)}}(\y_n), \x_n^*)\}_{n=1}^N$};\\
		Low-dose \BLK{sinograms} $\y_n$ and weights $\W_n$, $\forall n$;\\
		Unsupervised (iterative) module regularizer $\R$, e.g., $\R_{\textup{EP}}$ or $\R_{\textup{ULTRA}}$;\\
		number of SUPER training layers $L$, number of unsupervised (iterative) module iterations $I$, and number of inner iterations $P$ ($P$ is only used with $\R_{\textup{ULTRA}}$, and denotes the number of inner iterations in the image update step).
		\Ensure \BLK{A set of} layer-wise supervised (deep) model parameters \BLK{$\{\bm{\theta}^{(l)}\}_{l=1}^L$}.
		\setcounter{ALG@line}{0}
\makeatother
	    \For {$l=1,2,\cdots, L$}
		\State \textbf{(1) update $\bm{\theta}^{(l)}$ }: with fixed input \BLK{$\{\hat{\x}^{(l-1)}_{\bm{\theta}^{(l-1)}}(\y_n)\}_{n=1}^N$}, optimize \eqref{eq:super-all-sim} with SGD or Adam \cite{Adam15} to obtain $\bm{\theta}^{(l)}$; 
		\State \textbf{(2) update \BLK{$\hat{\x}^{(l)}_{\bm{\theta}^{(l)}}(\y_n)$}}: a) apply the updated network $G_{\bm{\theta^{(l)}}}(\cdot)$ to the previous layer reconstructions \BLK{$\{\hat{\x}^{(l-1)}_{\bm{\theta}^{(l-1)}}(\y_n)\}_{n=1}^N$}, i.e., obtain each \BLK{$G_{\bm{\theta^{(l)}}}(\hat{\x}^{(l-1)}_{\bm{\theta}^{(l-1)}}(\y_n))$};
		\State b) \BLK{update each image $\hat{\x}^{(l)}_{\bm{\theta}^{(l)}}(\y_n)$ by optimizing the PWLS cost in \eqref{eq:SUPER-reconstruction} with $G_{\bm{\theta^{(l)}}}(\hat{\x}^{(l-1)}_{\bm{\theta}^{(l-1)}}(\y_n))$ as the initial image, and using $I$ iterations of the relaxed LALM algorithm~\cite{nien:16:rla} for $\R_{\textup{EP}}$ based cost, or $I$ alternations and $P$ inner iterations \BLK{(with the relaxed LALM algorithm)} for $\R_{\textup{ULTRA}}$ based cost \cite{pwls-ultra2018}.}
		\EndFor	
	\end{algorithmic} 
\end{algorithm}	
\end{minipage}

\BLK{In updating $\hat{\x}^{(l)}_{\bm{\theta}^{(l)}}(\y)$, we adopt the relaxed LALM \cite{nien:16:rla} algorithm for both EP based and ULTRA based SUPER reconstruction costs. Particularly, when using the ULTRA regularizer, we alternate several times between updating $\x$ and $\{\z_j,C_k\}$. In the \emph{image} ($\x$) \emph{update step}, we fix the sparse coefficients $\{\z_j\}$ and cluster assignments $\{C_k\}$ and solve
\begin{equation*}\label{eq:unsup-ultra}
\begin{aligned}
&\hat{\x}^{(l)}_{\bm{\theta}^{(l)}}(\y) = \argmin_{\x} \|\y - \A \x\|_{\W_n}^2 + \\ 
&\beta \sum_{k=1}^{K} \sum_{j\in{C_k}} \tau_j \left\{\|\omg_k\P_j\x-\z_j\|^2_2\right\}
 + \mu\|\x - G_{\bm{\theta}^{(l)}}(\hat{\x}^{(l-1)}_{\bm{\theta}^{(l-1)}}(\y))\|_2^2,
\end{aligned}
\vspace{-0.05in}
\end{equation*}
via the efficient} relaxed LALM algorithm \cite{nien:16:rla}. \BLK{The initial image in the image update step of each SUPER layer is $G_{\bm{\theta}^{(l)}}(\hat{\x}^{(l-1)}_{\bm{\theta}^{(l-1)}}(\y))$.} We then fix the updated $\x$ and jointly optimize $\{\z_j\}$ and $\{C_k\}$ (\textit{sparse coding and clustering step}). In the resulting subproblem, the sparse vectors $\z_j$ can be replaced with their optimal values ${\z_j = H_\gamma(\omg_k\P_j \x)}$ in the cost, where $H_\gamma(\cdot)$ is a hard-thresholding function that sets vector elements with magnitudes smaller than $\gamma$ to $0$, and leaves other entries unchanged. The optimal clustering is then obtained patch-wise as
\begin{equation*}\label{eq:k-update}
    \begin{aligned}
&  \hat{k}_j =  \mathop{\arg\min}_{1\le k \le K} \|\omg_k\P_j\x-H_\gamma(\omg_k\P_j\x)\|^2_2+\gamma^2\|H_\gamma(\omg_k\P_j\x)\|_0,
    \end{aligned}
    \vspace{-0.06in}
\end{equation*}
and the corresponding optimal sparse coefficients are ${\hat{\z}_j = H_\gamma(\omg_{\hat{k}_j}\P_j \x)}$ \cite{pwls-ultra2018}. 

The SUPER learning algorithm based on \eqref{eq:super-all-sim} is illustrated in
\textbf{Algorithm}~\ref{alg: super-alg}.




\BLK{The SUPER reconstruction algorithm in each single ($l$th) SUPER layer is the same as that for updating $\hat{\x}^{(l)}_{\bm{\theta}^{(l)}}(\y)$ in the training, while using the trained $\bm{\theta}^{(l)}$ for the supervised penalty term. }

\section{Experiments and Discussions}\label{sec:experiments}
In this section, we first describe the experimental setup, training procedures, and evaluation metrics.
Then, we present the results for SUPER learning with different combinations of supervised and unsupervised components, and compare these results with multiple standalone supervised and iterative methods from the literature. 
Finally, we present several experiments to explore how the SUPER model works, including analysis of the impact of the supervised and unsupervised components in SUPER on the reconstruction performance and the convergence behavior of the proposed algorithms.

\subsection{Experimental Setup}
\subsubsection{Data and Imaging system} \label{sec:data-gen}
We used Mayo Clinics dataset established for “the 2016 NIH-AAPM-Mayo Clinic Low Dose CT Grand Challenge” \cite{McC-Mayo} in our experiments.
We randomly selected 520 slices from data \BLK{(of 3~mm thickness)} for six out of ten patients, from which 500 slices were used for training and 20 slices were used for  validation.
\BLUE{We extracted 18 regular-dose CT images from the 500 training slices to pre-train a union of five sparsifying transforms used for the ULTRA regularizer.
}
We tested on 20 slices that were randomly extracted from the remaining four patients' data.
We simulated low-dose CT sinograms $\y$ from the provided regular-dose images $\x^*$ using the Poisson-Gaussian noise model \cite{pre-post-log,ye2019TMI}:
\begin{equation*}
\centering
y_{i}=- \log \left( I_0^{-1}\max\big(\textup{Poisson}\{ I_0 e^{-[\A\x^*]_i}\} + \mathcal{N}\{0,\,\sigma^2\},\epsilon \big) \right), 
\end{equation*} 
where the number of incident photons per ray is $I_0=10^4$, the Gaussian noise variance is $\sigma^2=25$, and $\epsilon$ is a small positive number to avoid negative measurement data when taking the logarithm.
We used \BLK{the Michigan Image Reconstruction Toolbox\footnote{Jeffrey A Fessler, available at \url{http://web.eecs.umich.edu/~fessler/irt/irt}.} to construct} fan-beam CT geometry with 736 detectors~$\times$~1152 regularly spaced projection views, and a no-scatter monoenergetic source. 
\BLK{The width of each detector column is 1.2858~mm, the source to detector distance is 1085.6~mm, and the source to rotation center distance is 595~mm. We reconstruct}
images of size $512\times 512$ with the pixel size being 0.69~mm $\times$ 0.69~mm.

\subsubsection{Parameter Settings for Proposed and Compared Methods}\label{sec:para_method}
In the SUPER model, we adopted two architectures for the deep network, namely FBPConvNet \cite{jin:17:dcn} and WavResNet \cite{WavResNet18}. Specifically, during SUPER training, we ran $4$ epochs (over the training set) of the SGD optimizer for the FBPConvNet module in each SUPER layer to reduce overfitting risks. For the WavResNet case, we chose the faster feed-forward neural network architecture in \cite{WavResNet18} and employed contourlet transforms with 15 channels on input images. We used $256\times 256 \times 15$ wavelet domain patches to train WavResNet.
During SUPER training, we ran $50$ epochs of the SGD optimizer to update the WavResNet weights each time to capture sufficient wavelet domain features. 
The SUPER models with both WavResNet and FBPConvNet networks were trained with batch size 1 in each SUPER layer. 
When running iterative reconstruction methods in our algorithms, we used $20$ iterations (each time) of the iterative algorithms for \BLK{the EP regularization case, and $20$ alternations with 5 inner iterations (each time) for the} ULTRA regularization case.
For the EP regularizer, we set $\delta=20$ and regularization parameter $\beta=2^{15}$; for the ULTRA regularizer, we 
set the regularization parameters $\beta=5\times 10^{3}$ and $\gamma=20$ during training and reconstruction. Since the parameter $\mu$ controls the balance between the supervised and unsupervised modules in our formulation, we empirically set $\mu$ as $5\times 10^4$ and $5\times 10^5$ for EP and ULTRA based SUPER methods, respectively.

We compared our proposed model with the standalone supervised methods, i.e., FBPConvNet and WavResNet, and standalone unsupervised methods, i.e., PWLS methods with EP regularizer (PWLS-EP) and ULTRA regularizer (PWLS-ULTRA), respectively. We ran $100$ epochs and $200$ epochs of SGD for training standalone FBPConvNet and WavResNet, respectively, to sufficiently learn image features with low overfitting risks (also evaluated on validation data).
For the standalone PWLS-EP iterative approach, we set $\beta=2^{16}$ and $\delta=20$, and ran 100 iterations of the relaxed LALM algorithm \cite{nien:16:rla} to obtain convergent images.
For the standalone PWLS-ULTRA method, we set $\beta=10^4$ and $\gamma=25$, and ran 1000 alternations between the image update step (with $5$ relaxed LALM iterations), and sparse coding and clustering step. 
\BLUE{Apart from comparing with standalone supervised and unsupervised methods that we used for SUPER, we also compared SUPER methods with a state-of-the-art deep learning-based post-processing method MAP-NN~\cite{shan2019competitive}. We extracted image patches from the training data used for SUPER and created a training set consisting of similar amounts of image patches as used in~\cite{shan2019competitive} for MAP-NN training. 
Since MAP-NN involves output clipping in each of its modules, we increased the clipping window to $[0,2400]$~HU (shifted HU before normalization) such that most image structures can be preserved while the denoising capability is maintained. 
We also compared SUPER methods with a recent plug-and-play ADMM-Net method~\cite{PnP-Stanley-17}, which unrolls the ADMM algorithm and replaces the proximal operator inside the algorithm with off-the-shelf denoising methods. 
We used trained WavResNets as the off-the-shelf denoisers for the ADMM-Net method, and used the same training setup as for WavResNet in SUPER methods. The MBIR step in ADMM-Net only involves a network-based regularizer, which is similar to SUPER with $\beta=0$ while $\mu\neq 0$.
We set the regularization parameter for ADMM-Net as $1 \times 10^6$ and ran $20$ iterations for the MBIR step in each ADMM-Net iteration, which worked well in our experiments.}
\MG{We used the default data augmentation tool in WavResNet~\cite{WavResNet18} to train all the WavResNet-related networks in our experiments. There is no data augmentation in MAP-NN by default~\cite{shan2019competitive}, and since the amount of training data we used satisfies the required amount in FBPConvNet and MAP-NN, we did not use data augmentation for methods related to these two (to reduce the training time).}
All methods used the FBP reconstruction as initialization or network input.

\subsubsection{Evaluation metrics}
We chose root mean square error (RMSE), signal-to-noise ratio (SNR), and structural similarity index measure (SSIM) \cite{xu:12:ldx} to quantitatively evaluate the performance of reconstruction methods.
The RMSE in Hounsfield units (HU) is defined as $\textup{RMSE}=\sqrt{\sum_{j=1}^{N_p}(\hat{\x}_j-\x^*_j)^2/N_p}$, where $\x^*_j$ is the $j$th pixel of the reference regular-dose image $\x^*$, $\hat{\x}_j$ is the $j$th pixel of the reconstructed image $\hat{\x}$, and $N_p$ is the number of pixels.
The SNR in decibels (dB) is defined as $\textup{SNR}=10\log_{10} \frac{\|\x^*\|^2}{\|\hat{\x}-\x^*\|^2}$.

\subsection{Numerical Results and Comparisons}
\begin{figure}[!htp]\leftskip30pt
    \centering
\includegraphics[width=0.48\textwidth]{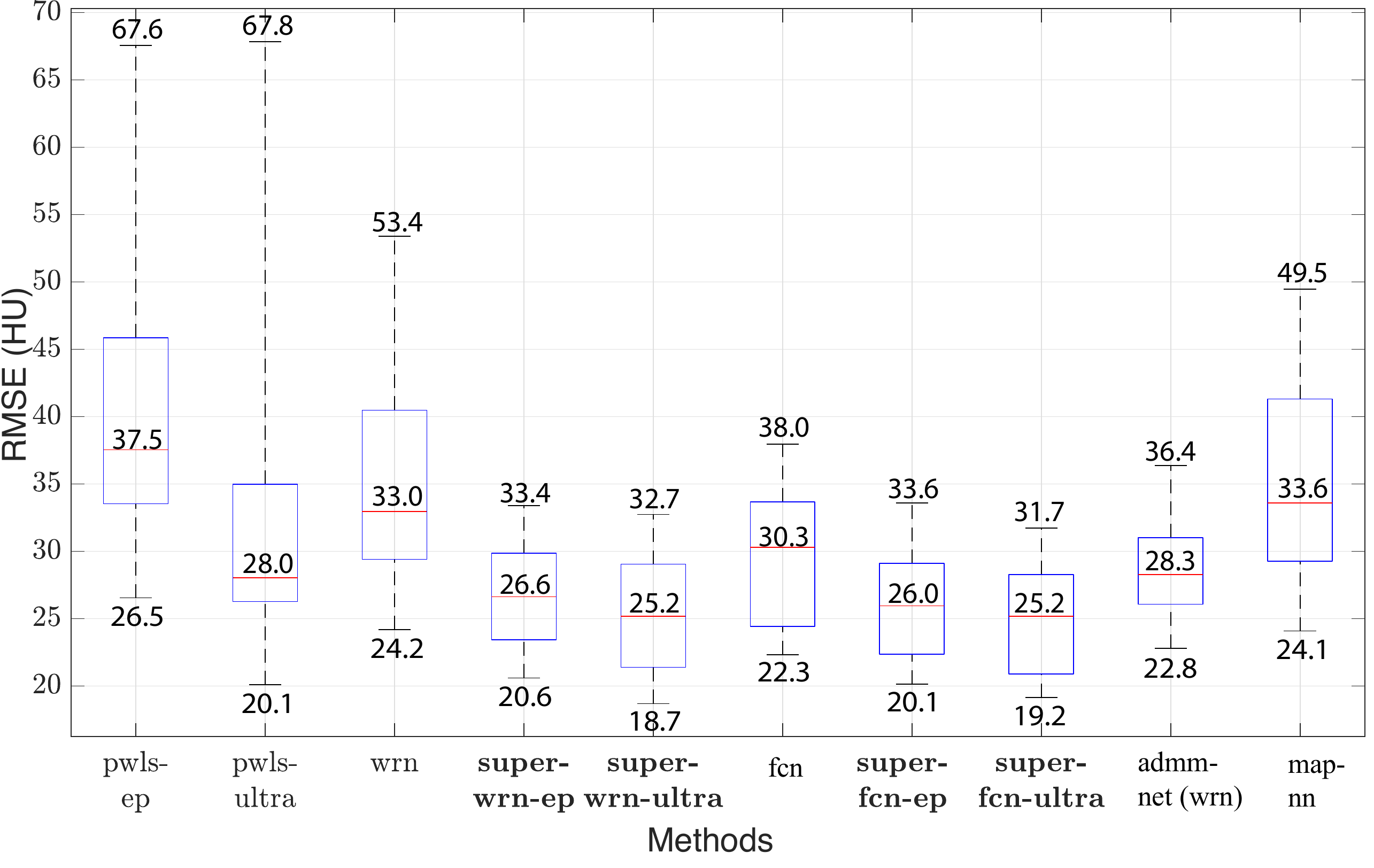}
   	 \vspace{-0.1in}
    \caption{RMSE spread (shown using box plots) over $20$ test cases using different methods. Here, ``wrn" stands for WavResNet and ``fcn" stands for FBPConvNet. Each box plot for a method describes the statistics of RMSE values over the 20 test slices: the central red line indicates the median; the bottom and top edges of the boxes indicate the 25th and 75th percentiles, respectively; and the whiskers represent the extreme values. \BLUE{We marked the median values and the extreme values for each method in the box plot.} The proposed SUPER combinations \BLUE{(highlighted with bold in x-axis)} outperform both their supervised and unsupervised counterparts. \BLUE{They also outperform the competing plug-and-play ADMM-Net method with WavResNet denoiser (admm-net (wrn)) and a state-of-the-art post-processing method MAP-NN.}}
    \label{fig:wrs_fcn_box}
    \vspace{-0.02in}
\end{figure}

Fig.~\ref{fig:wrs_fcn_box} shows box plots of RMSE values of (test) reconstructions with \BLUE{SUPER methods and their constituent supervised and unsupervised methods, as well as an external learned iterative reconstruction method, plug-and-play ADMM-Net~\cite{PnP-Stanley-17} with WavResNet denoiser (ADMM-Net (WRN)) and a state-of-the-art post-processing method MAP-NN~\cite{shan2019competitive}.}
From this figure, we observe that the proposed SUPER methods, namely SUPER-WRN-EP, SUPER-WRN-ULTRA, SUPER-FCN-EP and SUPER-FCN-ULTRA, decrease RMSE values dramatically in test slices compared to standalone iterative or deep learning methods such as PWLS-EP, PWLS-ULTRA, WavResNet, and FBPConvNet. Specifically, SUPER methods can effectively handle highly corrupted scans for which either the standalone supervised methods or unsupervised methods may have limited performance. For example, the SUPER methods further reduce the RMSE of the most corrupted (FBP) image (slice 150 from patient L067) by approximately 35~HU compared with the standalone PWLS-EP or PWLS-ULTRA method; the RMSE value is around 20~HU lower than the result using the feed-forwarded WavResNet method, and 6~HU lower than using the FBPConvNet scheme. The reconstructions of this most corrupted test image using different methods are shown in the supplement \BLUE{(Fig.~11).}
It is also obvious from Fig.~\ref{fig:wrs_fcn_box} that the methods exploiting SUPER learning reduce the interquartile ranges and shrink the gap between the maximum and minimum RMSE values. This indicates the robustness of the proposed schemes and their generalization for reconstructing images with various noise or artifacts levels. 
The SNR and SSIM comparisons also reflect the robustness of the proposed schemes. We show the box plots for these two metrics in the supplement \BLUE{(Figs.~8~and~9)}.

The proposed methods are also robust to the choices of the supervised and unsupervised/analytical parts. 
For example, the reconstruction performance of supervised learning schemes may be affected by the network architecture, number of training samples, hyper parameter tuning, etc. In our experiments, we used two distinct networks learned in a supervised manner, WavResNet \cite{WavResNet18} and FBPConvNet \cite{jin:17:dcn}. 
Although the performance of WavResNet is worse than FBPConvNet due to the architecture differences (and possibly the somewhat limited training data), the proposed SUPER methods based on these two methods have comparable reconstruction metrics. Among the unsupervised learning or analytical prior-based methods, although the PWLS-EP method is substantially inferior to the PWLS-ULTRA method, the performance (i.e., RMSE, SNR, SSIM) gap between the EP based SUPER and ULTRA based SUPER is much smaller than that between the standalone PWLS-EP and PWLS-ULTRA methods.
ULTRA-based SUPER schemes do outperform EP-based schemes overall, indicating that unsupervised learning approaches provide benefits over conventional mathematical priors.
\BLUE{Comparing WavResNet-based SUPER methods with the ADMM-Net method, both the extreme RMSE values and the median RMSE of the latter are higher than those of the former ones. As there is no unsupervised regularizer in the ADMM-Net framework, this phenomenon indicates the benefit of SUPER that incorporates the unsupervised regularizer, e.g., EP or ULTRA, in the MBIR module.
The numerical results of MAP-NN is similar to that of WavResNet, which under-performs the SUPER methods. MAP-NN also underperforms FBPConvNet, which may be caused by output clippings in MAP-NN such that pixel values outside the clipping window ($[0,2400]$~HU) are set to either $0$~HU (for values smaller than $0$~HU) or $2400$~HU (for values larger than $2400$~HU).
}

\begin{table}[!t]
	\centering
	\caption{Mean metrics of reconstructions of 20 test slices.}
	\vspace{-0.05in}
	\scalebox{0.82}{\begin{tabular}{cccc}
		\toprule
		\textbf{Method} &
		\shortstack{\textbf{ RMSE (HU)}}&
		\textbf{ SNR (dB)}&
		\textbf{ SSIM} \\  \midrule
		FBP & 128.8 & 16.7 & 0.347 \\ 
		PWLS-EP &41.4 &25.4 &0.673 \\ 
		PWLS-ULTRA &32.4 &27.8 &0.716 \\ 
		WavResNet   &35.3   &27.1     &0.646   \\ 
		SUPER-WRN-EP&  \textbf{26.7}	&\textbf{29.1}	&	\textbf{0.738}   \\ 
		SUPER-WRN-ULTRA &\textbf{25.4}	&\textbf{29.5} &	\textbf{0.744}\\ 
		FBPConvNet & 29.2 & 28.2 & 0.688 \\
		SUPER-FCN-EP  & \textbf{26.0} & \textbf{29.3} & \textbf{0.740}\\ 
		SUPER-FCN-ULTRA& \textbf{25.0} & \textbf{29.7} & \textbf{0.748}\\
		\BLUE{ADMM-Net (WRN)}&\BLUE{28.6} &\BLUE{28.4} &\BLUE{0.702}\\
		\BLUE{MAP-NN}&\BLUE{35.1} &\BLUE{26.7} &\BLUE{0.660}\\
		\bottomrule
	\end{tabular}}
	\label{tab:my_overview}
\vspace{-0.25in}
\end{table}
The superiority of SUPER learning is also reflected in the averaged (over all test slices) reconstruction quality metrics shown in Table~\ref{tab:my_overview}.
In Table~\ref{tab:my_overview}, we observe that among WavResNet based methods, SUPER-WRN-ULTRA achieves the best RMSE, SNR and SSIM values. Both SUPER-WRN-EP and SUPER-WRN-ULTRA provide significantly improved performance compared to the standalone (WavResNet, EP, ULTRA) components\BLUE{, and outperformed the competing ADMM-Net method.}
In particular, SUPER-WRN-ULTRA achieves 9.8~HU, 7.0~HU and 3.2~HU better average RMSE over \BLUE{its constituent standalone supervised method (WavResNet), unsupervised method (PWLS-ULTRA), and the plug-and-play ADMM-Net method, respectively;}
SUPER-WRN-EP \BLUE{improves the average RMSE by 8.6~HU and 15.8~HU respectively compared with its constituent standalone supervised and unsupervised methods (WavResNet and PWLS-EP), and improves by 1.9~HU compared with the plug-and-play ADMM-Net.}
A similar trend is observed with FBPConvNet-based SUPER methods. Specifically, SUPER-FCN-ULTRA achieves average RMSE improvements of 4.3~HU and 7.5~HU over \BLUE{its constituent standalone supervised method (FBPConvNet) and unsupervised method (PWLS-ULTRA), respectively;}
and SUPER-FCN-EP achieves average RMSE improvements of 3.3~HU and 15.5~HU respectively, compared to \BLUE{its constituent standalone FBPConvNet and PWLS-EP methods.}

\subsection{Visual Results and Comparisons}
\begin{figure*}[!t]
	\centering	
	\includegraphics[width=0.8\textwidth]{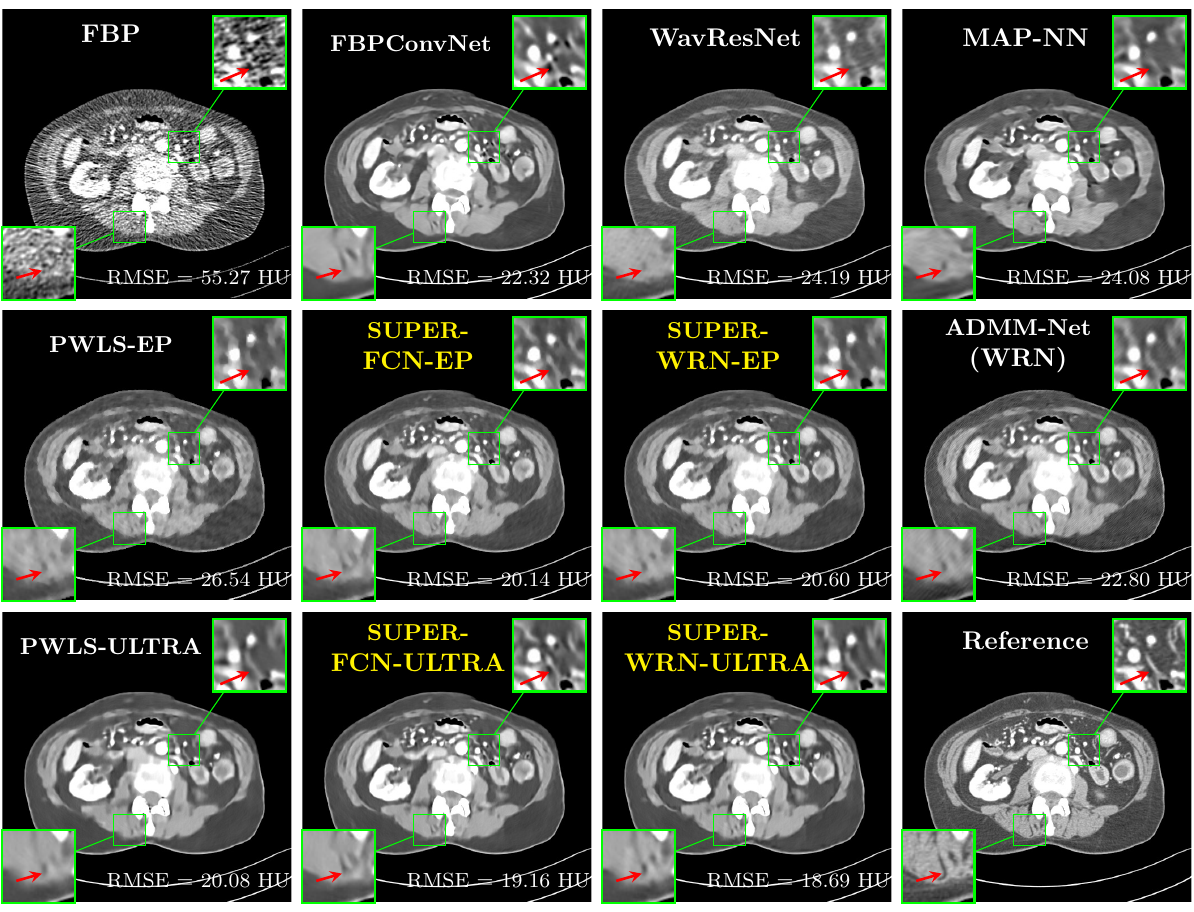}
	\caption{\BLUE{Reconstructions of slice~100 from patient L192 using various methods. The display window is [800 1200]~HU.}}
	\label{fig:super-all-L192s100}
	\vspace{-0.15in}
\end{figure*}

Fig.~\ref{fig:super-all-L192s100} shows a test example reconstructed using various methods.
We observe that PWLS-EP reduces the severe noise and streak artifacts observed in the low-dose FBP images, and the transform learning-based method PWLS-ULTRA further suppresses noise and reconstructs more details of the image such as the zoom-in areas. However, both methods have some blurry artifacts. The standalone FBPConvNet method heavily removes noise and streak artifacts, while introducing several artificial features \BLK{(e.g., feature indicated by the arrow in the top-right box in Fig.~\ref{fig:super-all-L192s100})}. WavResNet denoises the image without introducing artifical features, but still retains some streaks \BLK{around image boundaries and blurs some details (e.g., feature indicated by the arrow in the bottom-left box in Fig.~\ref{fig:super-all-L192s100})}.
\BLUE{The state-of-the-art MAP-NN method performs slightly better than WavResNet in terms of suppressing streak artifacts, while it still loses some details as indicated in the zoomed regions.
The competing plug-and-play unrolled method---ADMM-Net with WavResNet denoiser---outperforms the standalone WavResNet method, but still has some streak artifacts and blurred details.}
Compared to these methods, the proposed SUPER methods (SUPER-WRN-EP, SUPER-WRN-ULTRA, SUPER-FCN-EP, and SUPER-FCN-ULTRA)
improve the reconstruction quality in terms of removing noise and artifacts, and recovering details more precisely. 
\BLUE{Two other} example \BLUE{comparisons are} included in the supplement \BLUE{(Fig.~10 and Fig.~11}).

Fig.~\ref{fig:WRN_ultra_lyrs-L310s100} illustrates the image evolution over SUPER layers (i.e., with evolving network weights in the iterative reconstruction process) for one test case, when using SUPER-WRN-ULTRA. It is apparent that in the early SUPER layers, the proposed SUPER-WRN-ULTRA method mainly removes noise and artifacts, while later SUPER layers mainly reconstruct details such as the bone structures shown in the zoom-in box. A similar behaviour is observed with FBPConvNet-based SUPER methods, which are shown in the supplement (\BLUE{Figs.~13~and~14}).
    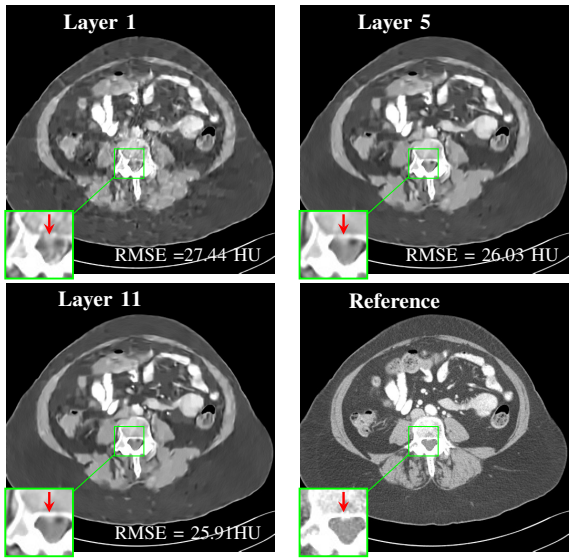
\begin{figure}[!htp]
    \centering 
    \vspace{-0.05in}
   		\begin{subfigure}{1\textwidth}\leftskip10pt
		\scalebox{0.9}{
		\centering 
		\begin{tikzpicture}
			\begin{scope}
			[spy using outlines={rectangle,green,magnification=2.3,size=10mm, connect spies}]
			\node {\includegraphics[width=0.22\textwidth]{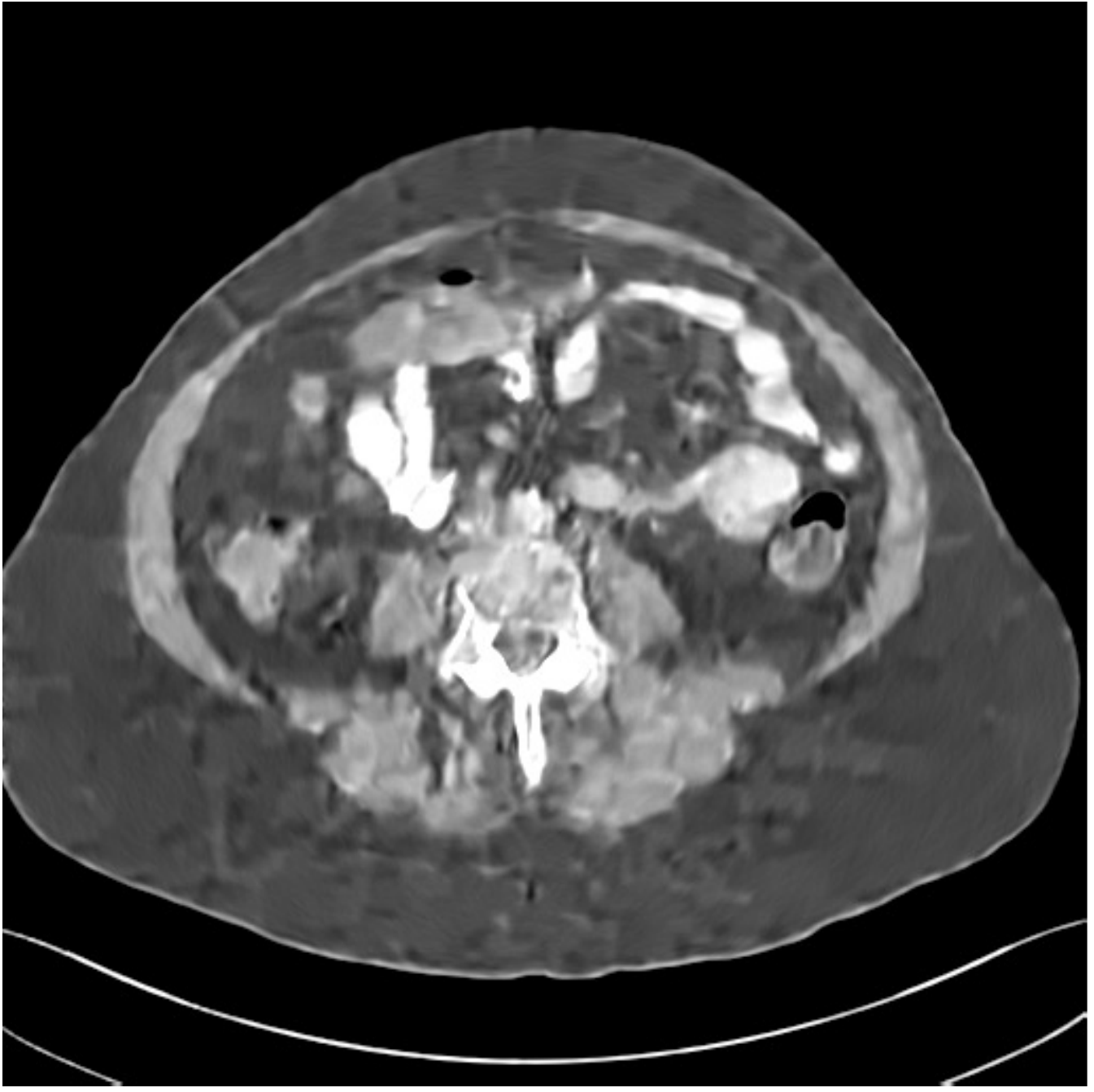} };
	        \spy on (-0.17,-0.35) in node [left] at (-1,-1.55);
	        \end{scope}
	 \draw [->,>=stealth,red,line width=1pt] (-1.35,-1.1) -- (-1.35,-1.4);
			\node [align = center,white, font=\bf] at (-0.6,1.72) {\small Layer 1};
	\node [white, font=\footnotesize] at (0.7,-1.7) {RMSE =27.44~HU};
	\end{tikzpicture}
	\begin{tikzpicture}
			\begin{scope}
			[spy using outlines={rectangle,green,magnification=2.3,size=10mm, connect spies}]
			\node {\includegraphics[width=0.22\textwidth]{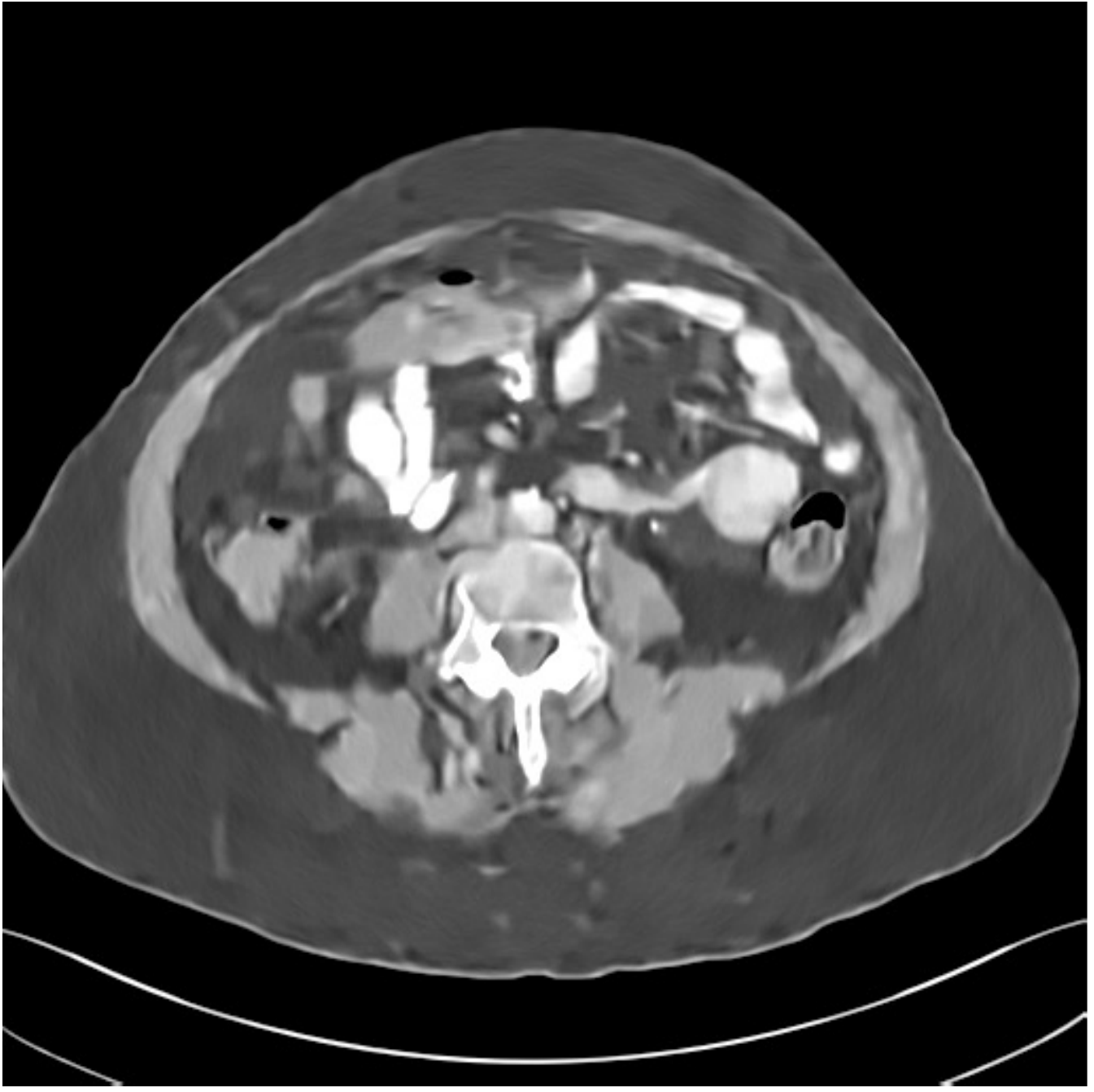} };
	        \spy on (-0.17,-0.35) in node [left] at (-1,-1.55);
	            \end{scope}
	 \draw [->,>=stealth,red,line width=1pt]  (-1.35,-1.1) -- (-1.35,-1.4);
			\node [align = center,white, font=\bf] at (-0.6,1.72) {\small Layer 5};
	\node [white, font=\footnotesize] at (0.7,-1.7) {RMSE = 26.03~HU};
	\end{tikzpicture}}
	\end{subfigure}
	\vfil \vspace{-0.04in}
	\begin{subfigure}{1\textwidth}\leftskip10pt
		\scalebox{0.9}{
		\centering 
		\begin{tikzpicture}
			\begin{scope}
			[spy using outlines={rectangle,green,magnification=2.3,size=10mm, connect spies}]
			\node {\includegraphics[width=0.22\textwidth]{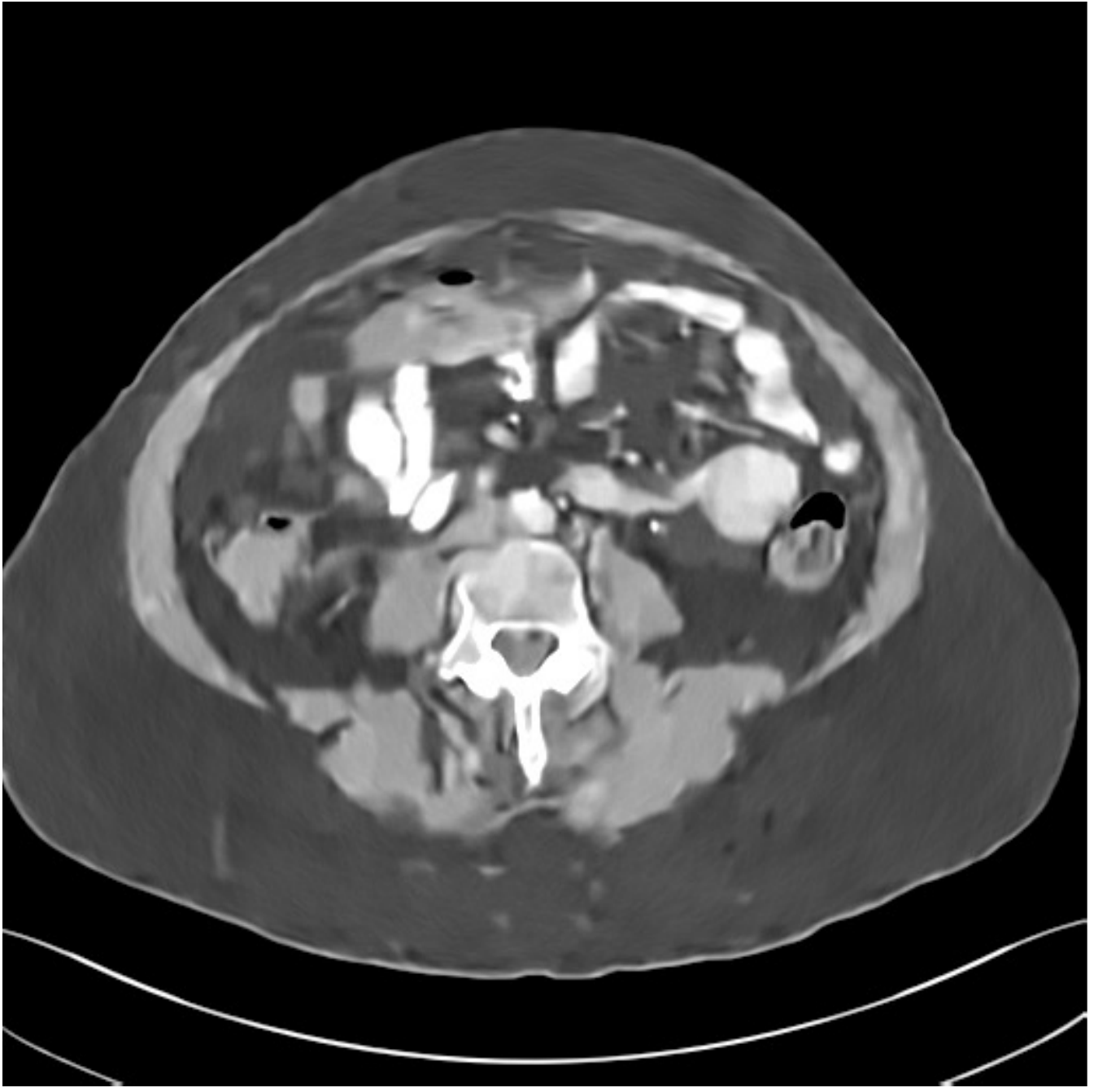} };
	         \spy on (-0.17,-0.35) in node [left] at (-1,-1.55);
	             \end{scope}
	 \draw [->,>=stealth,red,line width=1pt] (-1.35,-1.1) -- (-1.35,-1.4);
			\node [align = center,white, font=\bf] at (-0.6,1.72){\small Layer 11};
	\node [white, font=\footnotesize] at (0.7,-1.7) {RMSE = 25.91HU};
	\end{tikzpicture}
	\begin{tikzpicture}
			\begin{scope}
			[spy using outlines={rectangle,green,magnification=2.3,size=10mm, connect spies}]
			\node {\includegraphics[width=0.22\textwidth]{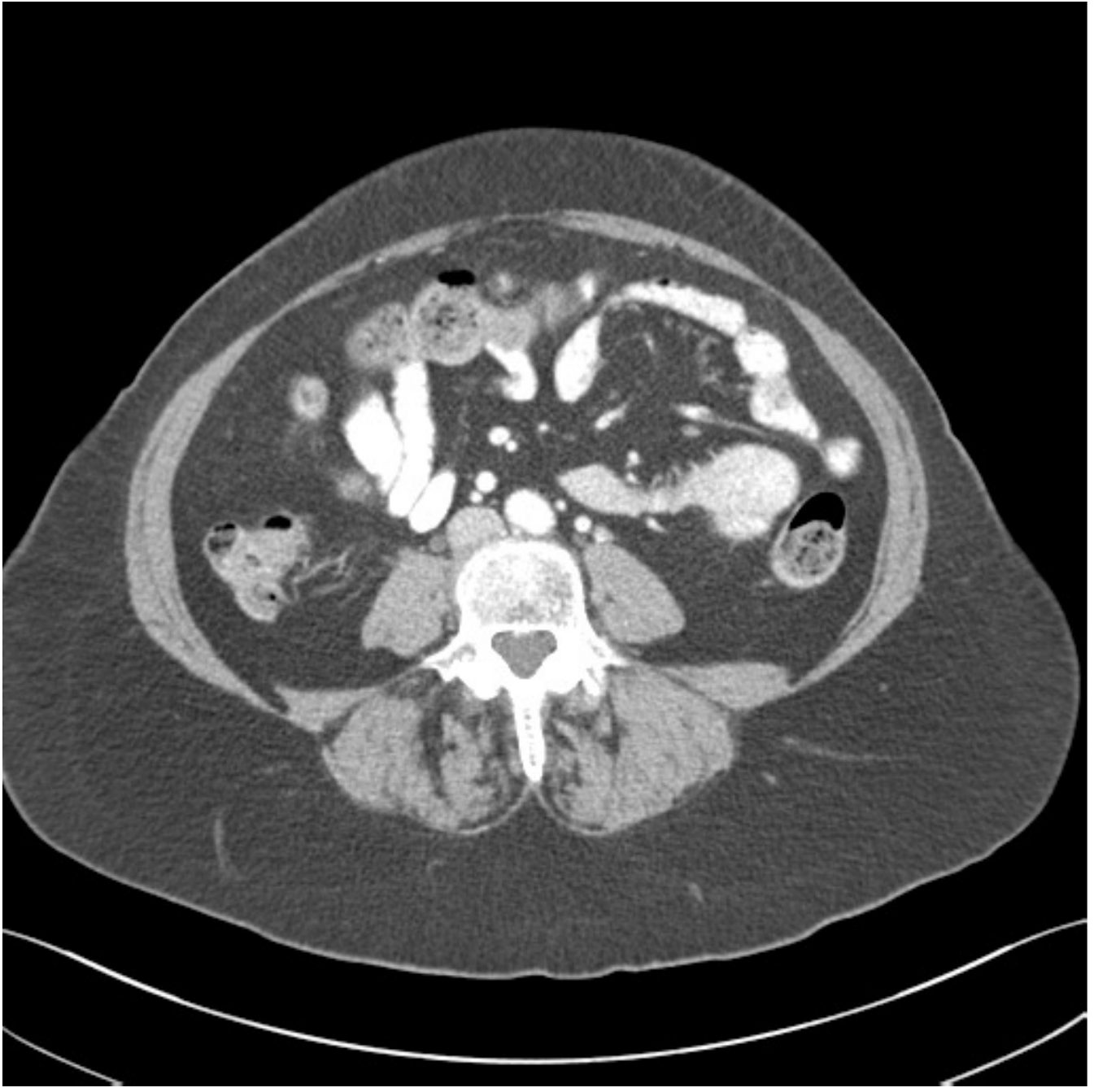} };
	         \spy on (-0.17,-0.35) in node [left] at (-1,-1.55);
	             \end{scope}
	 \draw [->,>=stealth,red,line width=1pt]  (-1.35,-1.1) -- (-1.35,-1.4);
			\node [align = center,white, font=\bf] at (-0.6,1.72) {\small Reference};
	\end{tikzpicture}}
	\end{subfigure}
	\vspace{-0.05in}
    \caption{Image evolution over SUPER layers using the SUPER-WRN-ULTRA method. RMSE values are also indicated.}
    \label{fig:WRN_ultra_lyrs-L310s100}
    \vspace{-0.25in}
\end{figure}

\subsection{Effect of Regularization Parameter $\mu$ in SUPER Models}
\label{subsec:mu}
Here, we study further the effect of the supervised learning-based regularizer weight $\mu$ on reconstruction performance. 
Table~\ref{tab:WRN-mu} shows the average RMSE (over test slices) of reconstructions for different choices of $\mu$ during training and during reconstruction. 
\BLUE{The $\mu=0$ case here corresponds to not having an explicit network-based regularizer term, but the iterative reconstruction algorithm is re-initialized with $\x = G_{\bm{\theta}^{(l)}} (\hat{\x}^{(l-1)}_{\bm{\theta}^{(l-1)}}(\y))$
in each SUPER layer.
This helps move the MBIR estimator towards a decent optimum when the cost of MBIR is not convex (with local optima), such as with ULTRA, or when the convex MBIR cost such as EP is partially optimized (i.e., not until algorithm convergence). Thus, this is different from just a conventional standalone PWLS algorithm run until convergence, wherein learned networks are not involved.
}
Our experimental results in Table~\ref{tab:WRN-mu} show that using the supervised learning-based regularizer in iterative reconstruction (i.e., $\mu \neq 0$) provides the best reconstruction performance. For example, in the SUPER-WRN-EP case, using $\mu=5\times 10^4$ in both training and testing leads to around 0.6~HU lower RMSE than using $\mu=0$ in training and testing (and using supervised learned network re-initializations \cite{super-19}). 
In the SUPER-WRN-ULTRA case, using $\mu=5\times 10^5$ in training and testing improves the RMSE by 0.1~HU compared to the aforementioned $\mu=0$ setting. 
Another observation is that using the same $\mu$ during training and testing usually works better than using mismatched $\mu$ values.
There is an exception when $\mu=0$ is used during training. In this case, using an explicit network regularizer with appropriate (positive) weighting at reconstruction time works better, i.e., it is better to work with the proposed combined priors during reconstruction.
In the SUPER-WRN-EP and SUPER-WRN-ULTRA cases, the mean RMSE over 20 test slices is 0.3 HU better with appropriate nonzero $\mu$ at testing time compared to the $\mu = 0$ setting used during training.
\BLUE{Section VII.E} in the supplement shows a similar behavior for FBPConvNet-based SUPER models. 
 \begin{table}[!htbp]
		\caption{Mean reconstruction RMSE (HU) over 20 test slices using different $\mu$ values during training/testing in WavResNet-based SUPER.}
		\vspace{-0.05in}
	\begin{subtable}[htbp]{0.5\textwidth}
		\centering
	\caption{SUPER-WRN-EP} \vspace{-0.1in}
\scalebox{0.8}{\begin{tabular}{c|c c c}
	\toprule
		\diagbox{Train}{Test}&\shortstack{$\mu = 0$}&\shortstack{$\mu = 5\times 10^4$}  &\shortstack{$\mu = 1\times 10^6$}  \\ 
    \midrule
	$\mu = 0$& 27.3	& \textbf{27.0}
	& 78.7
	\\ 
	$\mu = 5\times 10^4$& 27.8 	&  \textbf{26.7}
	&45.3
	\\ 
	$\mu = 1\times 10^6$&31.2  	&  30.3
	& \textbf{26.6}\\	\bottomrule
\end{tabular} }
\label{tab:WRN-EP-mu}
\end{subtable}
\hfil
	\begin{subtable}[htbp]{0.5\textwidth}
		\centering
	\caption{SUPER-WRN-ULTRA} \vspace{-0.1in}
\scalebox{0.8}{\begin{tabular}{c|c c c}
\toprule
	\diagbox{Train}{Test}&\shortstack{$\mu = 0$}&\shortstack{$\mu = 5\times 10^5$}  &\shortstack{$\mu = 1\times 10^8$}  \\ 
	\midrule
		$\mu = 0$& 25.5	& \textbf{25.2}
		& 44.5
		\\ 
		$\mu = 5\times 10^5$& 26.0 	& \textbf{25.4}
		& 41.2
		\\ 
		$\mu = 1\times 10^8$&29.6  	& 28.3
		& \textbf{26.3}\\ \bottomrule
\end{tabular} }
\end{subtable}
\label{tab:WRN-mu}
\vspace{-0.15in}
\end{table}


Fig.~\ref{fig:ablation} plots the mean RMSE values over the number of SUPER layers for various choices of (common) $\mu$ during training and testing.
The RMSE values converge quickly in all cases, with nonzero values of $\mu$ leading to lower RMSE values than $\mu=0$ (which concurs with Table~\ref{tab:WRN-mu}). ULTRA-based SUPER especially achieves bigger drops in RMSE in early SUPER layers compared to EP-based SUPER.



\subsection{\BLUE{Effect of Regularization Parameter $\beta$ in SUPER Models}}\label{subsec:beta}
\BLUE{We investigated the effect of the unsupervised regularization parameter $\beta$ in SUPER by taking the SUPER-FCN-ULTRA method as an example. We fixed the regularization parameter $\mu=5\times 10^5$ for ULTRA-based SUPER methods. Table~\ref{tab:fcn-beta} shows mean RMSE results of SUPER-FCN-ULTRA using different $\beta$ values during training and testing. We notice that for the cases where $\beta\neq 0$ during training,
reconstructing images in the testing set with the same $\beta$ value as used during training achieved the best RMSE values. 
Training and testing with nonzero $\beta$ values (e.g., $5 \times 10^{3}$) achieved the best mean RMSE values overall in Table~\ref{tab:fcn-beta}. 
Moreover, when SUPER is trained with $\beta=0$, which means no unsupervised learning-based prior was involved during training, one can improve the reconstruction quality by selecting a proper $\beta$ value during the testing stage. This suggests the benefits of including the unsupervised learned union of transforms prior with a proper weight during testing time.
For example, training SUPER with $\beta=0$ and then testing with $\beta=5\times 10^3$ achieved around 17~HU RMSE improvement compared to testing with $\beta=0$.
}

\begin{table}[!htbp]
		\caption{\BLUE{Mean RMSE (HU) of $20$ test slices using different $\beta$ values in SUPER-FCN-ULTRA.}}
		\centering
		\color{black}{
    	\begin{tabular}{c|c c c}
	     \toprule
	    	\diagbox{train}{test}&\shortstack{$\beta = 0$}&\shortstack{$\beta = 5\times10^3$}  &\shortstack{$\beta = 1\times 10^4$}  \\ 
        	\hline
        	$\beta = 0$ & 45.6 & \textbf{27.9}  	&  30.7
        	\\ 
        	$\beta = 5\times 10^3$ & 133.5 &\textbf{25.0} 	&  25.9
        	\\ 
        	$\beta = 1\times 10^4$ & 172.7 &  26.9	& \textbf{25.4} 
             \\	\bottomrule
        \end{tabular} }
\label{tab:fcn-beta}
\end{table}
\subsection{Special Cases of SUPER Models}
\BLK{We now present experiments on some special cases of SUPER.}
\subsubsection{Sequential Supervised Networks}
\label{subsec:Seq_SUPER}
\begin{figure}[!t]
    \centering
   		\begin{subfigure}[h]{0.22\textwidth}
		\centering
		\includegraphics[width=1\textwidth]{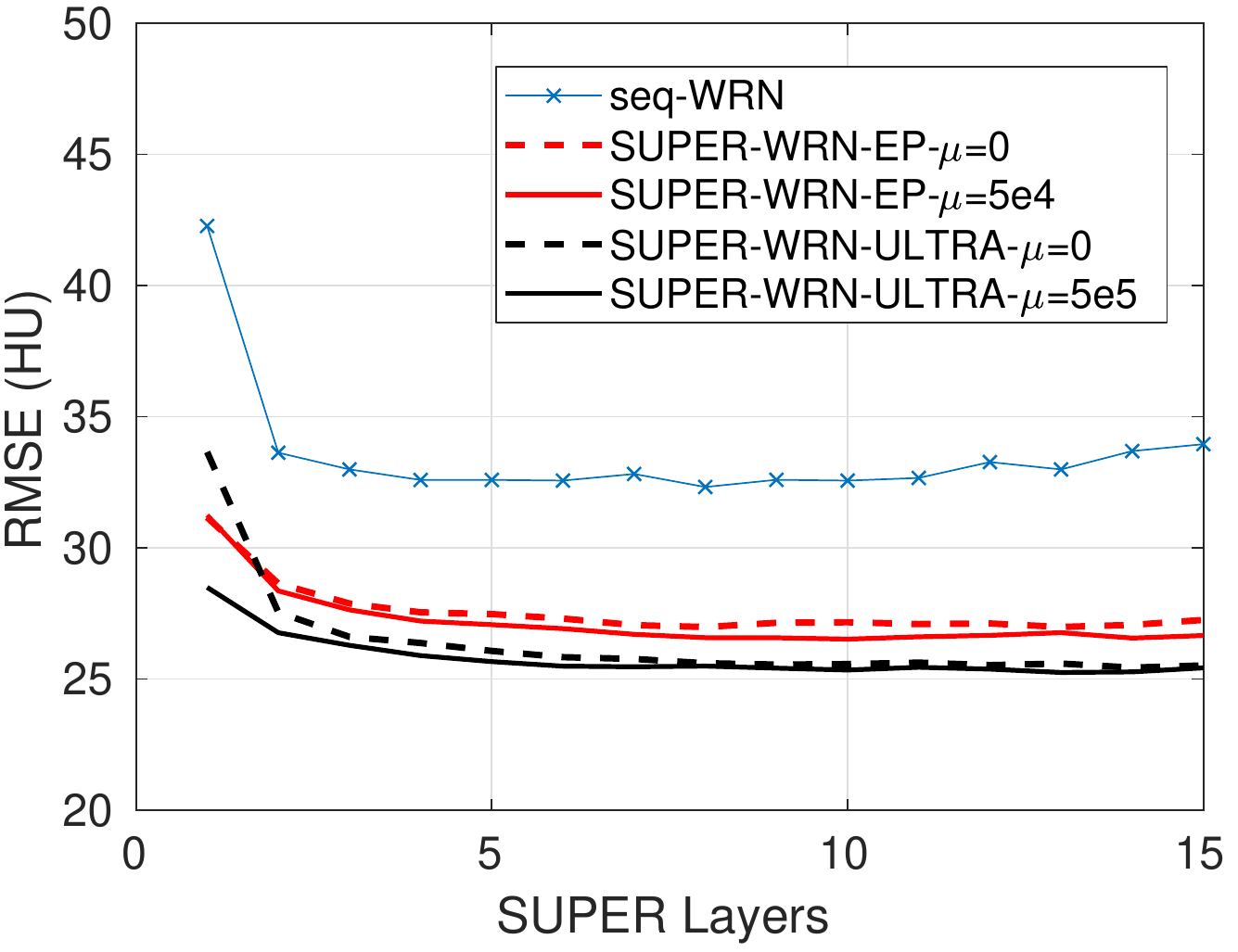}
		\caption{WavResNet based SUPER}
		\label{fig:ablation-wrs}
	\end{subfigure}
	\hfil
	\begin{subfigure}[h]{0.22\textwidth}
		\centering
	\includegraphics[width=1\textwidth]{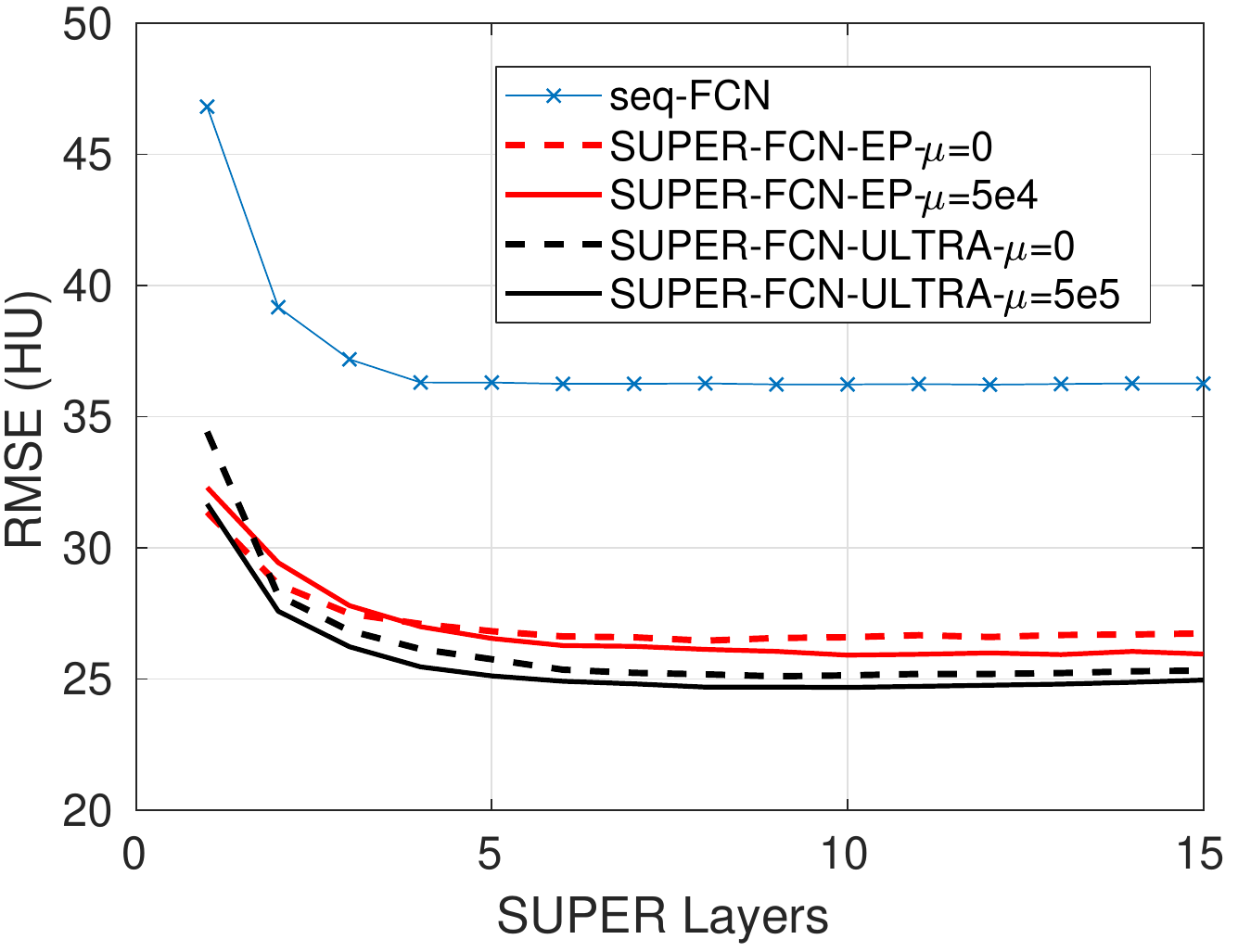}
		\caption{FBPConvNet based SUPER}
		\label{fig:ablation-fcn}
	\end{subfigure}
	\caption{Mean RMSE (over test slices) comparisons among the proposed SUPER methods, and seq-WRN or seq-FCN, where the WavRestNet and FBPConvNet architectures are repeated (x-axis indicates number of times repeated) or connected in sequence.}
	\label{fig:ablation}
	\vspace{-0.22in}
\end{figure}

\BLUE{Sequential supervised networks are a special case of the SUPER model with $J(\x,\y)=0$ in \eqref{eq:SUPER-reconstruction}.
}
This is equivalent to deep networks connected in sequence and learned in a supervised and greedy manner, with the initial image passed through the sequence of deep models to obtain a reconstruction, \BLK{with no MBIR components involved.}
We refer to the sequential supervised networks formed with the FBPConvNet and WavResNet architectures for the individual networks as seq-FCN and seq-WRN, respectively. 
Fig.~\ref{fig:ablation} shows the evolution of mean RMSE (over test slices) in seq-FCN and seq-WRN  over the number of networks (SUPER layers) connected sequentially. 
The sequential supervised networks underperform the proposed SUPER methods (with WRN or FCN, and EP or ULTRA) by around 6~HU, which indicates that the unified optimization approach incorporating data-fidelity terms and various priors can dramatically improve the reconstruction quality over deep networks in sequence.

\subsubsection{SUPER with Data-fidelity only cost}
To further explore the different special cases of SUPER, here we empirically validate the relative effect of the data-fidelity term used in \BLK{\eqref{eq:SUPER-reconstruction}} by setting $\beta=\mu=0$.
In particular, at reconstruction time, the initial FBP image is passed through networks learned in a supervised manner, each time followed by few iterations of descent on the data-fidelity cost, which enforces data consistency in conjunction with the supervised learning-based network.
\begin{figure}[!t]
	\centering
   	\begin{subfigure}{1\textwidth}\leftskip10pt
   		\scalebox{0.9}{
   			\centering  
	    \begin{tikzpicture}
			[spy using outlines={rectangle,green,magnification=2.8,size=10mm, connect spies}]
			\node {\includegraphics[width=0.22\textwidth]{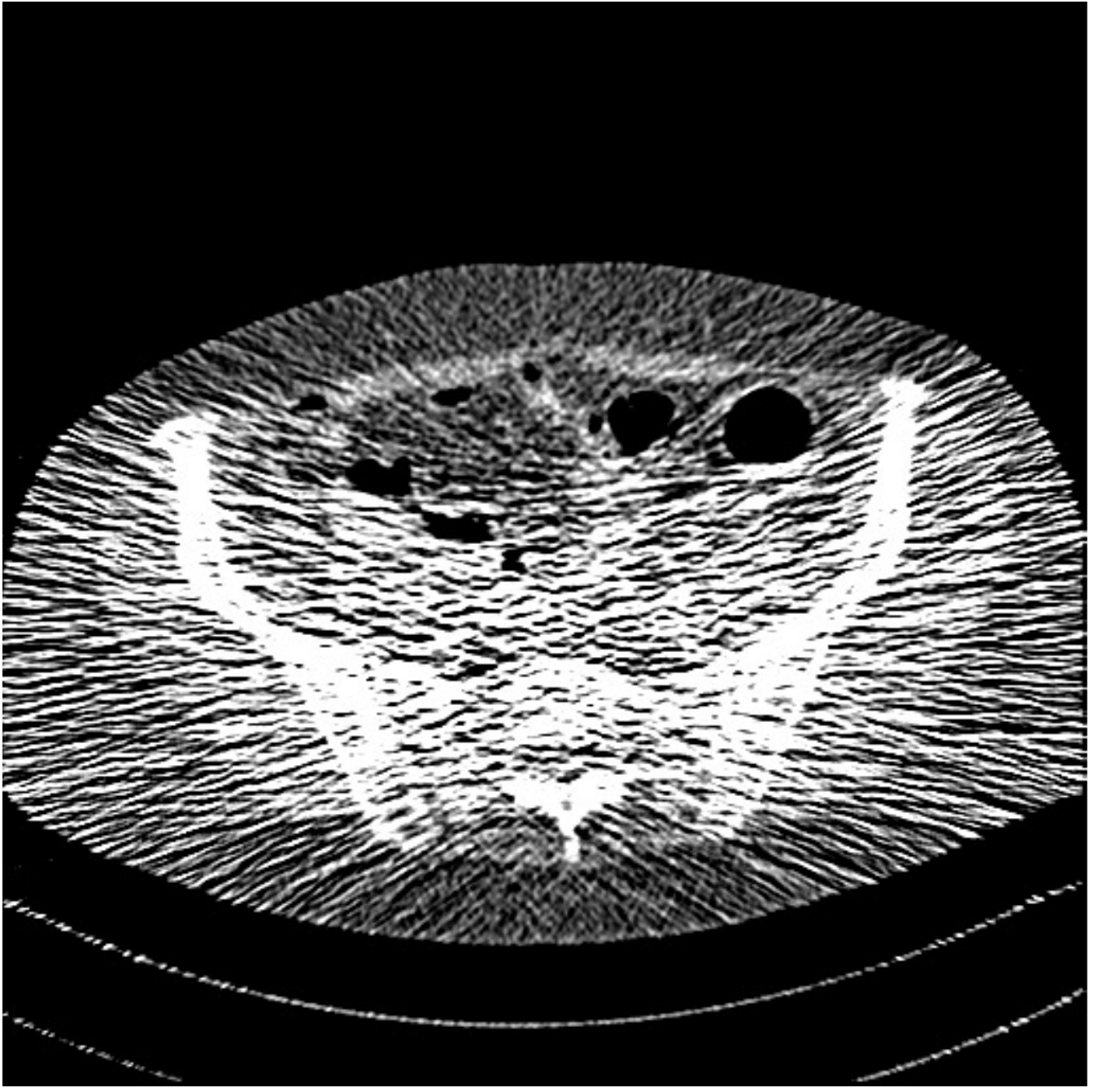} };
			\spy on (0.5,-0.1) in node [right] at (1.1,1.5);
			\node [align = center,white, font=\bf] at (-0.9,1.55) {\small FBP};
	\node [white, font=\footnotesize] at (0.7,-1.7) {RMSE = 194.09~HU};
	\end{tikzpicture}
	\begin{tikzpicture}
			[spy using outlines={rectangle,green,magnification=2.8,size=10mm, connect spies}]
			\node {\includegraphics[width=0.22\textwidth]{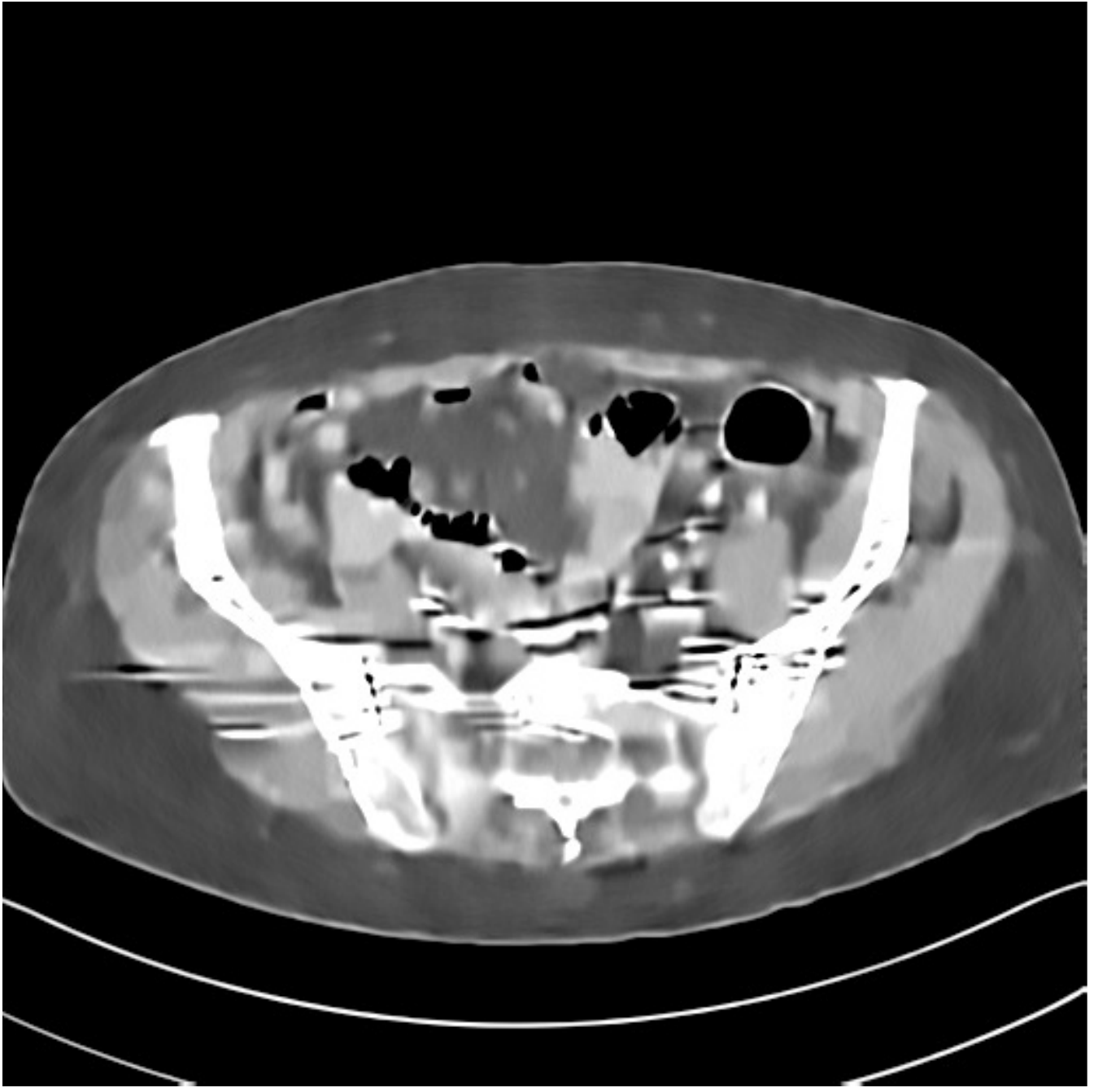} };
			\spy on (0.5,-0.1) in node [right] at (1.1,1.5);
			\node [align = center,white, font=\bf] at (-0.9,1.55) {\small PWLS-ULTRA};
		    \node [white, font=\footnotesize] at (0.7,-1.7) {RMSE = 43.40~HU};
	\end{tikzpicture}}
	\end{subfigure}
 \vfil \vspace{-0.05in}
		\begin{subfigure}{1\textwidth}\leftskip10pt
		\scalebox{0.9}{
		\centering  \begin{tikzpicture}
			[spy using outlines={rectangle,green,magnification=2.8,size=10mm, connect spies}]
			\node {\includegraphics[width=0.22\textwidth]{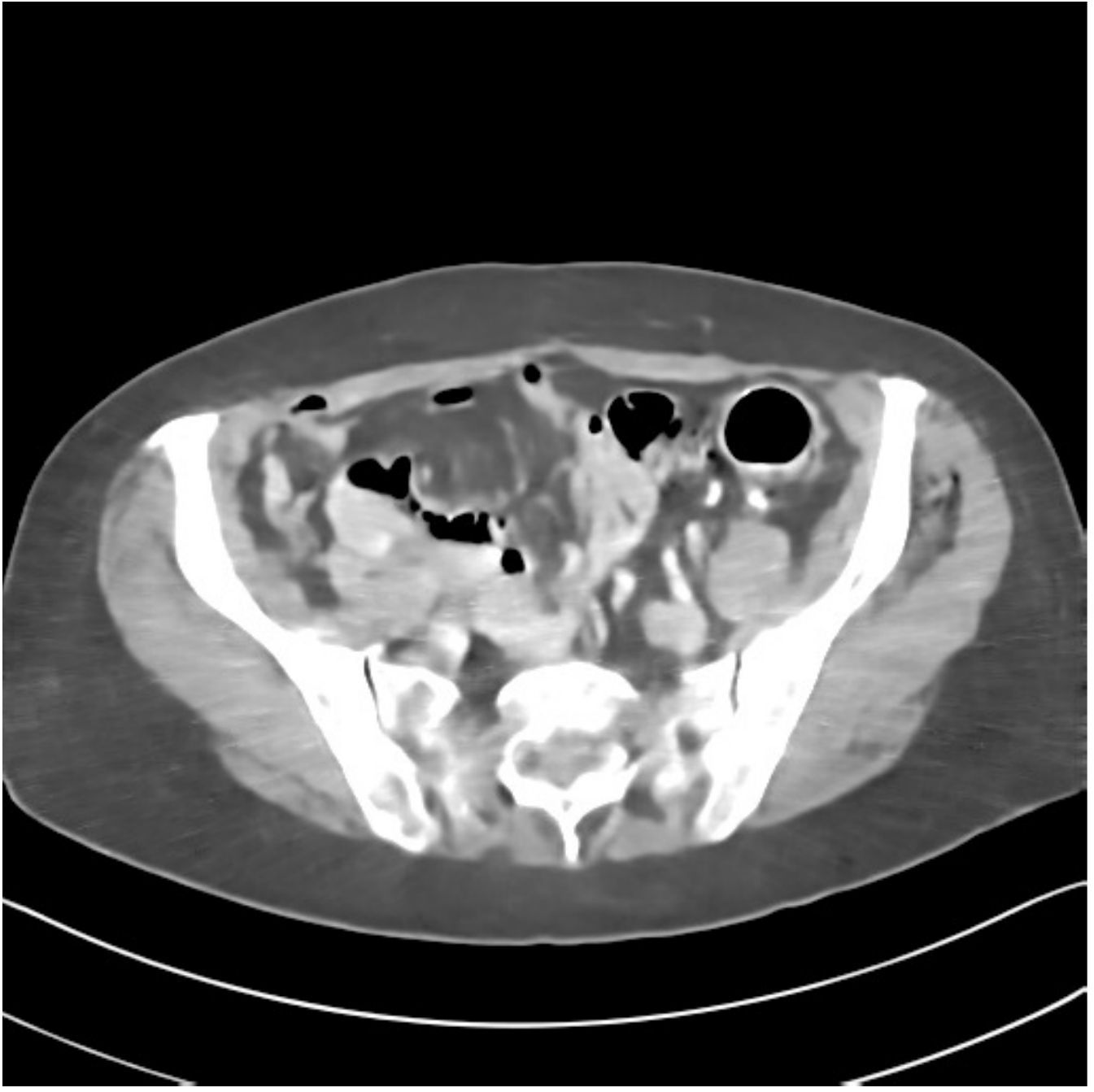} };
			\spy on (0.5,-0.1) in node [right] at (1.1,1.5);
			\node [align = center,white, font=\bf] at (-0.9,1.7){\small FBPConvNet};
			\node [white, font=\footnotesize] at (0.7,-1.7) {RMSE = 34.24~HU};
	\end{tikzpicture}
    \begin{tikzpicture}
    			[spy using outlines={rectangle,green,magnification=2.8,size=10mm, connect spies}]
    			\node {\includegraphics[width=0.22\textwidth]{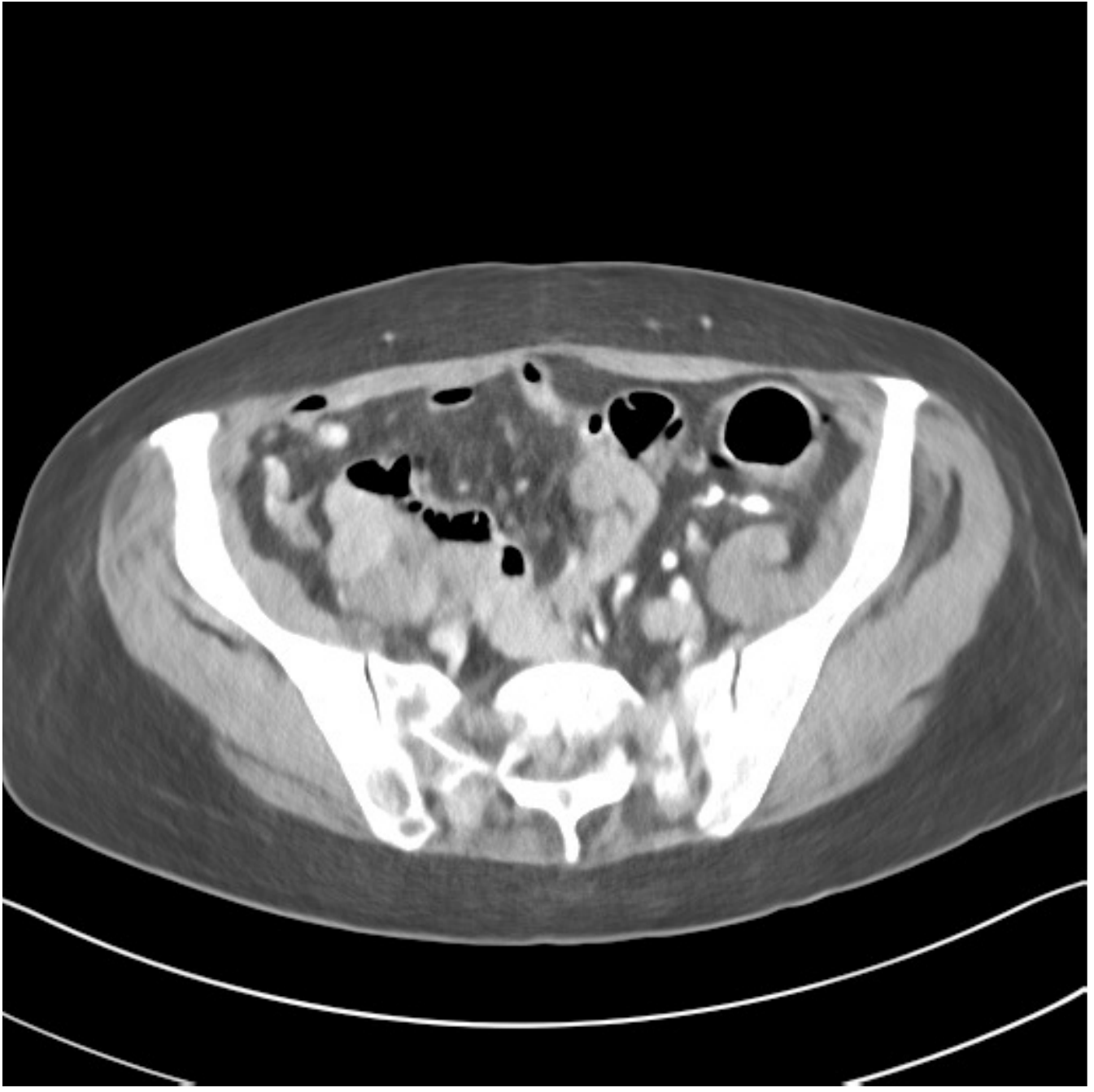} };
    			\spy on (0.5,-0.1) in node [right] at (1.1,1.5);
    			\node [align = center,white, font=\bf] at (-0.9,1.6){\small SUPER-FCN-\\DataTerm};
    			\node [white, font=\footnotesize] at (0.7,-1.7) {RMSE = 31.21~HU};
    \end{tikzpicture} }
    \end{subfigure}
    \vfil \vspace{-0.05in}
    	\begin{subfigure}{1\textwidth} \leftskip10pt
		\scalebox{0.9}{
		\centering \begin{tikzpicture}
			[spy using outlines={rectangle,green,magnification=2.8,size=10mm, connect spies}]
			\node {\includegraphics[width=0.22\textwidth]{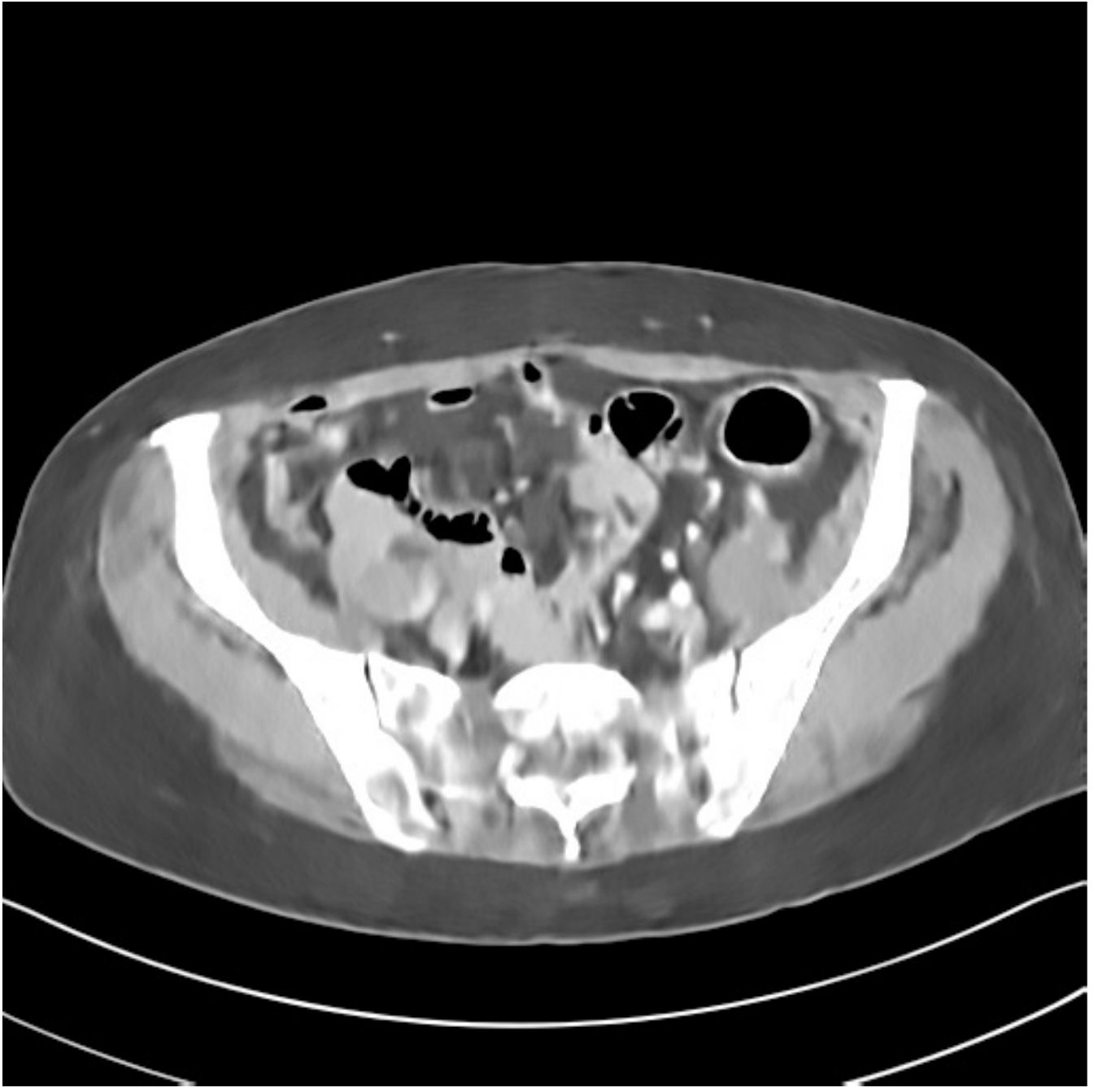} };
			\spy on (0.5,-0.1) in node [right] at (1.1,1.5);
			\node [align = center,white, font=\bf] at (-0.9,1.55) {\small SUPER-FCN-\\ULTRA};
	\node [white, font=\footnotesize] at (0.7,-1.7) {RMSE = 28.82~HU};
	\end{tikzpicture}
    \begin{tikzpicture}
			[spy using outlines={rectangle,green,magnification=2.8,size=10mm, connect spies}]
			\node {\includegraphics[width=0.22\textwidth]{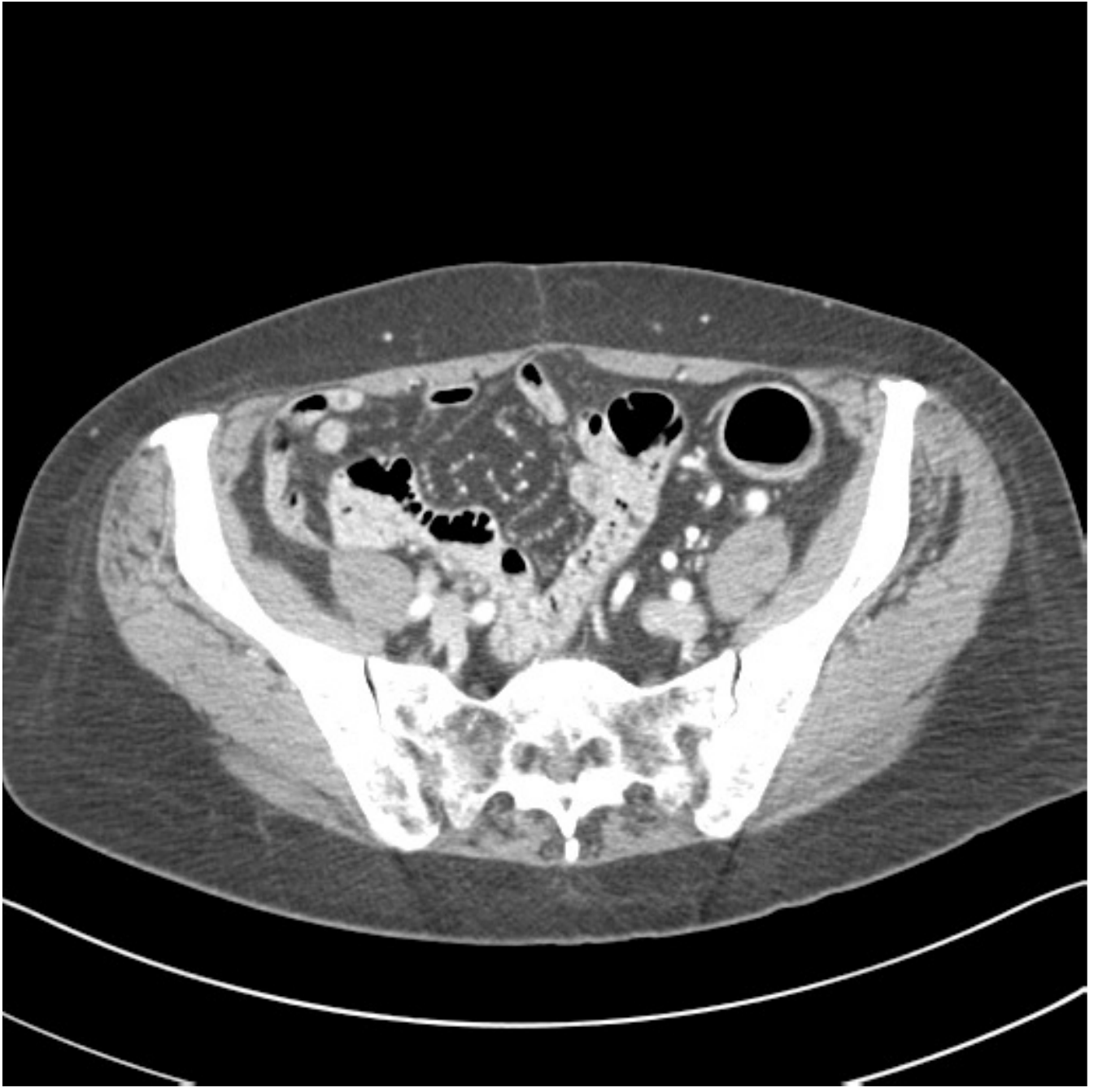} };
			\spy on (0.5,-0.1) in node [right] at (1.1,1.5);
			\node [align = center,white, font=\bf] at (-0.9,1.55) {\small Reference};
	\end{tikzpicture}}
	\end{subfigure}
	\caption{Reconstructed images of slice 150 of patient L192 using of FBP, PWLS-ULTRA, FBPConvNet, SUPER-FCN-DataTerm, and SUPER-FCN-ULTRA, respectively, shown along with the reference.}
	\label{fig:Data-term-L192s150}
	\vspace{-0.18in}
\end{figure}
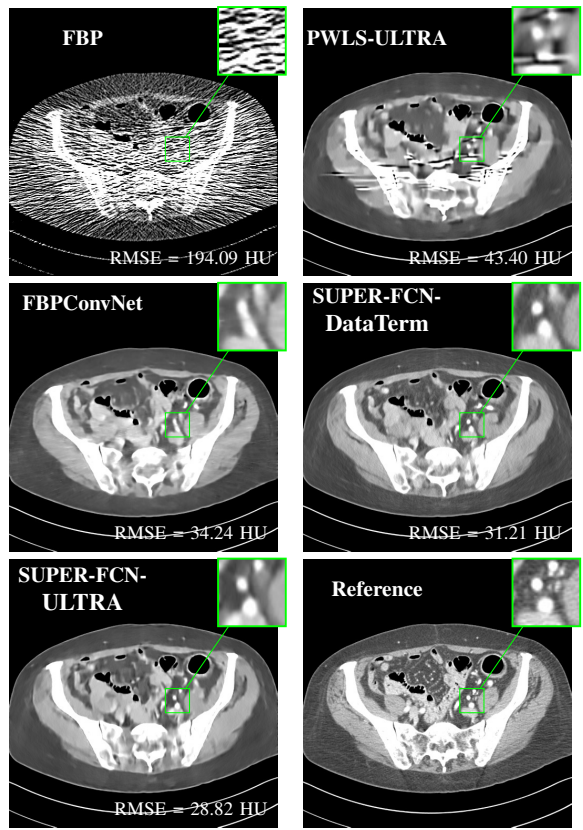
Fig.~\ref{fig:Data-term-L192s150} shows reconstructions using FBPConvNet, SUPER-FCN-DataTerm (i.e., $\beta=\mu=0$), PWLS-ULTRA, and SUPER-FCN-ULTRA, respectively. 
For SUPER-FCN-DataTerm, when optimizing the data-fidelity term, we start with the deep network's output and ran 5 iterations for the data-fidelity term to avoid overfitting to the analytical FBP images.
In Fig.~\ref{fig:Data-term-L192s150}, obviously, FBPConvNet significantly suppresses noise and artifacts compared to PWLS-ULTRA, but it also over-smooths many details (e.g., \BLK{features in the zoom-in box) in the reconstruction.}
SUPER-FCN-DataTerm, by enforcing data consistency, helps reduce overfitting issues and reconstructs image details and tissue boundaries better compared to the standalone FBPConvNet. Our SUPER-FCN-ULTRA method, however, exploits richer prior information (via the union of learned sparsifying transforms) and explicit network regularizer and outperforms the SUPER-FCN-DataTerm approach.
Additional such comparisons for other selected test slices are included in the supplement \BLUE{(Fig.~15)}.

\begin{figure}[!t]
    \centering
   		\begin{subfigure}[h]{0.22\textwidth}
		\centering
		\includegraphics[width=1\textwidth]{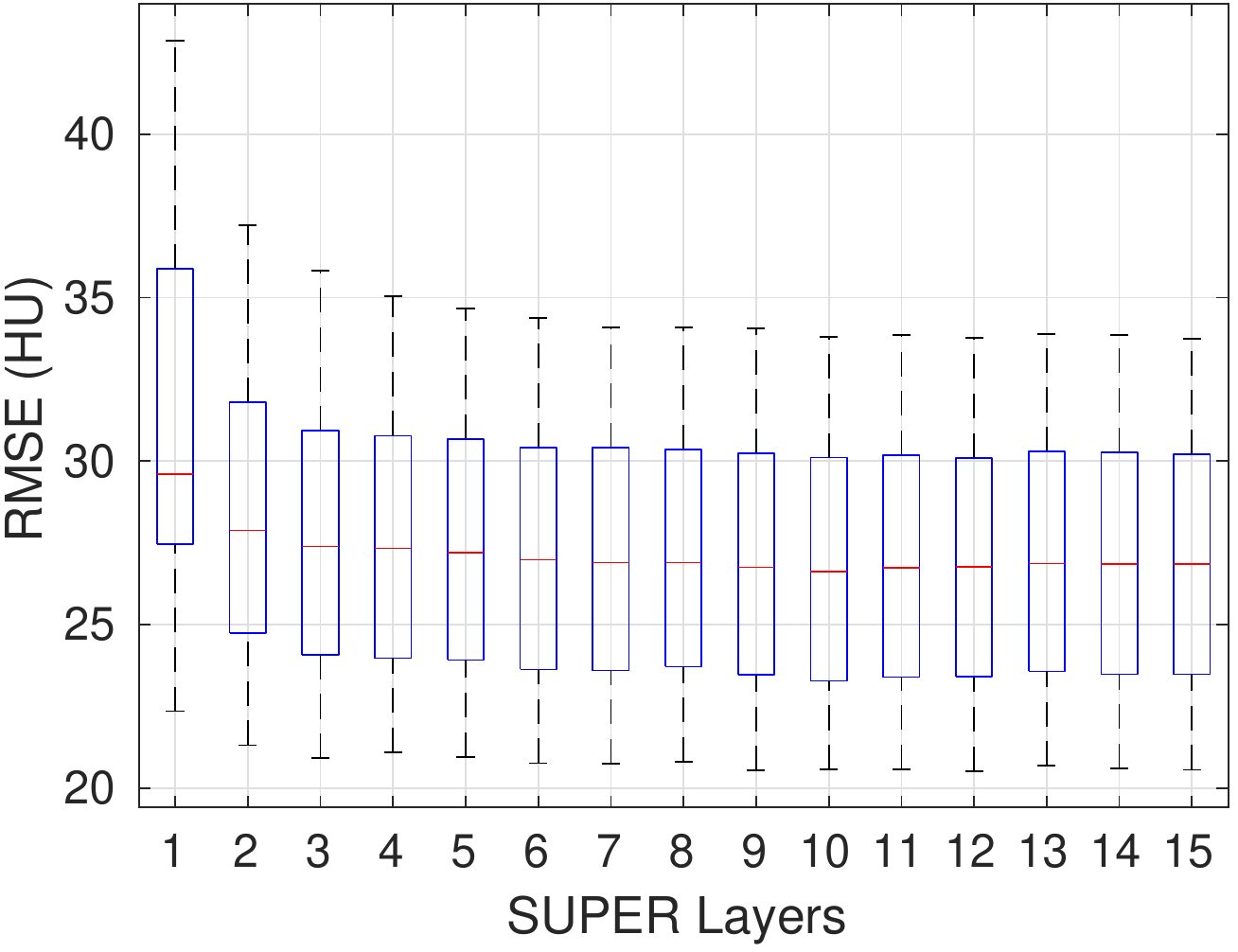}
		\caption{SUPER-WRN-EP}
	\end{subfigure}
	\hfil
	\begin{subfigure}[h]{0.22\textwidth}
		\centering
	\includegraphics[width=1\textwidth]{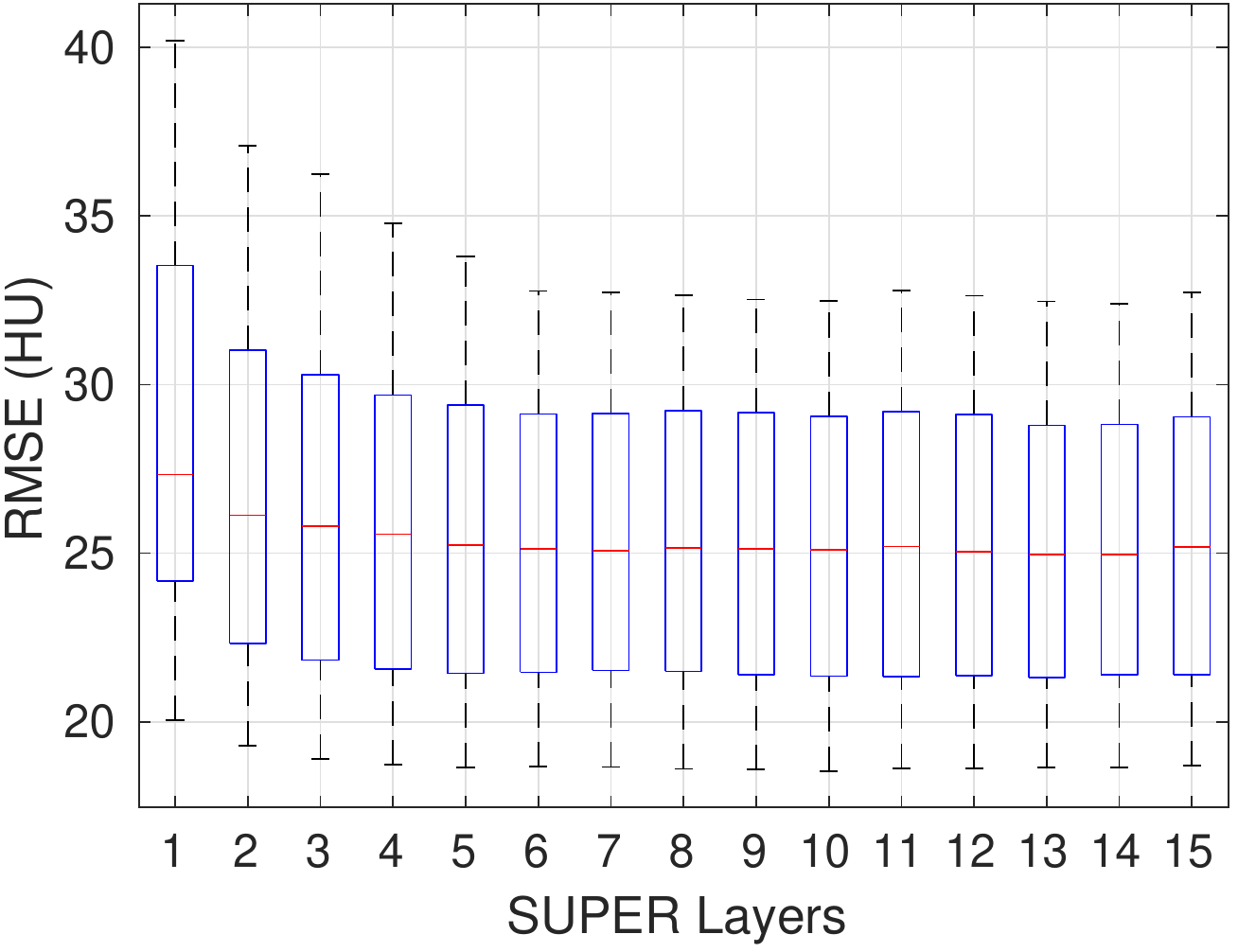}
		\caption{SUPER-WRN-ULTRA}
	\end{subfigure}
    \caption{RMSE (box plots showing the spread over 20 test slices) over super layers for SUPER-WRN-EP and SUPER-WRN-ULTRA.}
    \label{fig:WRN-rmse-layers}
    \vspace{-0.25in}
\end{figure}

\subsubsection{\BLUE{SUPER with only a Supervised Regularizer}}
\begin{figure}[!htb]
	\centering
	\begin{tikzpicture}
    			[spy using outlines={rectangle,green,magnification=2.3,size=10mm, connect spies}]
    			\node {\includegraphics[width=0.22\textwidth]{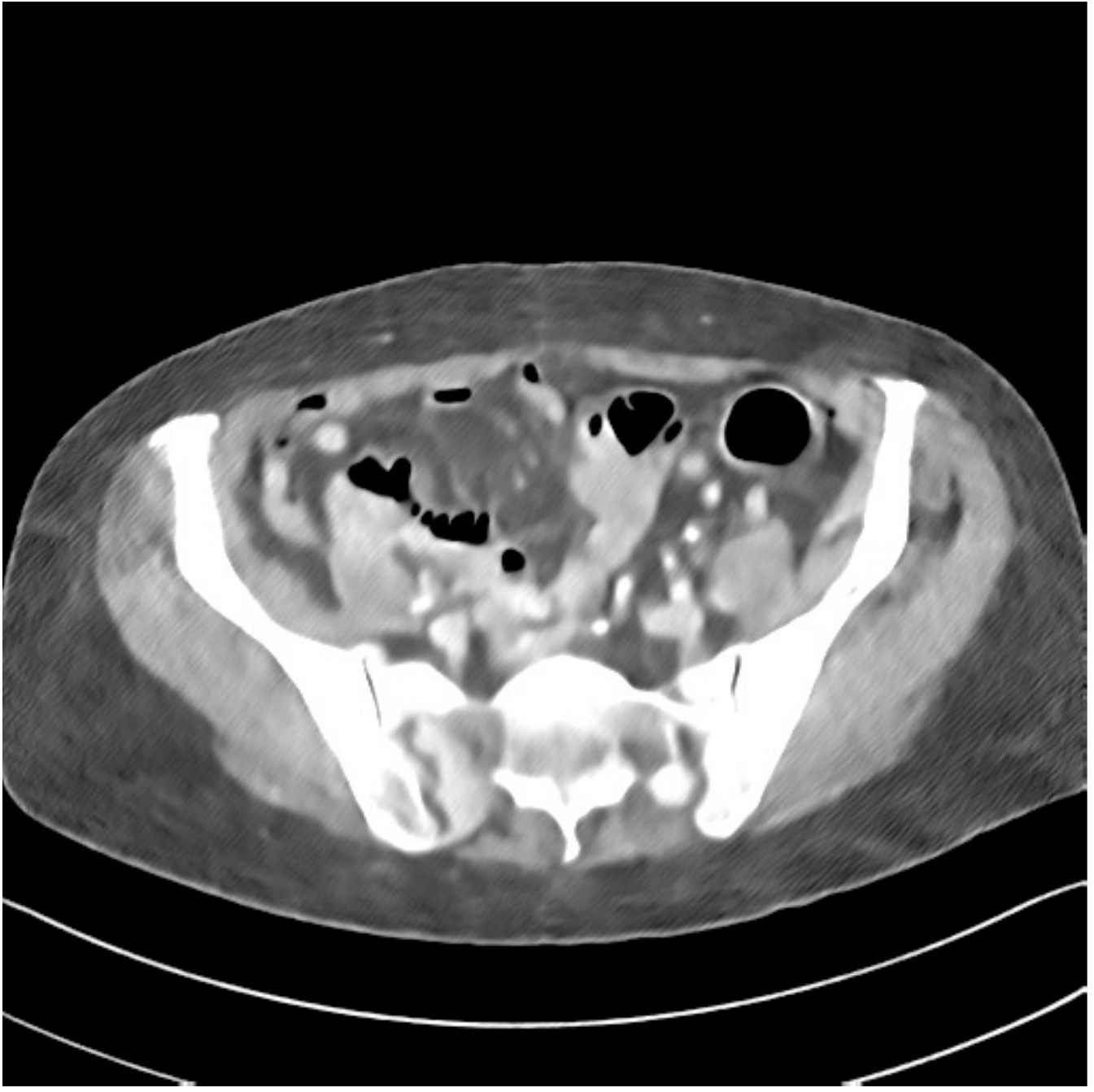}};
    			\spy on (0.05,-0.8) in node [right] at (0.85,1.3);
    			\node [align = center,white, font=\bf] at (-0.5,1.55){\footnotesize ADMM-Net (WRN)};
    			\node [white, font=\footnotesize] at (0.7,-1.6) {RMSE = 32.90~HU};
    \end{tikzpicture} 
    \hfil
    \begin{tikzpicture}
    			[spy using outlines={rectangle,green,magnification=2.3,size=10mm, connect spies}]
    			\node {\includegraphics[width=0.22\textwidth]{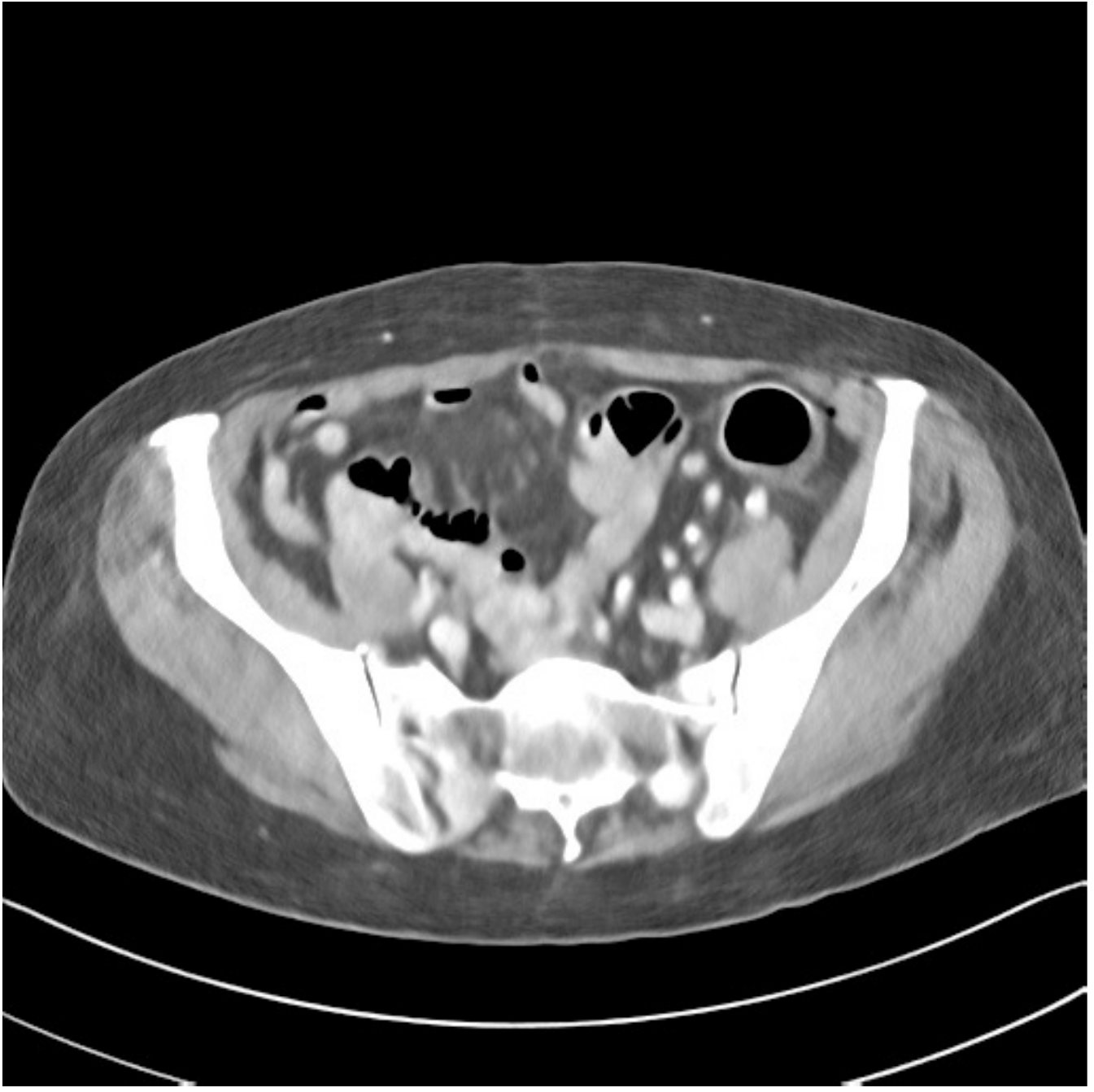}};
    			\spy on (0.05,-0.8) in node [right] at (0.85,1.3);
    			\node [align = center,white, font=\bf] at (-0.5,1.55){\footnotesize SUPER-WRN-$\beta=0$};
    			\node [white, font=\footnotesize] at (0.7,-1.6) {RMSE = 31.32~HU};
    \end{tikzpicture} 
    \vfil
    \begin{tikzpicture}
			[spy using outlines={rectangle,green,magnification=2.3,size=10mm, connect spies}]
			\node {\includegraphics[width=0.22\textwidth]{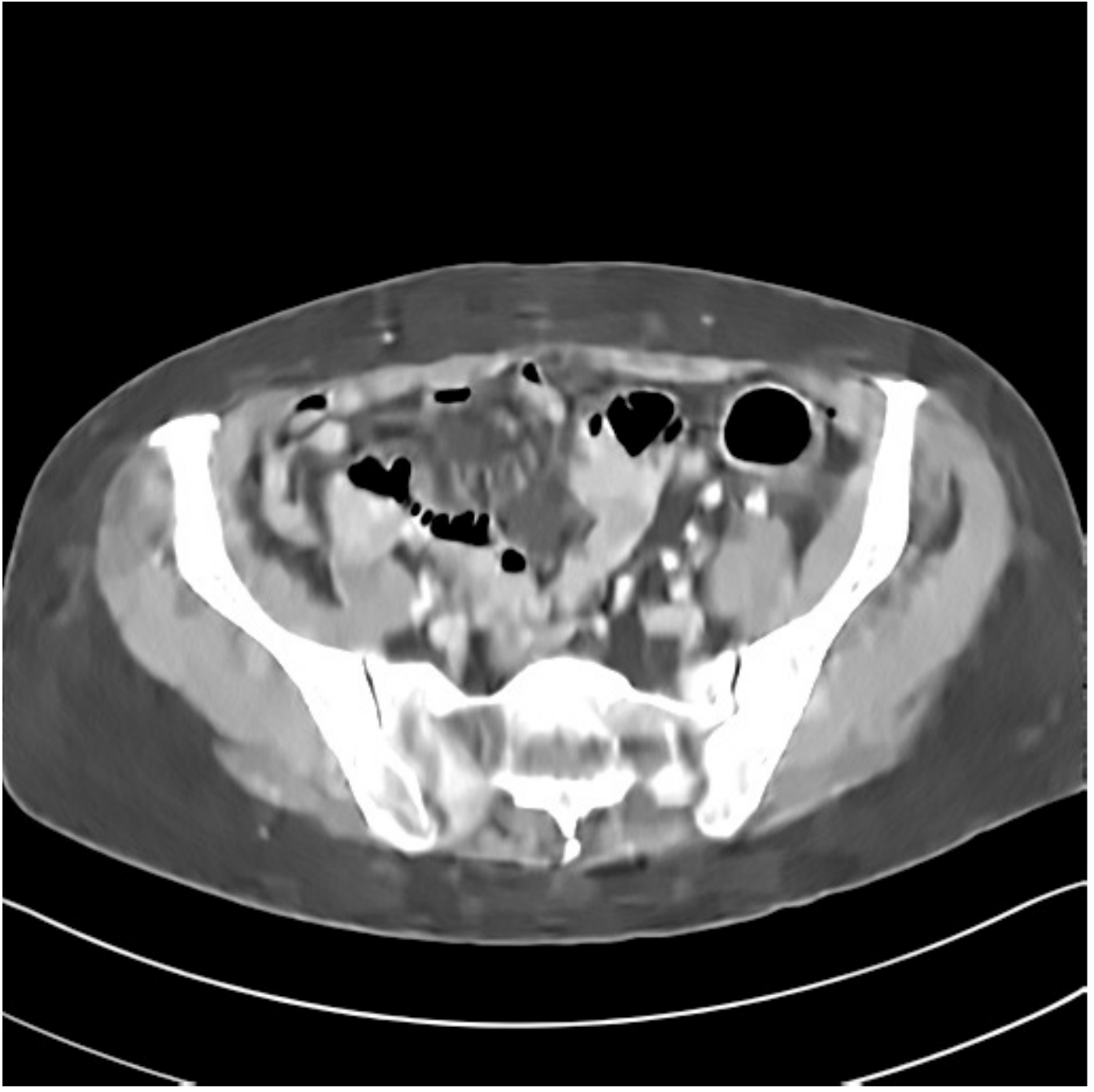} };
			\spy on (0.05,-0.8) in node [right] at (0.85,1.3);
			\node [align = center,white, font=\bf] at (-0.4,1.45) {\footnotesize SUPER-WRN\\\footnotesize -ULTRA};
	\node [white, font=\footnotesize] at (0.7,-1.6) {RMSE = 29.74~HU};
	\end{tikzpicture}
	\hfil
    \begin{tikzpicture}
			[spy using outlines={rectangle,green,magnification=2.3,size=10mm, connect spies}]
			\node {\includegraphics[width=0.22\textwidth]{figures/fbpconvnet/L192slice150_fbp_normaldose_recon.pdf} };
			\spy on (0.05,-0.8) in node [right] at (0.85,1.3);
			\node [align = center,white, font=\bf] at (-0.9,1.55) {\small Reference};
	\end{tikzpicture}
	\caption{\BLUE{Reconstructed images of L192 slice 150 with WavResNet plugged ADMM-Net, SUPER-WRN-$\beta=0$ (both training and testing), and SUPER-WRN-ULTRA ($\beta=5\times 10^3$ in both training and testing), respectively, shown with the reference image.}}
	\label{fig:beta0-wrn-L192s150}
\end{figure}

\BLUE{This special case of SUPER corresponds to the case where SUPER excludes the unsupervised regularizer, while only involves the data-fidelity and a supervised regularizer in the MBIR cost, i.e., $\beta=0$ and $\mu\neq 0$.
In this case, the proposed SUPER model is similar to a generalized block coordinate descent-based network by replacing a simple denoising autoencoder~\cite{ravchfess17,chfess18,Chun&etal:19MICCAI} with a general CNN that forms our supervised regularizer. 
This SUPER is also similar to the plug-and-play ADMM-Net method
except that the inputs to each supervised network are the preceding reconstructions, while plug-and-play ADMM-Net updates inputs to the network (denoiser) based on auxiliary variables in the ADMM algorithm.
Here, we used $\mu=1\times 10^6$, which worked well for the plug-and-play ADMM-Net method, for this special case of SUPER. 
Fig.~\ref{fig:beta0-wrn-L192s150} shows a comparison between plug-and-play ADMM-Net, SUPER without unsupervised regularizers (in both training and testing), and the full SUPER version with ULTRA regularization ($\beta=5\times 10^3$, $\mu=5\times 10^5$). All these methods used WavResNet as their denoisers/supervised networks.
In this example, SUPER with only a supervised regularizer (SUPER-WRN-$\beta=0$) outperforms the plug-and-play ADMM-Net (ADMM-Net (WRN)) by 1.6~HU RMSE and provides sharper image details.
Comparing SUPER-WRN-$\beta=0$ and the full SUPER-WRN-ULTRA scheme, we observe that the latter provides a lower RMSE and higher contrast image features than the former that excludes the unsupervised component.
This again shows the effect of the unsupervised ULTRA model (in capturing local image details better with a union of learned transforms) in the SUPER scheme.}

\subsection{Convergence Behavior of SUPER Methods}
\BLUE{From Fig.~\ref{fig:ablation}, we observe that our proposed SUPER methods with proper choice of the regularization parameter $\mu$ provide convergent mean RMSE curves within 15 SUPER layers. Moreover,}
Fig.~\ref{fig:WRN-rmse-layers} shows that the RMSE values of SUPER-WRN-EP and SUPER-WRN-ULTRA decrease and converge quickly over SUPER layers for reconstructing the test slices.
\BLUE{We observe a similar behavior for FBPConvNet based SUPER methods and show it in the supplement (Section VII.G).}
We also show \BLUE{in the supplement (Section VII.G)} the convergence of the unsupervised iterative module's cost at reconstruction time (i.e., \BLK{\eqref{eq:SUPER-reconstruction}}) for two test slices, \BLUE{as well as the training losses corresponding to (P1) and the upper-level cost of the bilevel problem in~\eqref{eq:bilevel}. 
The experimental results indicate that our proposed algorithm for SUPER can give practical convergent reconstructions with several layers (e.g., 15 layers).}


\subsection{\BLUE{Computational Cost of SUPER Methods}}
\BLUE{The computational cost of SUPER methods closely relates to the run time of each SUPER layer and the total number of SUPER layers.
In general, SUPER training takes longer time compared to training the corresponding standalone supervised network because of the sequential greedy training process involving iterative MBIR updates. 
In the testing stage, the computational cost of SUPER mainly depends on the number of iterations in each iterative module and the number of SUPER layers. }

\BLUE{
Our experiments were performed with a 2.40 GHz Intel Xeon Gold-6148 processor with (maximum) 40 threads and a graphics processing unit Tesla P40.
We report the runtime of training and testing for SUPER methods and their supervised and unsupervised counterparts in Table~\ref{tab:runtime}.
In particular, for SUPER training, we report the approximate runtime of training a single SUPER layer by averaging the runtime of training 15 layers.
From this table, we observe that SUPER training (with 15 layers) is more time consuming than training the standalone supervised methods. During testing, the EP-based SUPER methods ran more slowly than PWLS-EP because of overall more iterations through all the SUPER layers (300 iterations for 15 layers in SUPER versus 100 iterations in standalone PWLS-EP). The ULTRA-based SUPER methods can work faster than the typical standalone PWLS-ULTRA method because of fewer overall required iterations (600 iterations for 15 layers in SUPER versus 1000 iterations in PWLS-ULTRA).
}
\begin{table}[!htbp]
	\centering \leftskip-5pt
	\caption{\BLUE{Runtime of SUPER methods and the constituent standalone supervised and unsupervised methods during training and testing.}}
	\vspace{-0.05in}
	\scalebox{0.8}{
	\color{black}{
	\begin{tabular}{ccc} 
		\toprule
		\textbf{Method} &
		\shortstack{\textbf{ Train}}&
		\textbf{ Test (per image)} \\  \midrule
		PWLS-EP &- & 1 minute (100 iters)\\ 
		PWLS-ULTRA &5.5 hours & 1.5 hours (1000 iters) \\ 
		WavResNet   &20 hours (200 epochs)   &5 seconds   \\ 
		SUPER-WRN-EP&  3 hours/layer	& 5 minutes (15 layers)	  \\ 
		SUPER-WRN-ULTRA &5.5 hours/layer	&30 minutes (15 layers)\\ 
		FBPConvNet & 10 hours (100 epochs) &5 seconds \\
		SUPER-FCN-EP  & 1 hour/layer & 5 minutes (15 layers)\\ 
		SUPER-FCN-ULTRA& 4 hours/layer & 30 minutes (15 layers)\\
		\bottomrule
	\end{tabular}}
	}
	\label{tab:runtime}
\end{table}	
	


\subsection{\BLUE{Performance with Various Dose Levels and Generalization}}
\BLUE{Here, we trained SUPER methods under another dose level with $I_0=1\times 10^5$.
The training set was created using the same regular-dose images that were used to generate the training data with $I_0=1\times 10^4$ in the preceding experiments. The hyper-parameters for training the SUPER methods as well as the corresponding standalone supervised methods under $I_0~=~1\times 10^5$ were the same as those under $I_0 = 1\times 10^4$ which worked well.
To evaluate the trained networks, we used the same 20 test slices but generated their low-dose realizations with $I_0 = 2\times 10^5$, $I_0 = 1\times 10^5$, $I_0 = 8\times 10^4$, and $I_0 = 2\times 10^4$, respectively. We used the same parameters as those for $I_0 = 1\times 10^4$ to reconstruct testing images under various dose levels to study model generalization.
}

\BLUE{Table~\ref{tab:multi-dose} reports averaged RMSE, SNR, and SSIM over the $20$ test samples under different dose levels using SUPER-FCN-ULTRA and its constituent unsupervised and supervised methods, i.e., PWLS-ULTRA and FBPConvNet. For testing samples under $I_0=1\times 10^5$, where the dose levels of training and testing are matched, SUPER-FCN-ULTRA achieves the best results compared to PWLS-ULTRA and FBPConvNet. 
\MG{We show an example test image reconstructed by different methods in Section VII.H in the supplement.}
When applying the trained networks with $I_0=1\times 10^5$ to reconstruct images under neighboring doses, e.g., $I_0~=~2\times 10^5$ and $I_0 = 8 \times 10^4$, SUPER-FCN-ULTRA still performs better than PWLS-ULTRA and FBPConvNet. However, when the dose level of the testing samples differs too much from that of training samples, results with SUPER methods may not beat unsupervised learning-based methods such as PWLS-ULTRA, but they still outperform the corresponding supervised methods, 
because the MBIR components in SUPER
moderate overfitting issues that easily happen to supervised methods. 
The bottom row of Table~\ref{tab:multi-dose} shows that when reconstructing images under $I_0=2\times 10^4$ using the trained networks under $I_0 = 1\times 10^5$, the mean RMSE and SNR values of SUPER-FCN-ULTRA are in between of those of PWLS-ULTRA and FBPConvNet. The mean SSIM of SUPER-FCN-ULTRA is slightly better than (or comparable to) that of the unsupervised PWLS-ULTRA, and is much better than that of FBPConvNet, indicating that the MBIR component in SUPER helps preserve image structures and improves the generalization property compared with the standalone supervised FBPConvNet method.
\MG{The supplement (Section VII.H) shows example reconstructions at $I_0=2\times 10^4$, where SUPER-FCN-ULTRA outperforms (generalizes better than) PWLS-ULTRA and FBPConvNet.}
}

\begin{table}[!htp]
	\centering \leftskip-5pt
	\caption{\BLUE{Averaged numerical results of $20$ test slices under different dose levels using networks trained with data under $I_0 = 1\times 10^5$. The units of RMSE and SNR are HU and dB, respectively.}}
	\vspace{-0.05in}
	\scalebox{0.75}{
	\color{black}{
	\begin{tabular}{cccc}
			\toprule
			\textbf{$I_0$} &
			\shortstack{\begin{tabular}[c]{@{}c@{}}\textbf{PWLS-ULTRA}\\ (RMSE/SNR/SSIM)\end{tabular}}&
			\begin{tabular}[c]{@{}c@{}}\textbf{FBPConvNet}\\ (RMSE/SNR/SSIM)\end{tabular}&
			\begin{tabular}[c]{@{}c@{}}\textbf{SUPER-FCN-ULTRA}\\ (RMSE/SNR/SSIM)\end{tabular}\\  \midrule
			$2\times 10^5$ & 15.0 / 34.1 / 0.84 &   14.8 / 34.1 / 0.83&     \textbf{13.3} / \textbf{35.0} / \textbf{0.84}\\ 
			$1\times 10^5$ & 15.9 / 33.2 / 0.81 &   15.7 / 33.6 / 0.82&     \textbf{14.7} / \textbf{34.1} / \textbf{0.82}\\ 
			$8\times 10^4$ &16.7 / 32.9 / 0.78 &    16.3 / 33.2 / 0.82 &    \textbf{15.2} / \textbf{33.8} / \textbf{0.82} \\ 
			$2\times 10^4$ &\textbf{23.3} / \textbf{30.6} / 0.72 &  47.1 / 25.4 / 0.60 &    31.6 / 28.5 / \textbf{0.74} \\ 
			\bottomrule
	\end{tabular}}
	}
	\label{tab:multi-dose}
\end{table}

\section{Conclusions}\label{sec:conclusion} 
This paper proposed a mathematical framework for unified supervised and unsupervised learning, dubbed SUPER, for low-dose X-ray CT image reconstruction.
The proposed SUPER framework combines physics-based forward models, statistical models of measurements and noise, machine learned models, and analytical image models in a common framework. 
Regularizers based on both supervised and unsupervised learning are jointly incorporated into model-based image reconstruction formulations and algorithms.
We proposed an efficient \BLK{approximate algorithm for} learning the proposed SUPER
model.
We studied four example SUPER methods by combining the FBPConvNet and WavResNet architectures for the supervised learning-based regularizer, and edge-preserving and union of learned transforms models for the analytical or unsupervised learning-based regularizer.
We also performed ablation studies of the 
SUPER model and demonstrated that using \BLK{the proposed mixture of models and priors} in SUPER helps improve the reconstruction quality in terms of reducing noise and artifacts and reconstructing structural details.
In future work, we plan to study \BLUE{acceleration of the algorithm}
as well as convergence theory for the proposed unified (fixed point or bilevel optimization based) training schemes,
and will apply SUPER methods in other imaging modalities.
\MG{Additional data augmentation tools will also be considered to further improve the performance especially in limited training data setups.}

\section{Acknowledgment}
The authors thank Dr. Cynthia McCollough, the Mayo Clinic, the American Association of Physicists in Medicine, and the National Institute of Biomedical Imaging and Bioengineering for providing the Mayo Clinic data.

\bibliographystyle{IEEEbib}
\bibliography{egbib}

 \clearpage

 {\twocolumn[
 	\begin{center}
 		\Huge Unified Supervised-Unsupervised (SUPER) Learning for X-ray CT Image Reconstruction \\-- Supplementary Materials
 		\vspace{0.4in}
 	\end{center}]}
 This supplement provides additional details and results to accompany our manuscript.
\setcounter{section}{6}

\section{Additional Experimental Results}
\setcounter{figure}{7}
\setcounter{table}{4}
\subsection{SNR and SSIM Comparisons}
\begin{figure}[!htbp]
    \centering
   	\includegraphics[width=0.45\textwidth]{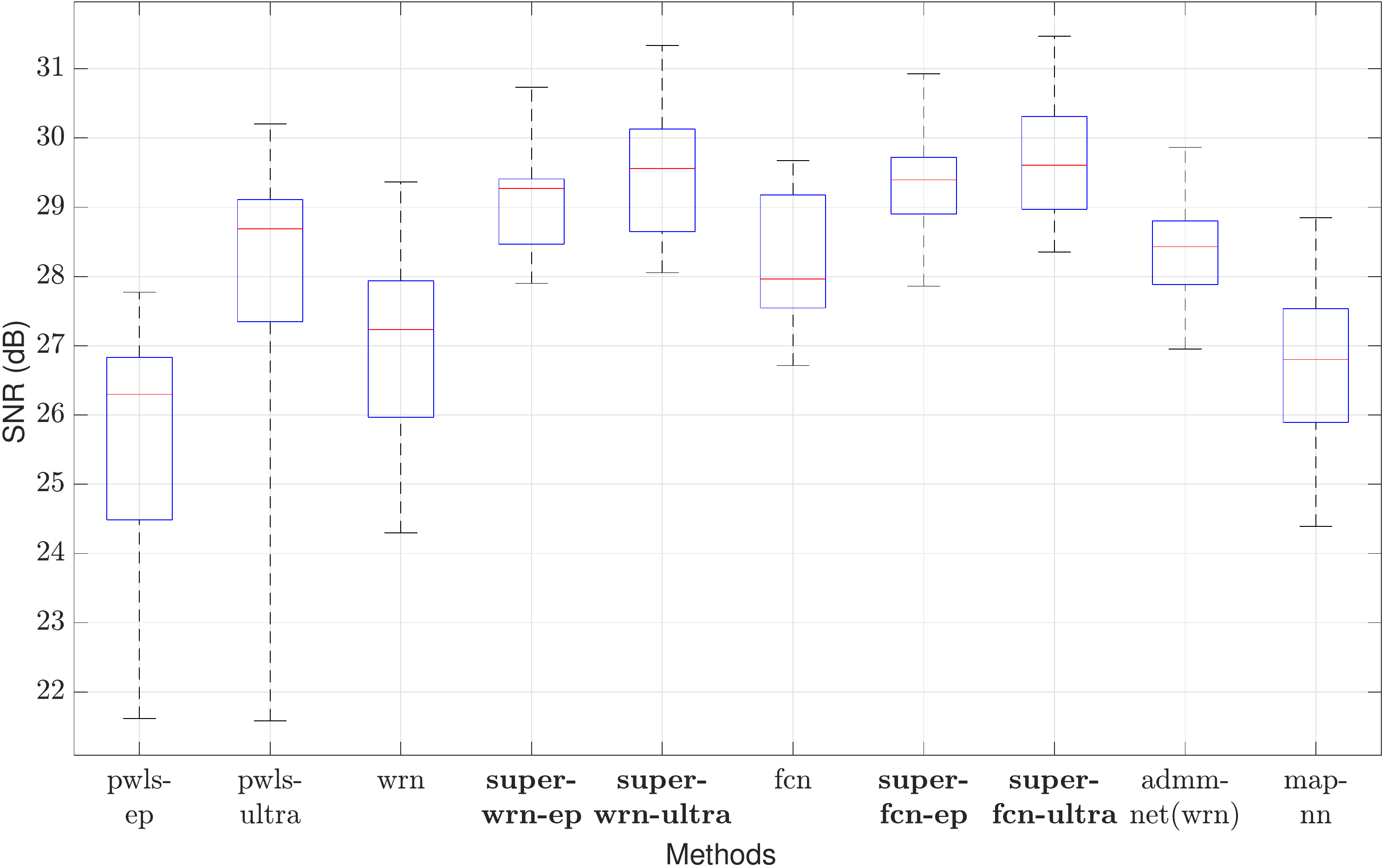}
    \caption{\BLUE{SNR spread (shown using box plots) over $20$ test cases using different methods. Here, ``wrn" stands for WavResNet, ``fcn" stands for FBPConvNet.}}
    \label{fig:wrs_fcn_box_snr}
    \vspace{-0.15in}
\end{figure}

\begin{figure}[!htbp]
    \centering
   	\includegraphics[width=0.45\textwidth]{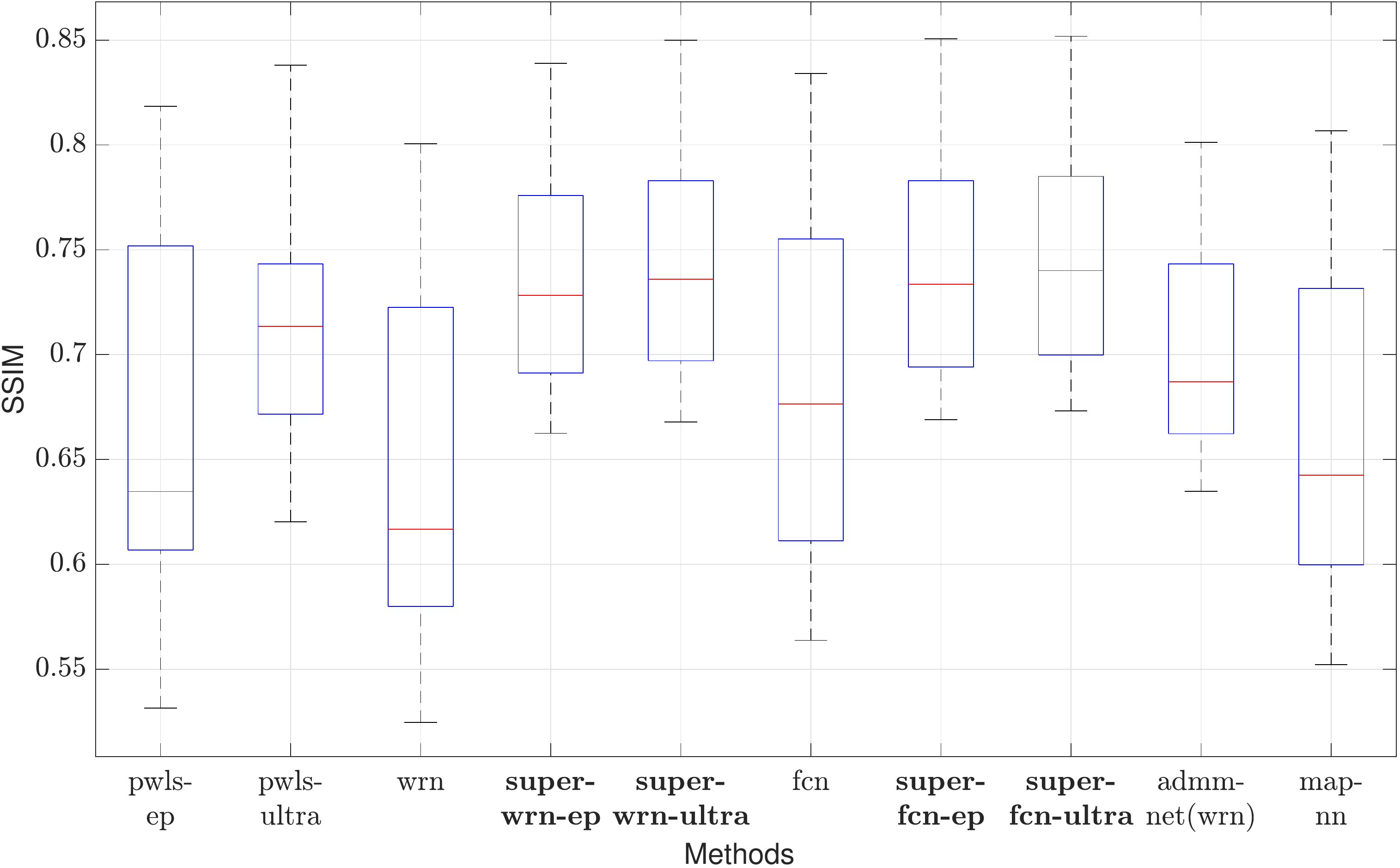}
    \caption{\BLUE{SSIM spread (shown using box plots) over $20$ test cases using different methods. Here, ``wrn" stands for WavResNet, ``fcn" stands for FBPConvNet.}}
    \label{fig:wrs_fcn_box_ssim}
    \vspace{-0.15in}
\end{figure}

We have shown the RMSE spread over $20$ test cases using different methods in the main paper (Fig.~1). Here, we show the SNR and SSIM variations over these test cases in Fig.~\ref{fig:wrs_fcn_box_snr} and Fig.~\ref{fig:wrs_fcn_box_ssim}, respectively. Both figures show the superiority of the proposed SUPER methods compared to standalone supervised and unsupervised methods in terms of robustness and numerical improvements.
\BLUE{SUPER methods also outperform the plug-and-play ADMM-Net method (ADMM-Net (WRN)).}

\subsection{Visual Results of SUPER}
\begin{figure*}[!t]
	\centering	
	\includegraphics[width=0.8\textwidth]{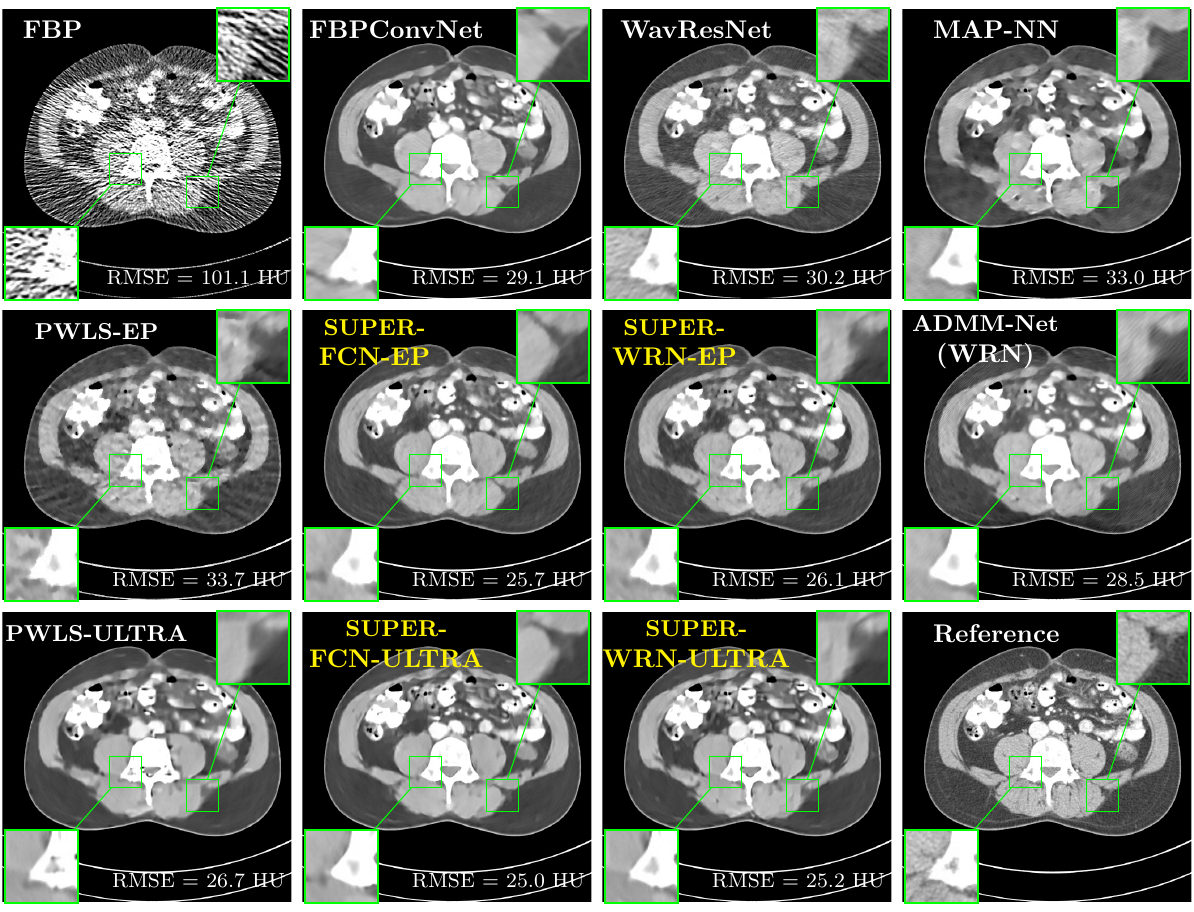}
	\caption{\BLUE{Reconstructions of slice~100 from patient L067 using various methods. The display window is [800 1200]~HU.}}
	\label{fig:super-all-L067s100}
	\vspace{-0.15in}
\end{figure*}

\begin{figure*}[!t]
	\centering	
	\includegraphics[width=0.8\textwidth]{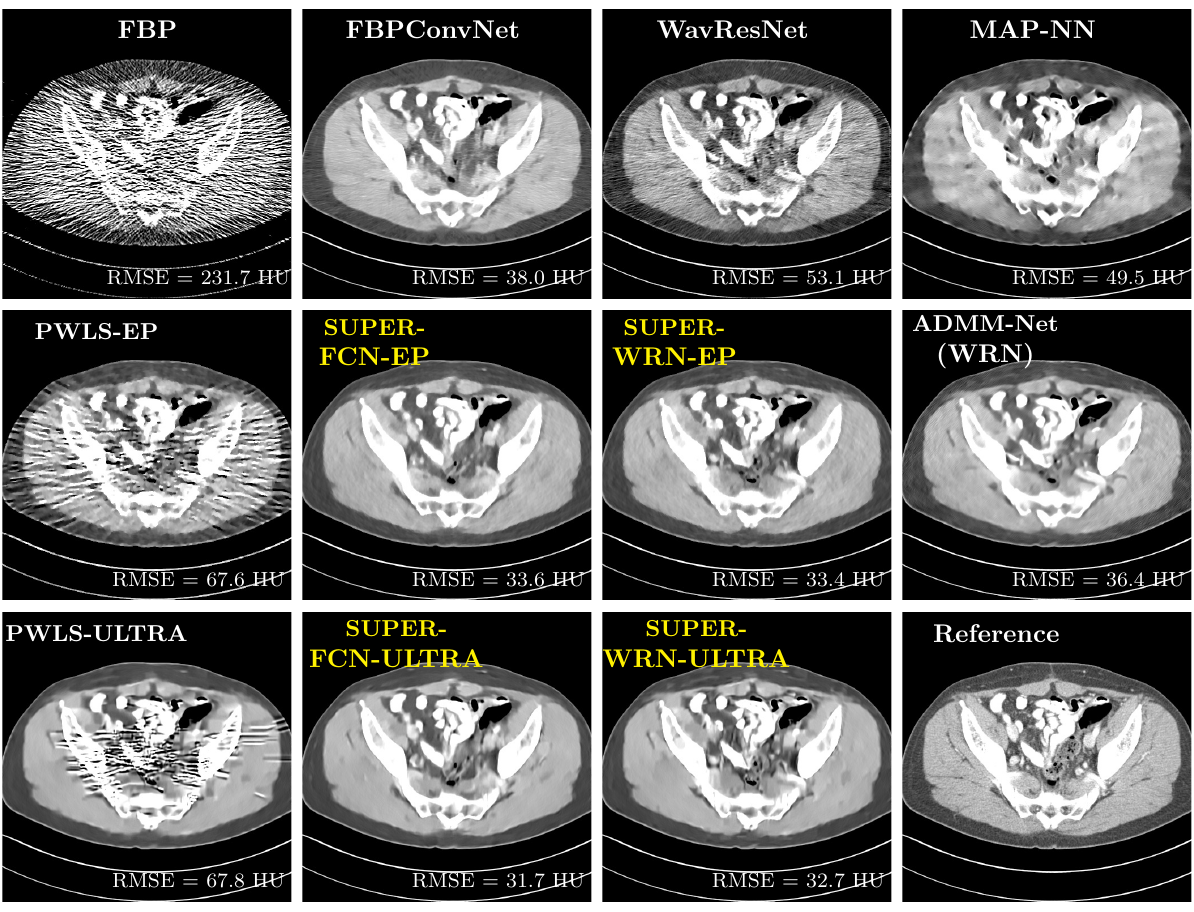}
	\caption{\BLUE{Reconstructions of slice~150 from patient L067 using various methods. The display window is [800 1200]~HU. This sample is with the most corrupted measurement data in the test set, and SUPER methods obviously reconstruct better images than the compared methods.}}
	\label{fig:super-all-L067s150}
	\vspace{-0.15in}
\end{figure*}

Fig.~2 of our main paper compared the reconstructions with various methods for one test sample.
Here, Fig.~\ref{fig:super-all-L067s100} and Fig.~\ref{fig:super-all-L067s150} show another two sets of comparisons for slice 100 and slice 150 of patient L067. Particularly, Fig.~\ref{fig:super-all-L067s150} shows reconstructions with the most corrupted measurement data in the test set, wherein both standalone iterative (unsupervised) methods PWLS-EP and PWLS-ULTRA have limited performance.
The proposed methods (SUPER-FCN-EP, SUPER-WRN-EP, SUPER-FCN-ULTRA, and SUPER-WRN-ULTRA) significantly reduce noise and artifacts, and improve the edge sharpness in the soft tissues and bones compared to other competing methods.

\subsection{\BLUE{Comparison of Bias and Standard Deviation Trade-offs}}
\begin{figure}[!htp]
    \centering
	\begin{subfigure}[h]{0.38\textwidth}
		\centering
		\includegraphics[width=0.8\textwidth]{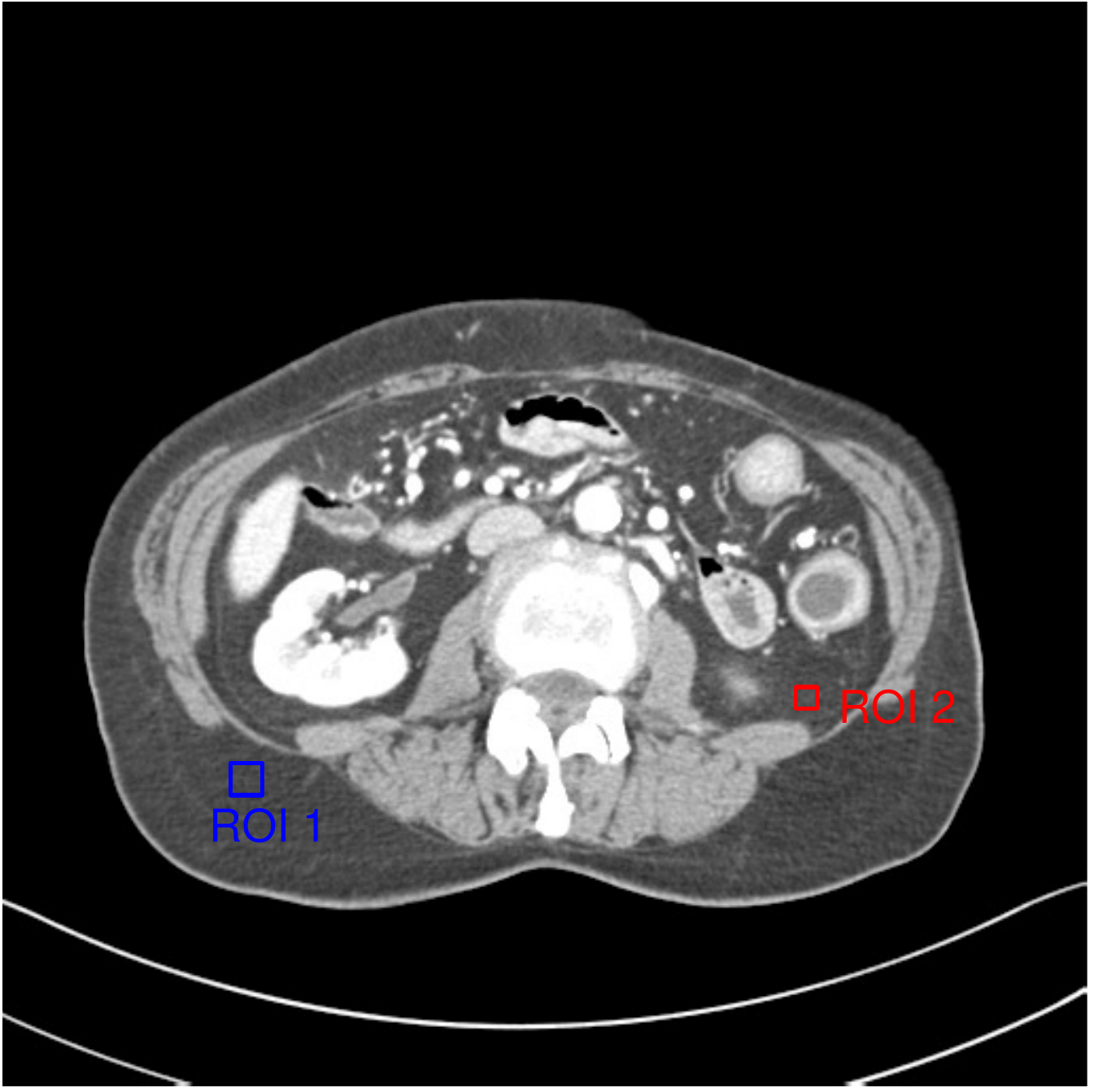}
		\caption{Reference Image (L192 slice 100)}
	\end{subfigure}
	\vfil
	\begin{subfigure}[h]{0.4\textwidth}
		\centering
		\includegraphics[width=1\textwidth]{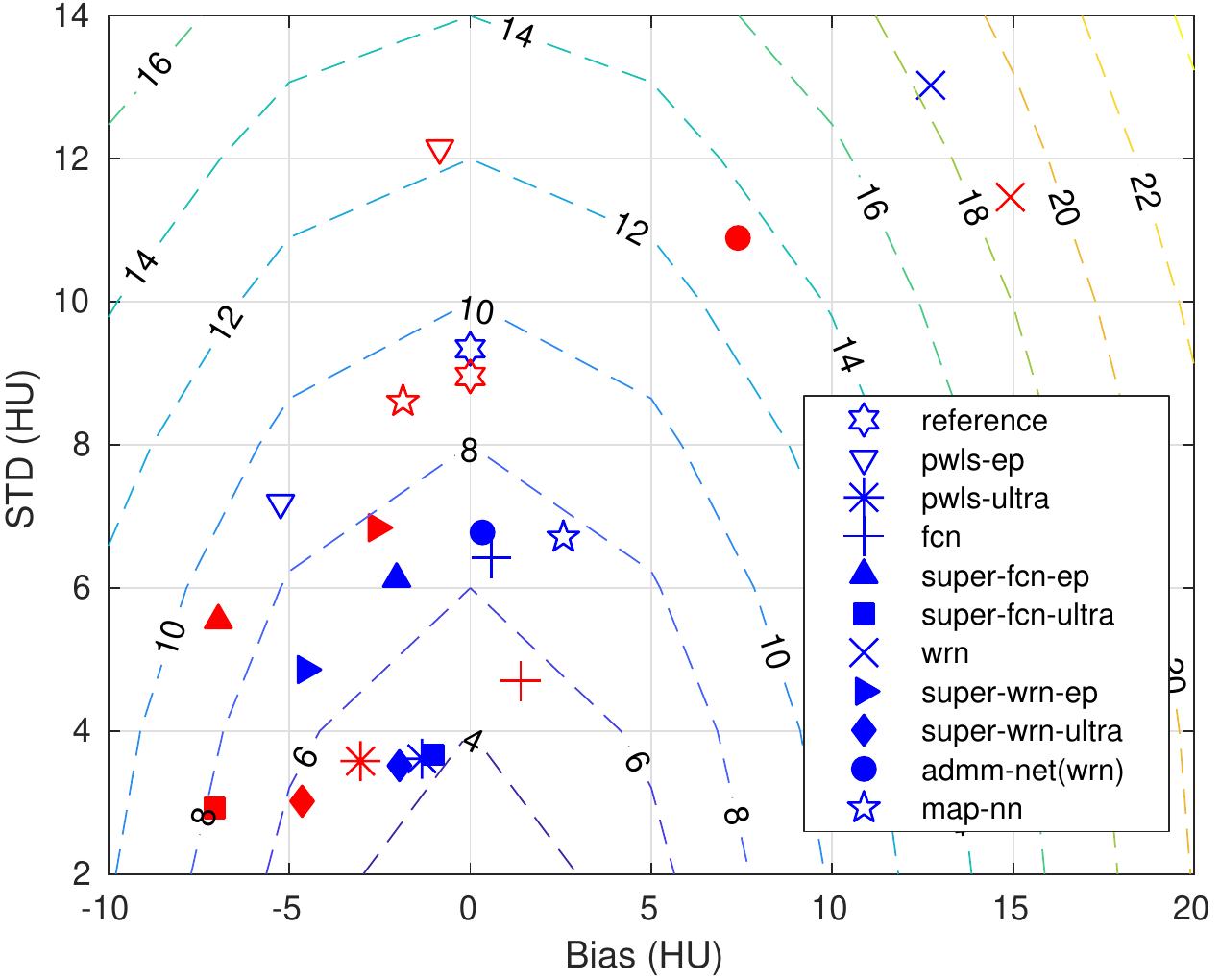}
		\caption{Bias-STD Plot}
		\label{fig:bias-std-plot}
	\end{subfigure}
    \caption{\BLUE{Bias-STD plots for two uniform ROIs indicated in (a) with different reconstruction methods. 
    In (b), the blue markers represent bias-std pairs for ROI~1 and the red markers represent bias-std pairs for ROI~2.}}
    \label{fig:bias-std}
    \vspace{-0.1in}
\end{figure}

\BLUE{We investigated the bias and standard deviation (STD) trade-offs of SUPER methods and the compared methods based on several uniform regions (ROIs) of the reconstructed images using such methods. 
The bias of a uniform ROI was computed as the mean error between the reconstructed ROI ($\hat{\x}_{ROI}$) and the reference ROI ($\x^*_{ROI}$), i.e., $\text{Bias} = \text{mean}(\hat{\x}_{ROI} - \x^*_{ROI})$, while the STD was calculated as STD($\x_{ROI}$), where $\x=\hat{\x}$ for reconstructed images and $\x=\x^*$ for the reference image.
To better illustrate the bias-STD trade-offs, we define a ``B-S Index" as $\sqrt{\text{Bias}^2 + \text{STD}^2}$. Smaller values of this metric imply better bias-STD trade-offs~\cite{zeng2011approximations}.
In Fig.~\ref{fig:bias-std}, we show two uniform regions (ROI~1 and ROI~2) of the test sample L192 slice 100 in the reference image and plot (bias,~STD) pairs \BLUE{(blue markers for ROI~1 and red markers for ROI~2} for these two ROIs reconstructed by different methods. 
From Fig.~\ref{fig:bias-std-plot}, we notice that although the two ROIs are located in different parts of the image, they have similar STD values (close to 10~HU) in the reference image~\BLUE{(which also contains noise).}
\BLUE{For both ROIs, we observe that SUPER methods (indicated in solid markers) have relatively stable and good bias-STD trade-offs with B-S indexes scattered around isocontours of $6\sim 8$~HU. The bias-STD trade-off of the plug-and-play ADMM method with WavResNet denoiser varies a lot for the different ROIs, while that of PWLS-ULTRA is small and stable for both ROIs. The supervised methods, MAP-NN, FBPConvNet, and WavResNet, have relatively stable bias-STD trade-offs in both ROIs, while FBPConvNet has the smallest B-S indexes and WavResNet has the largest B-S indexes among these three methods.
}
}


\subsection{Image Evolution over SUPER Layers with FBPConvNet-based Methods}
\begin{figure}[!htb]
    \centering
	\begin{tikzpicture}
			[spy using outlines={rectangle,green,magnification=2.3,size=10mm, connect spies}]
			\node {\includegraphics[width=0.22\textwidth]{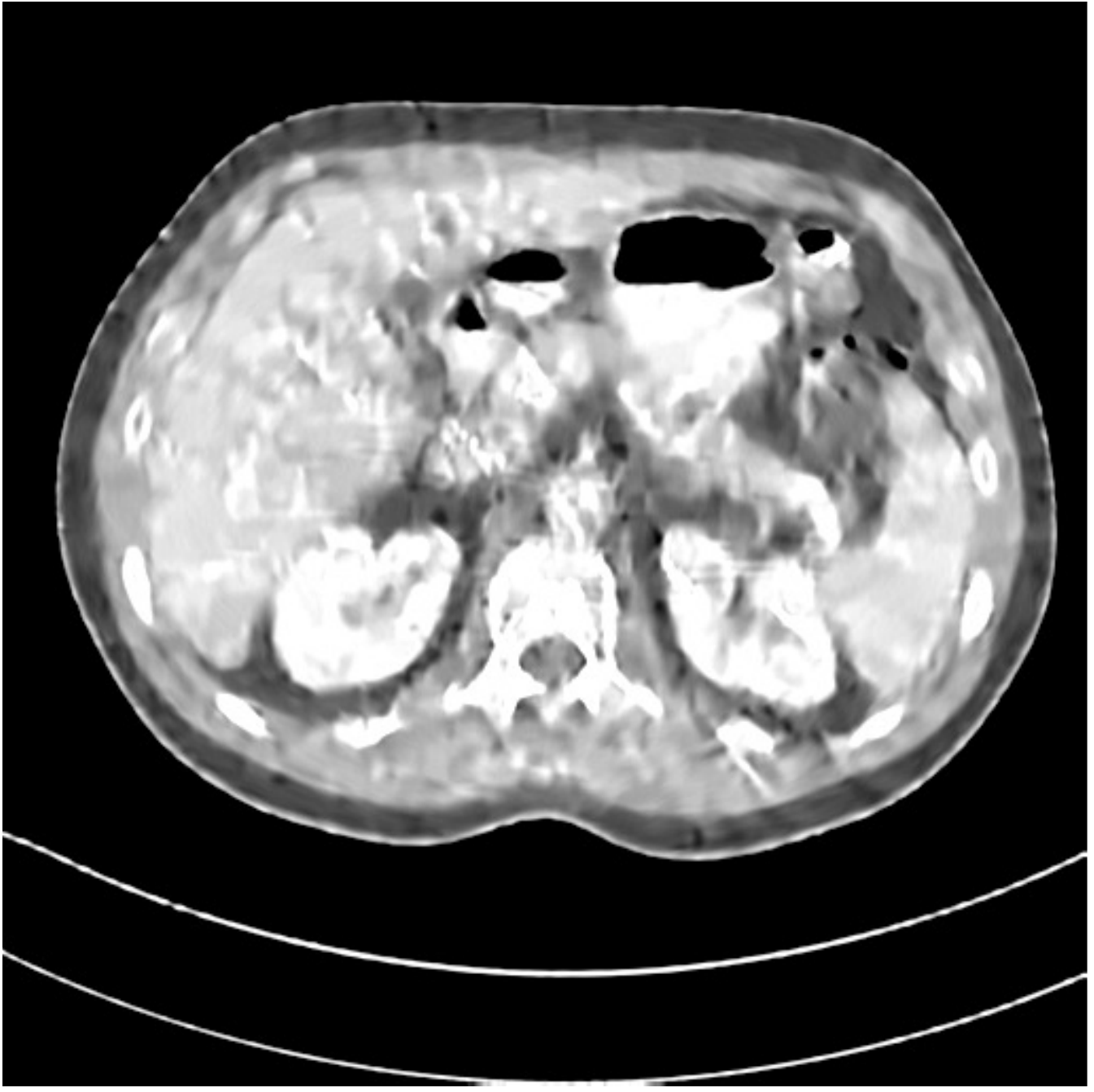} };
	        \spy on (0,0.6) in node [left] at (-1,-1.65);
			\node [align = center,white, font=\bf] at (-0.6,1.72) {\small Layer 1};
	\node [white, font=\footnotesize] at (0.7,-1.7) {RMSE = 37.00~HU};
	\end{tikzpicture}
	\begin{tikzpicture}
			[spy using outlines={rectangle,green,magnification=2.3,size=10mm, connect spies}]
			\node {\includegraphics[width=0.22\textwidth]{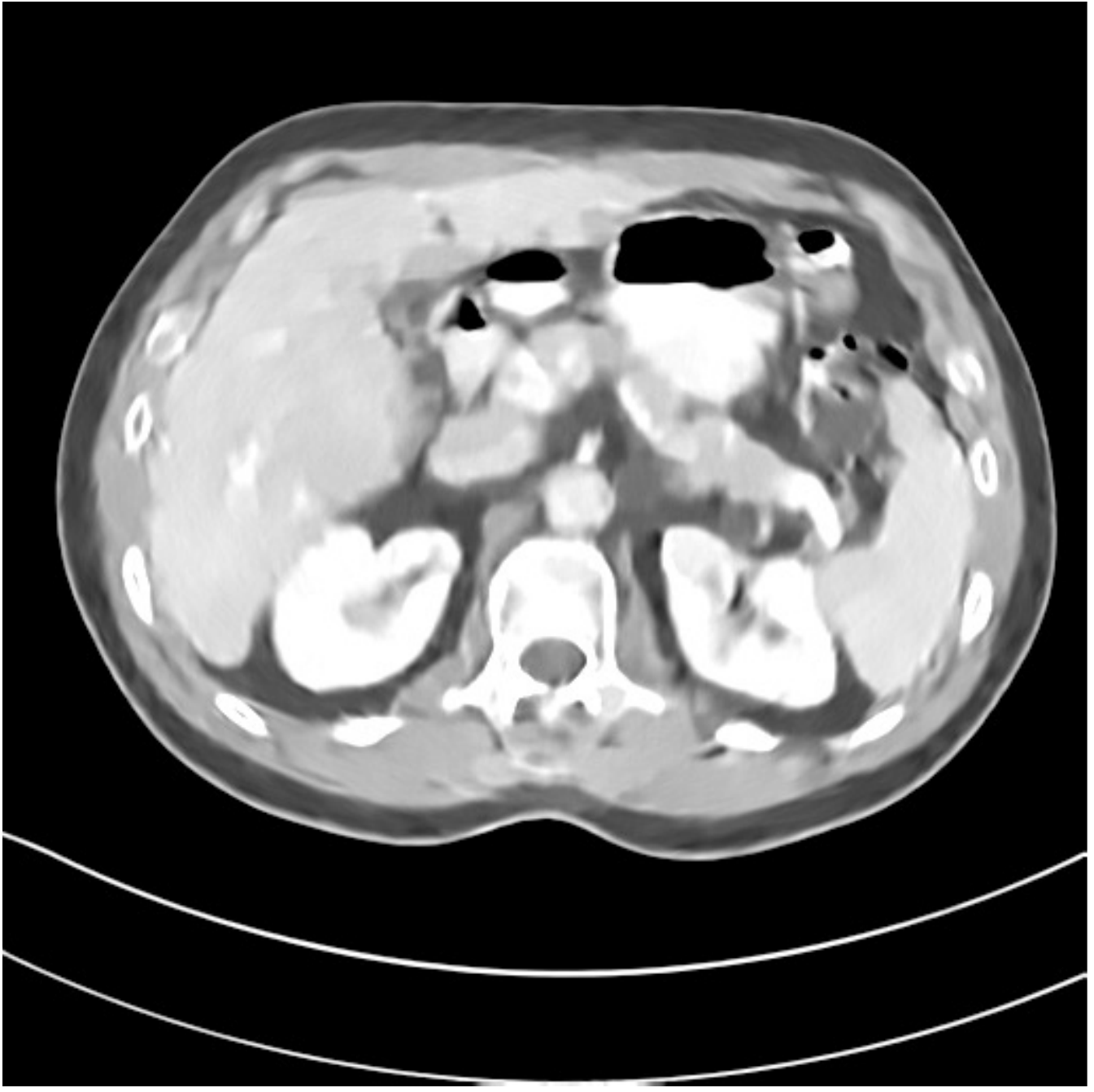} };
	        \spy on (0,0.6) in node [left] at (-1,-1.65);
			\node [align = center,white, font=\bf] at (-0.6,1.72) {\small Layer 5};
	\node [white, font=\footnotesize] at (0.7,-1.7) {RMSE = 29.78~HU};
	\end{tikzpicture} \\
	\begin{tikzpicture}
			[spy using outlines={rectangle,green,magnification=2.3,size=10mm, connect spies}]
			\node {\includegraphics[width=0.22\textwidth]{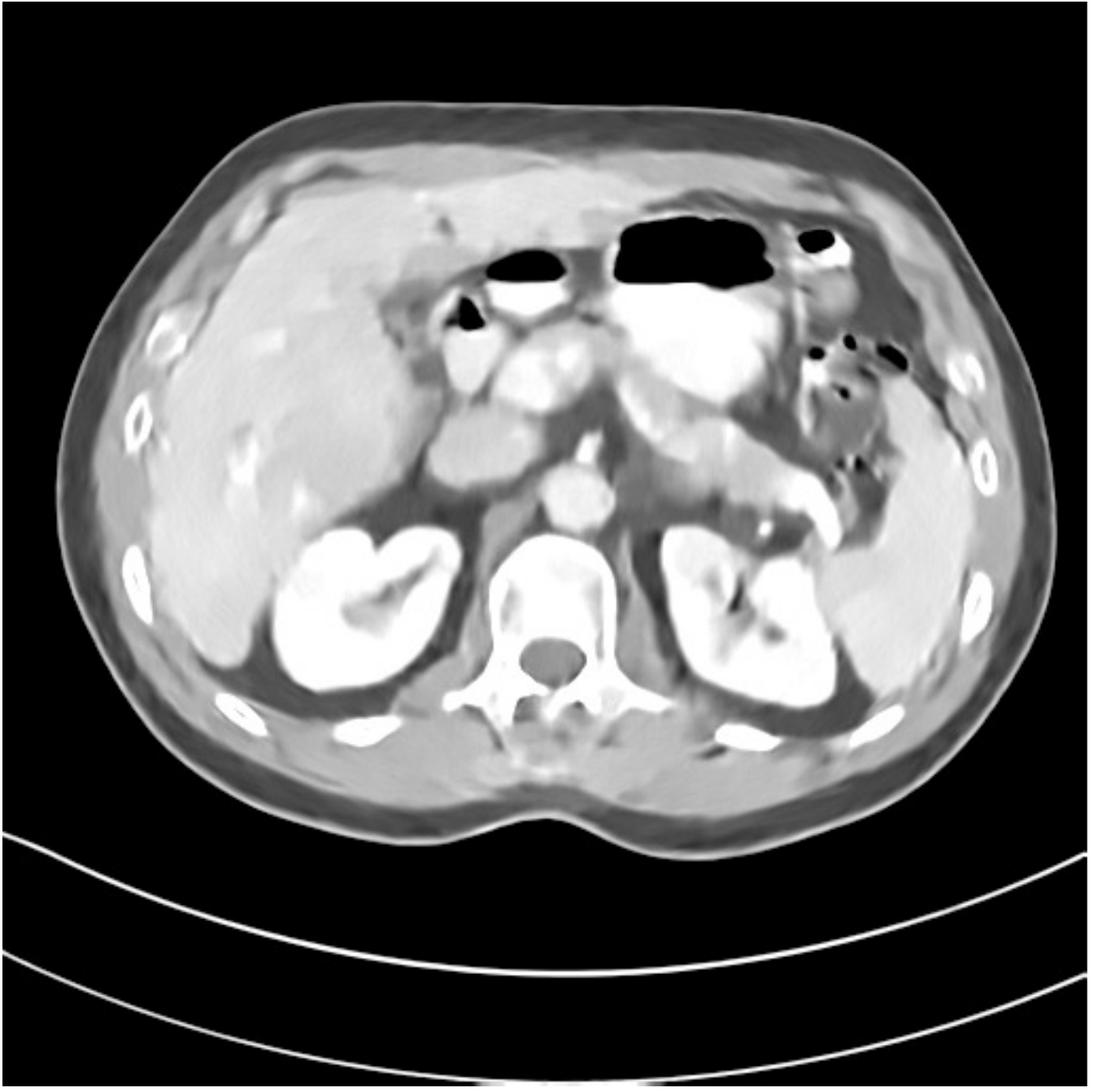} };
	        \spy on (0,0.6) in node [left] at (-1,-1.65);
			\node [align = center,white, font=\bf] at (-0.6,1.72) {\small Layer 11};
	\node [white, font=\footnotesize] at (0.7,-1.7) {RMSE = 29.14~HU};
	\end{tikzpicture}
	\begin{tikzpicture}
			[spy using outlines={rectangle,green,magnification=2.3,size=10mm, connect spies}]
			\node {\includegraphics[width=0.22\textwidth]{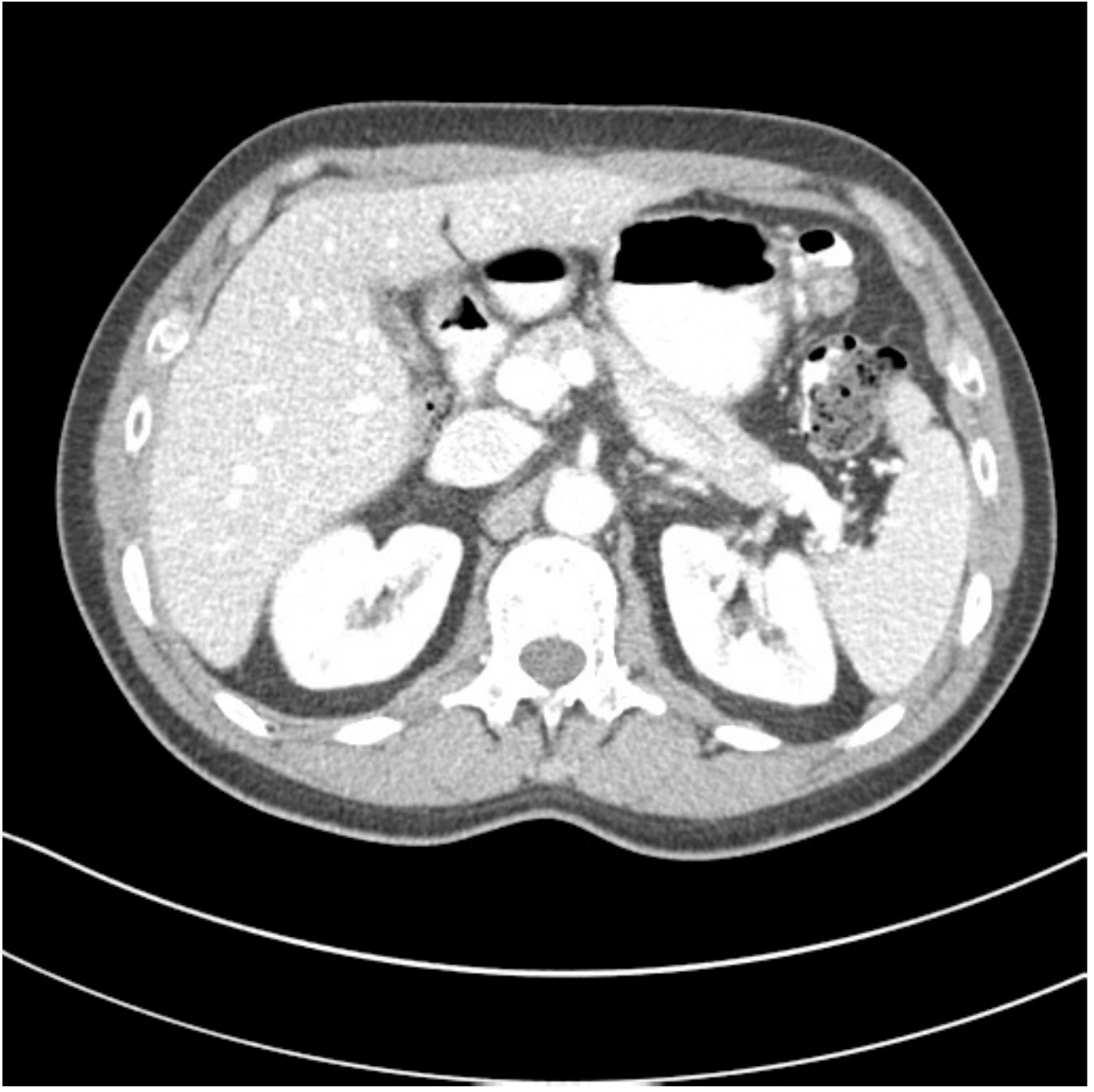} };
	        \spy on (0,0.6) in node [left] at (-1,-1.65);
			\node [align = center,white, font=\bf] at (-0.6,1.72) {\small Reference};
	\end{tikzpicture}
    \caption{Image evolution over SUPER layers using SUPER-FCN-ULTRA method. }
    \label{fig:FCN_ultra_lyrs-L067s50}
\end{figure}
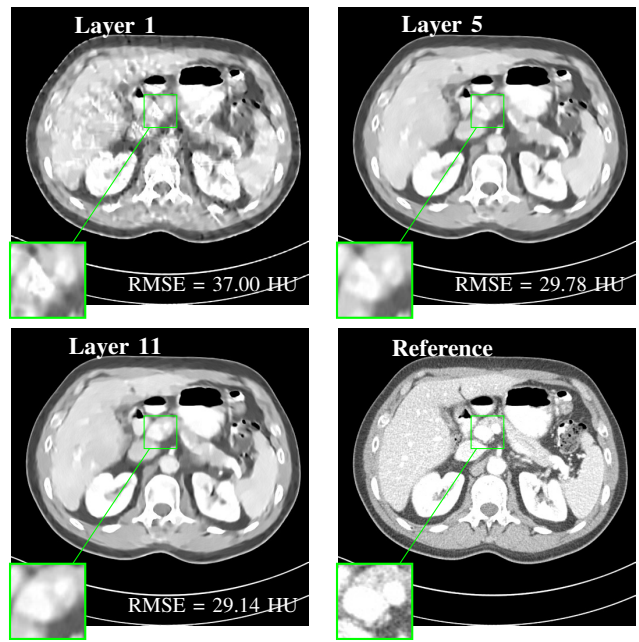

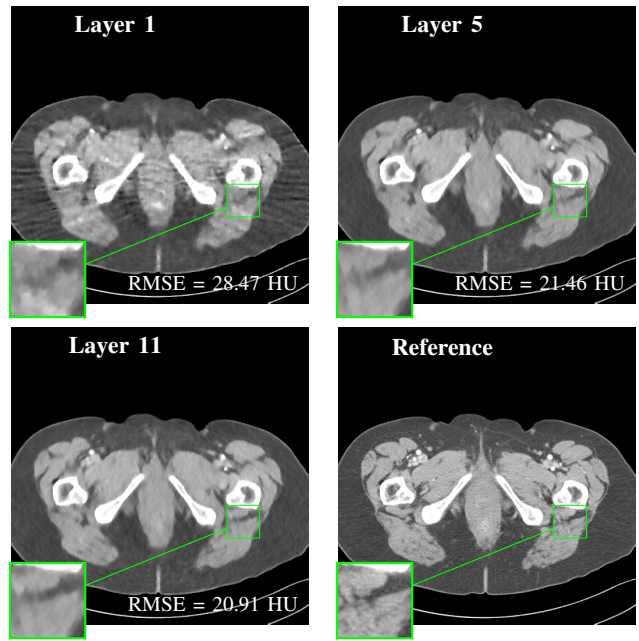
\begin{figure}[!htb]
    \centering
	\begin{tikzpicture}
			[spy using outlines={rectangle,green,magnification=2.3,size=10mm, connect spies}]
			\node {\includegraphics[width=0.22\textwidth]{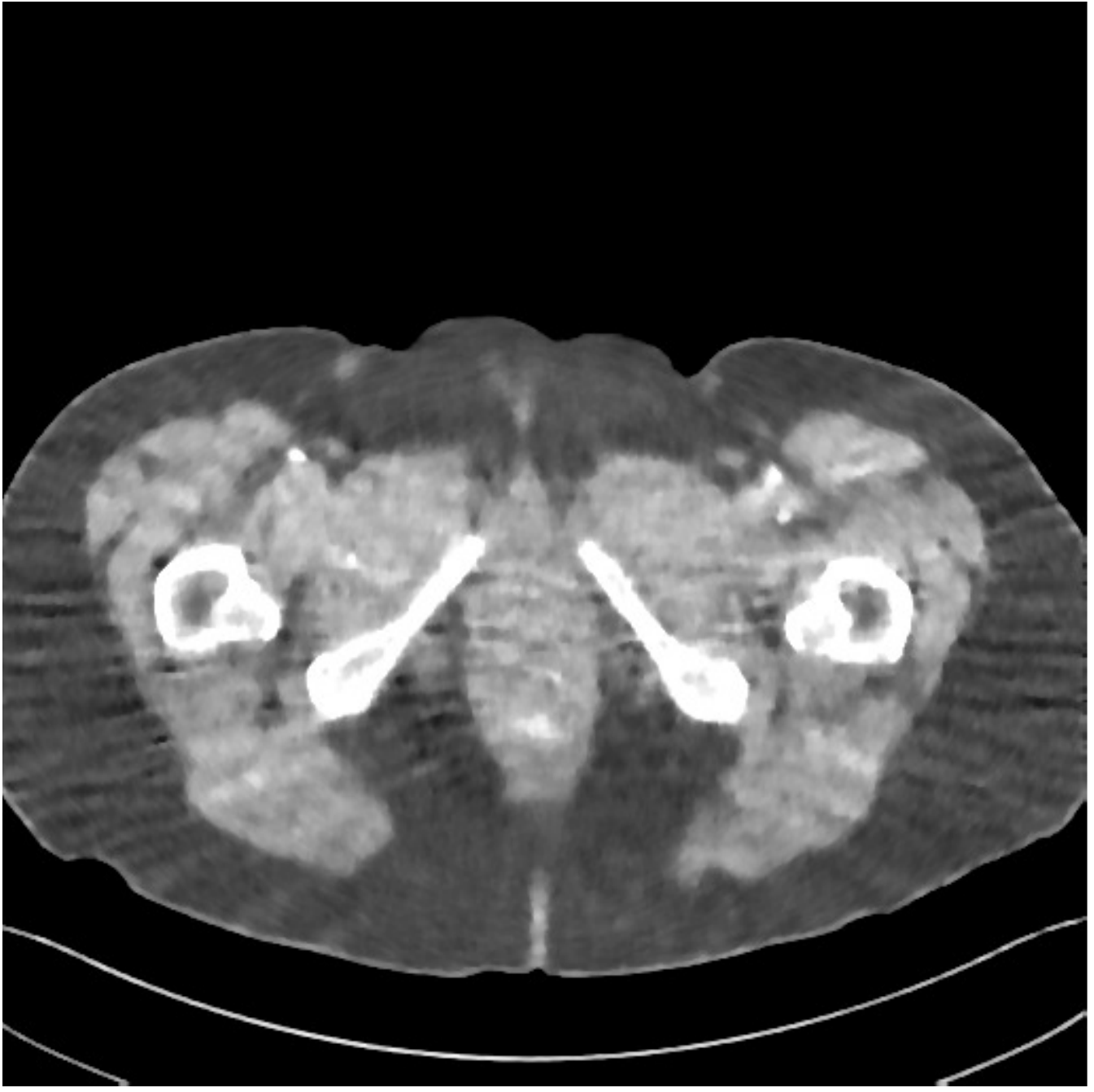} };
	        \spy on (1.1,-0.6) in node [left] at (-1,-1.65);
			\node [align = center,white, font=\bf] at (-0.6,1.72) {\small Layer 1};
	\node [white, font=\footnotesize] at (0.7,-1.7) {RMSE = 28.47~HU};
	\end{tikzpicture}
	\begin{tikzpicture}
			[spy using outlines={rectangle,green,magnification=2.3,size=10mm, connect spies}]
			\node {\includegraphics[width=0.22\textwidth]{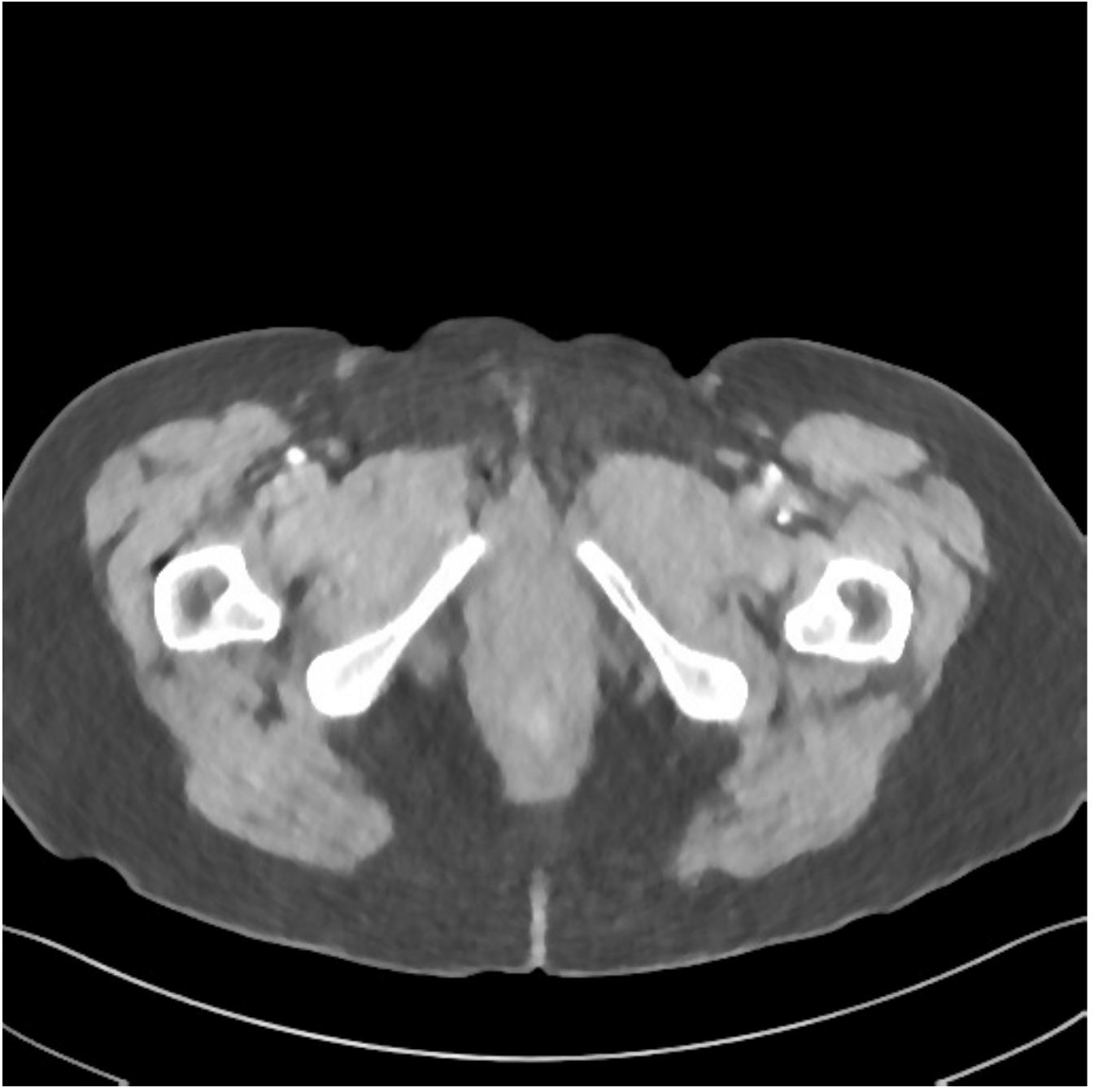} };
	        \spy on (1.1,-0.6) in node [left] at (-1,-1.65);
			\node [align = center,white, font=\bf] at (-0.6,1.72) {\small Layer 5};
	\node [white, font=\footnotesize] at (0.7,-1.7) {RMSE = 21.46~HU};
	\end{tikzpicture} \\
	\begin{tikzpicture}
			[spy using outlines={rectangle,green,magnification=2.3,size=10mm, connect spies}]
			\node {\includegraphics[width=0.22\textwidth]{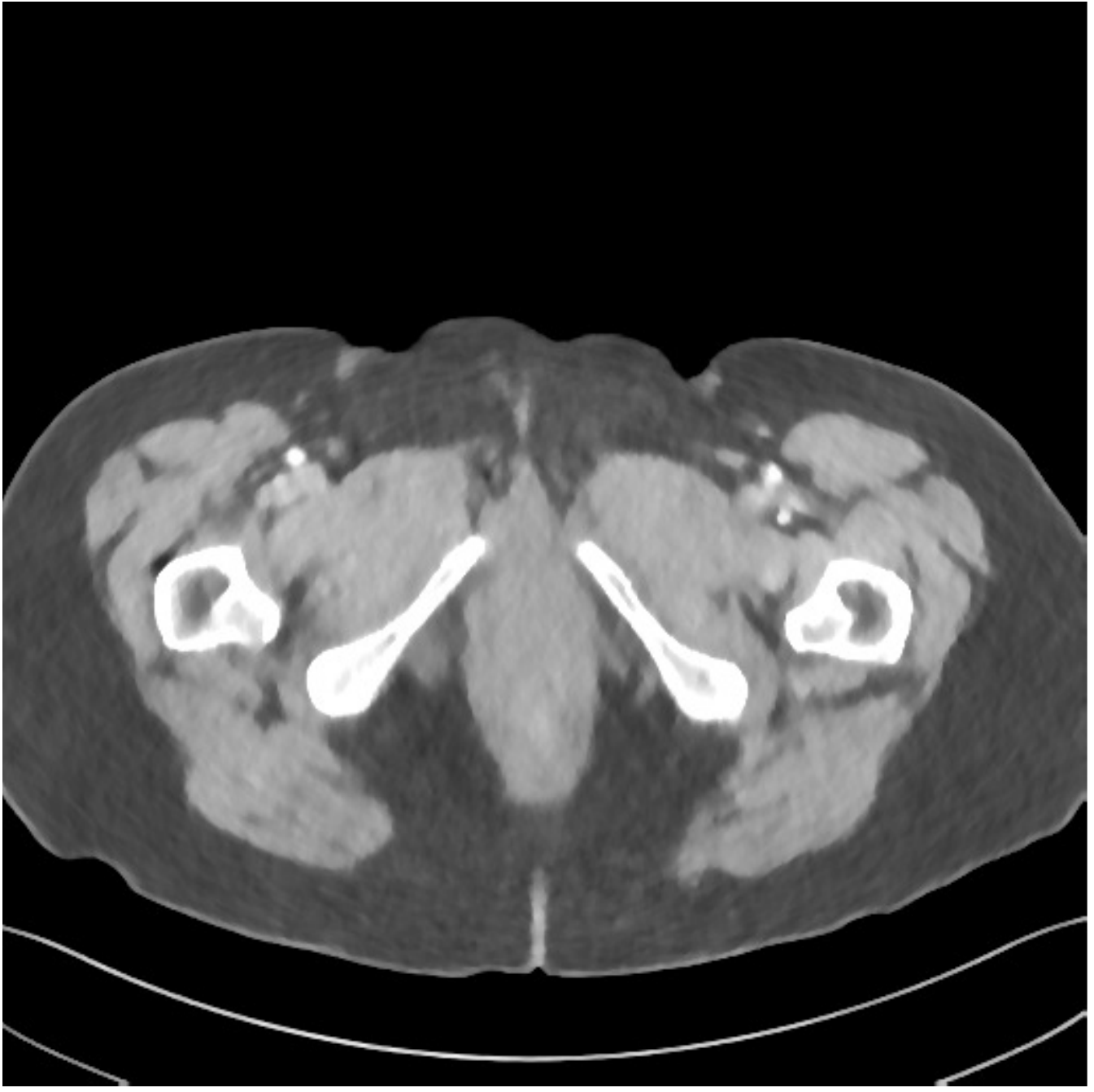} };
	        \spy on (1.1,-0.6) in node [left] at (-1,-1.65);
			\node [align = center,white, font=\bf] at (-0.6,1.72) {\small Layer 11};
	\node [white, font=\footnotesize] at (0.7,-1.7) {RMSE = 20.91~HU};
	\end{tikzpicture}
	\begin{tikzpicture}
			[spy using outlines={rectangle,green,magnification=2.3,size=10mm, connect spies}]
			\node {\includegraphics[width=0.22\textwidth]{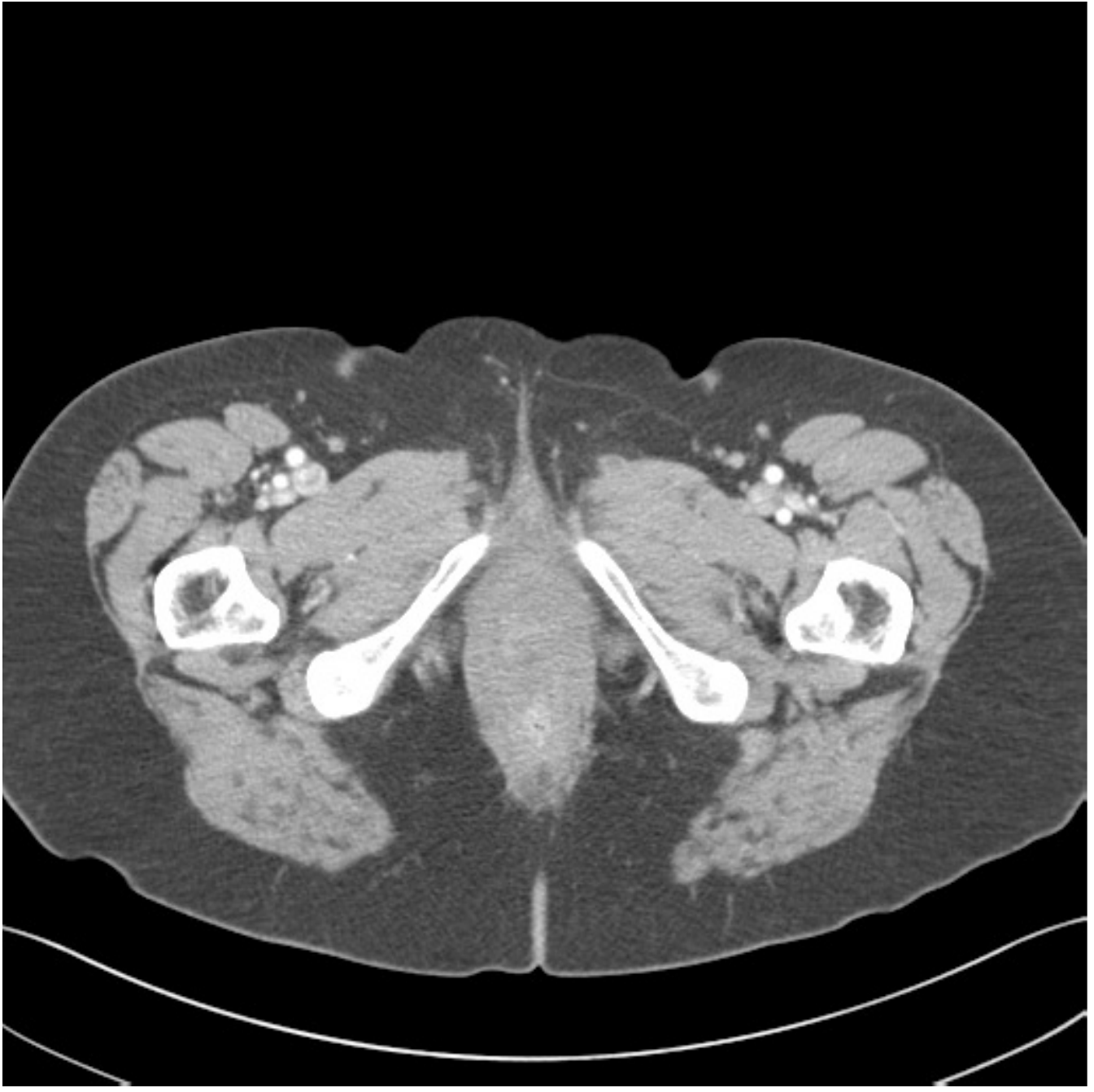} };
	        \spy on (1.1,-0.6) in node [left] at (-1,-1.65);
			\node [align = center,white, font=\bf] at (-0.6,1.72) {\small Reference};
	\end{tikzpicture}
    \caption{Image evolution over SUPER layers using SUPER-FCN-EP method.}
    \label{fig:FCN_ep_lyrs-L310s200}
\end{figure}

We show examples of image evolution across SUPER layers using SUPER-FCN-ULTRA in Fig.~\ref{fig:FCN_ultra_lyrs-L067s50}, and using SUPER-FCN-EP in Fig.~\ref{fig:FCN_ep_lyrs-L310s200}, respectively. Both examples indicate that early SUPER layers strongly suppress noise and artifacts, while later SUPER layers help with reconstructing detailed structures.

\subsection{Influence of $\mu$ Choice for FBPConvNet-based SUPER}
Table~\ref{tab:fcn-mu} shows that choosing the same $\mu$ value (i.e., the weight for the regularizer involving the deep CNN learned in a supervised manner) during training and testing is quite effective.

  \begin{table}[!htbp]
		\caption{Mean RMSE (HU) of 20 test slices using different $\mu$ values in FBPConvNet based SUPER.}
	\begin{subtable}[htbp]{0.48\textwidth}
		\centering
	    \caption{SUPER-FCN-EP}
    	\begin{tabular}{c|c c c}
	     \toprule
	    	\diagbox{train}{test}&\shortstack{$\mu = 0$}&\shortstack{$\mu = 5\times 10^4$}  &\shortstack{$\mu = 5\times 10^5$}  \\ 
        	\hline
        	$\mu = 0$ & \textbf{26.7} & 27.0  	&  59.9 
        	\\ 
        	$\mu = 5\times 10^4$ & 27.3 &\textbf{26.0}  	&  41.0
        	\\ 
        	$\mu = 5\times 10^5$ & 30.2 &29.3  	&  \textbf{26.3}
             \\	\bottomrule
        \end{tabular} 
    \end{subtable}
    \hspace{\fill}
	\begin{subtable}[htbp]{0.48\textwidth}
		\centering
	    \caption{SUPER-FCN-ULTRA}
        \begin{tabular}{c|c c c}
        \toprule
    	\diagbox{train}{test}&\shortstack{$\mu = 0$}&\shortstack{$\mu = 5\times 10^5$}&\shortstack{$\mu = 5\times 10^6$}   \\ 
		\hline 
		$\mu = 0$&  \textbf{25.3} &  \textbf{25.3}	&  29.5
		\\
		$\mu = 5\times 10^5$& 25.3 & \textbf{25.0} 	& 28.3
		\\ 
		$\mu = 5\times 10^6$& 26.3 &26.2  	& \textbf{25.1}
        \\	\bottomrule
\end{tabular} 
\end{subtable}
\label{tab:fcn-mu}
\end{table}

\subsection{SUPER with Only Data-Fidelity Cost}
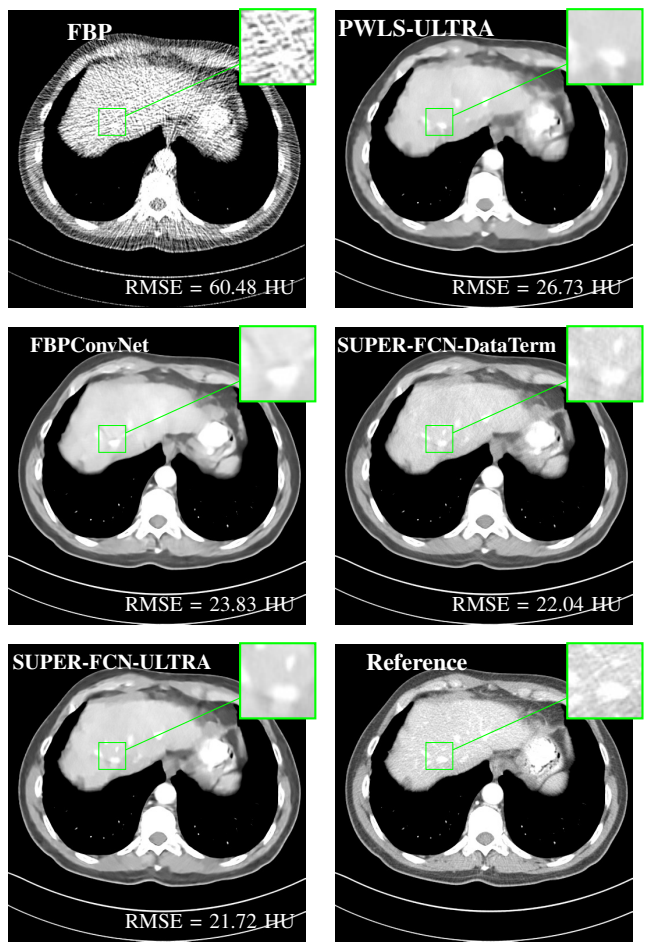
\begin{figure}[!htb]
	\centering
	\begin{tikzpicture}
			[spy using outlines={rectangle,green,magnification=2.8,size=10mm, connect spies}]
			\node {\includegraphics[width=0.22\textwidth]{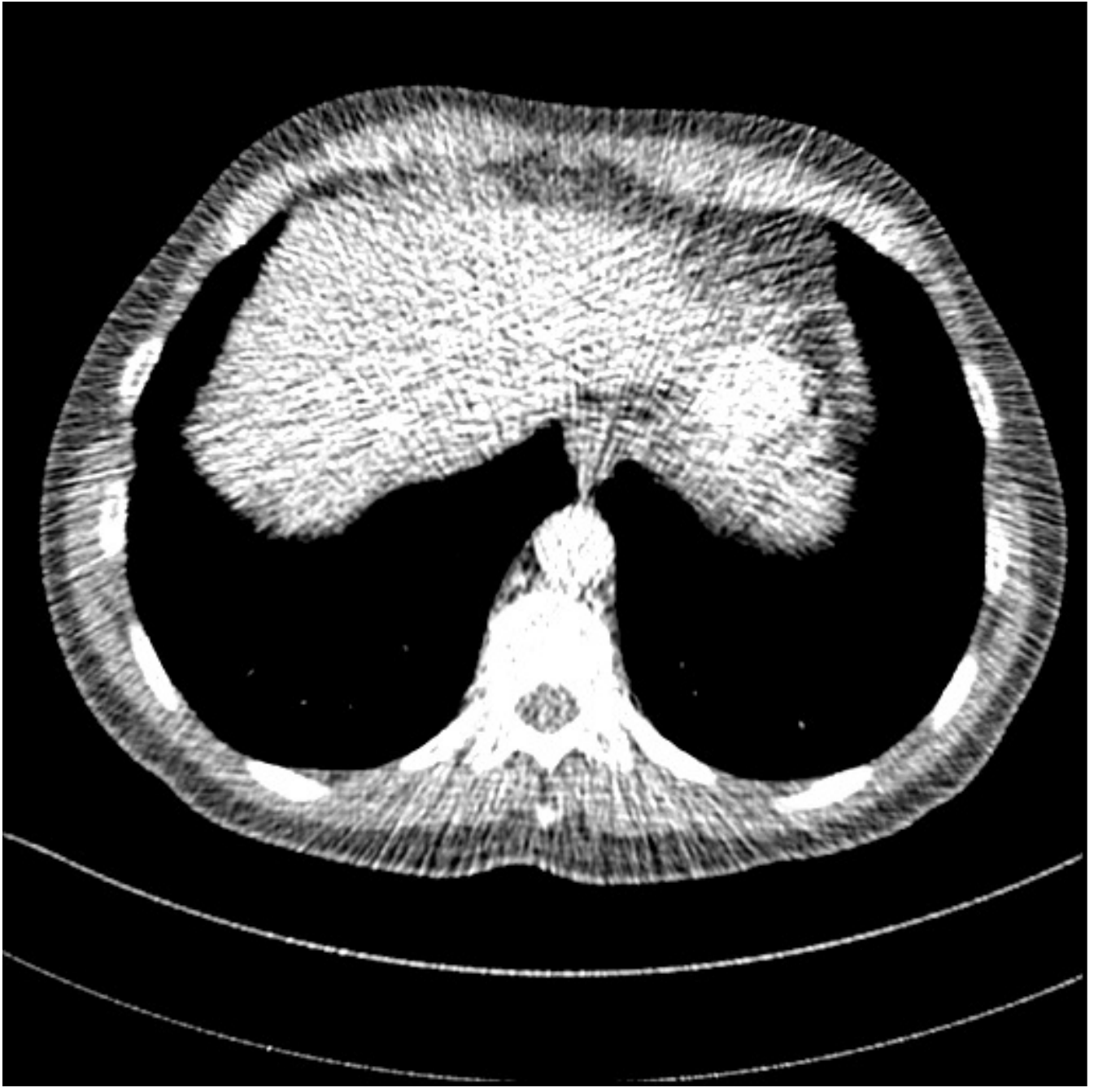} };
			\spy on (-0.6,0.5) in node [right] at (1.1,1.5);
			\node [align = center,white, font=\bf] at (-0.9,1.7){\small FBP};
			\node [white, font=\footnotesize] at (0.7,-1.7) {RMSE = 60.48~HU};
	\end{tikzpicture}
	\begin{tikzpicture}
			[spy using outlines={rectangle,green,magnification=2.8,size=10mm, connect spies}]
			\node {\includegraphics[width=0.22\textwidth]{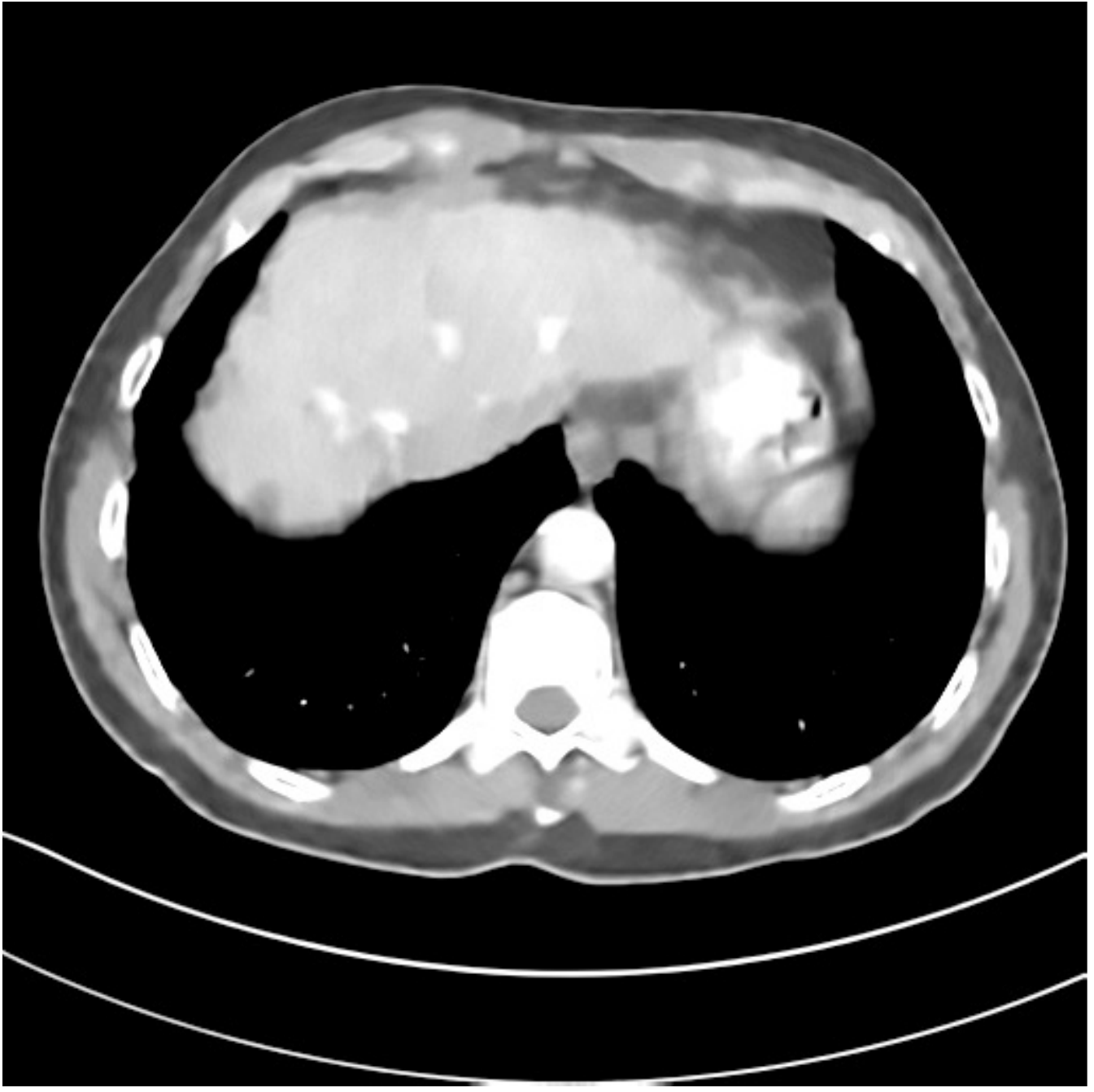} };
			\spy on (-0.6,0.5) in node [right] at (1.1,1.5);
			\node [align = center,white, font=\bf] at (-0.9,1.75) {\small PWLS-ULTRA};
		    \node [white, font=\footnotesize] at (0.7,-1.7) {RMSE = 26.73~HU};
	\end{tikzpicture}  \\
	\begin{tikzpicture}
			[spy using outlines={rectangle,green,magnification=2.8,size=10mm, connect spies}]
			\node {\includegraphics[width=0.22\textwidth]{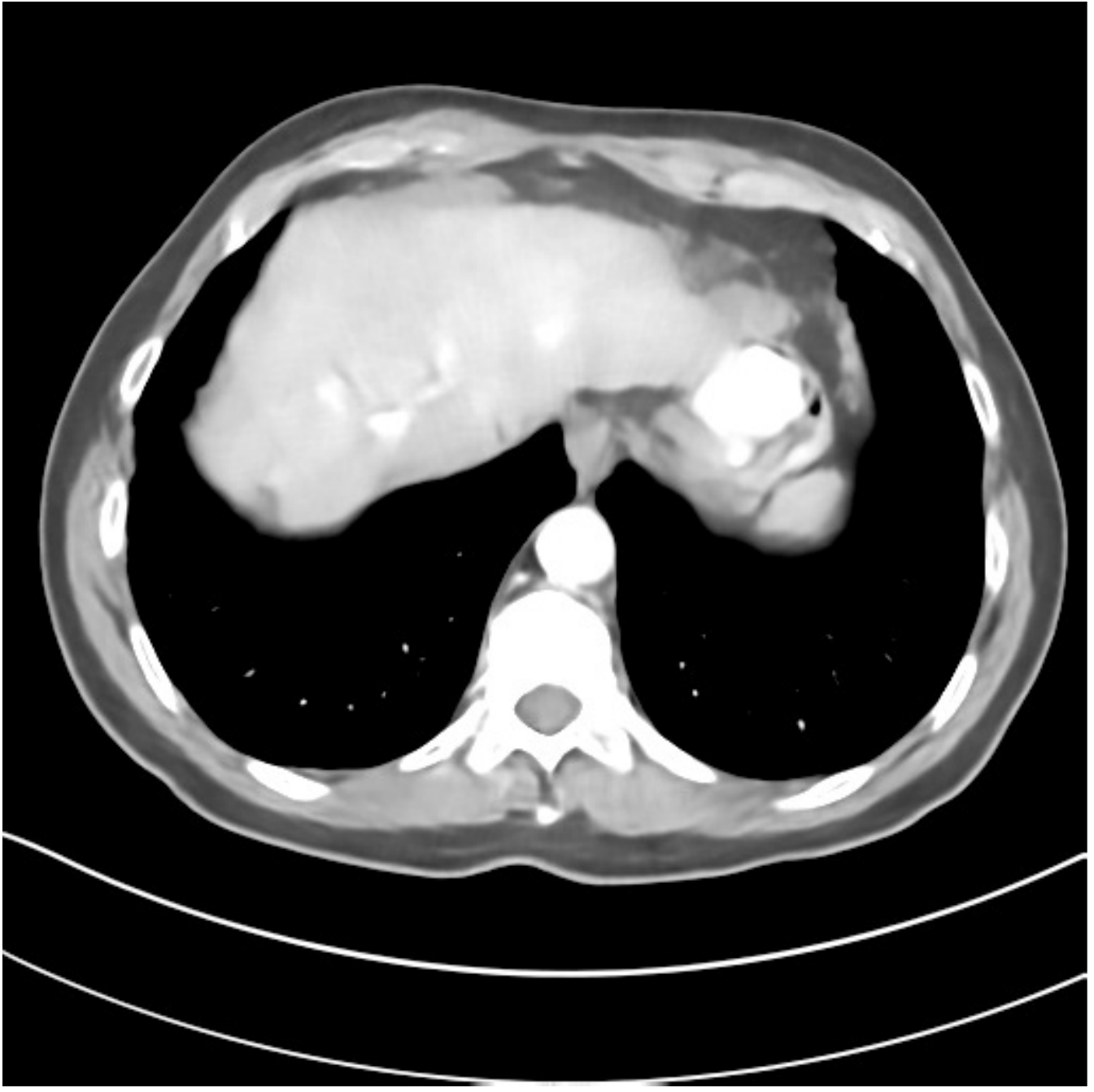} };
			\spy on (-0.6,0.5) in node [right] at (1.1,1.5);
			\node [align = center,white, font=\bf] at (-0.9,1.75){\footnotesize FBPConvNet};
			\node [white, font=\footnotesize] at (0.7,-1.7) {RMSE = 23.83~HU};
	\end{tikzpicture}
    \begin{tikzpicture}
    			[spy using outlines={rectangle,green,magnification=2.8,size=10mm, connect spies}]
    			\node {\includegraphics[width=0.22\textwidth]{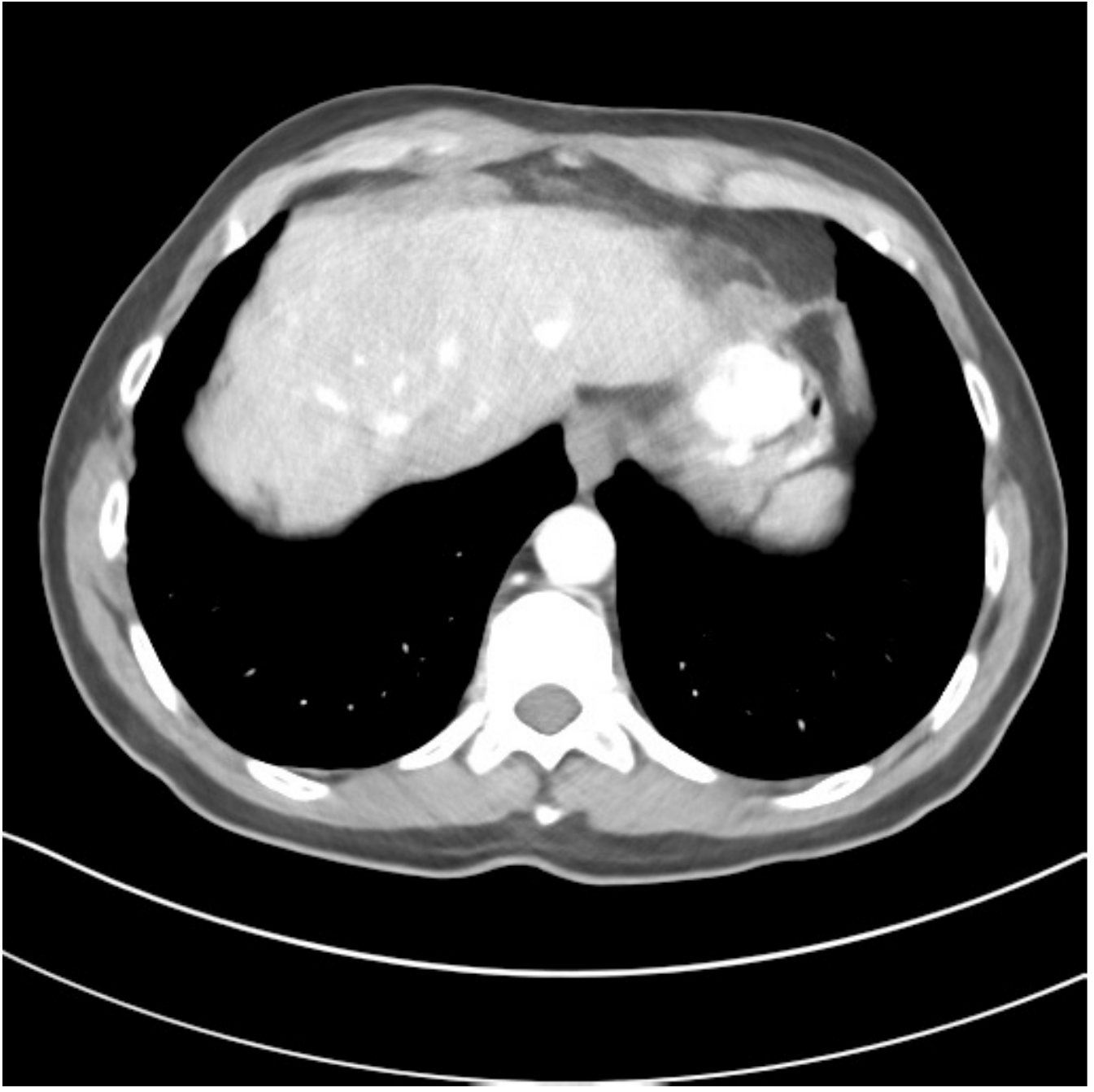} };
    			\spy on (-0.6,0.5) in node [right] at (1.1,1.5);
    			\node [align = center,white, font=\bf] at (-0.5,1.75){\footnotesize SUPER-FCN-DataTerm};
    			\node [white, font=\footnotesize] at (0.7,-1.7) {RMSE = 22.04~HU};
    	\end{tikzpicture}\\
    \begin{tikzpicture}
			[spy using outlines={rectangle,green,magnification=2.8,size=10mm, connect spies}]
			\node {\includegraphics[width=0.22\textwidth]{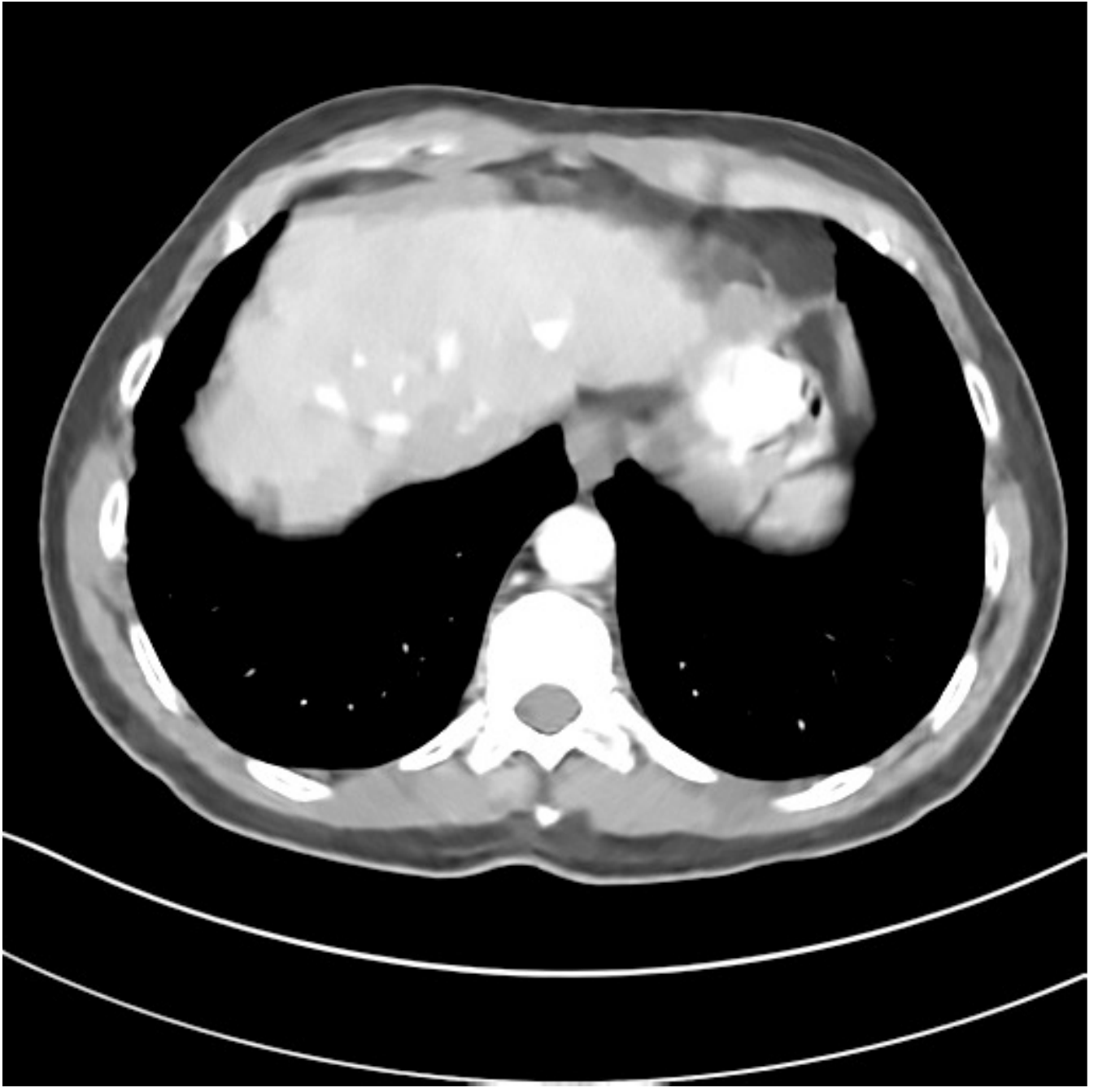} };
			\spy on (-0.6,0.5) in node [right] at (1.1,1.5);
			\node [align = center,white, font=\bf] at (-0.6,1.75) {\footnotesize SUPER-FCN-ULTRA};
	\node [white, font=\footnotesize] at (0.7,-1.7) {RMSE = 21.72~HU};
	\end{tikzpicture}
    \begin{tikzpicture}
			[spy using outlines={rectangle,green,magnification=2.8,size=10mm, connect spies}]
			\node {\includegraphics[width=0.22\textwidth]{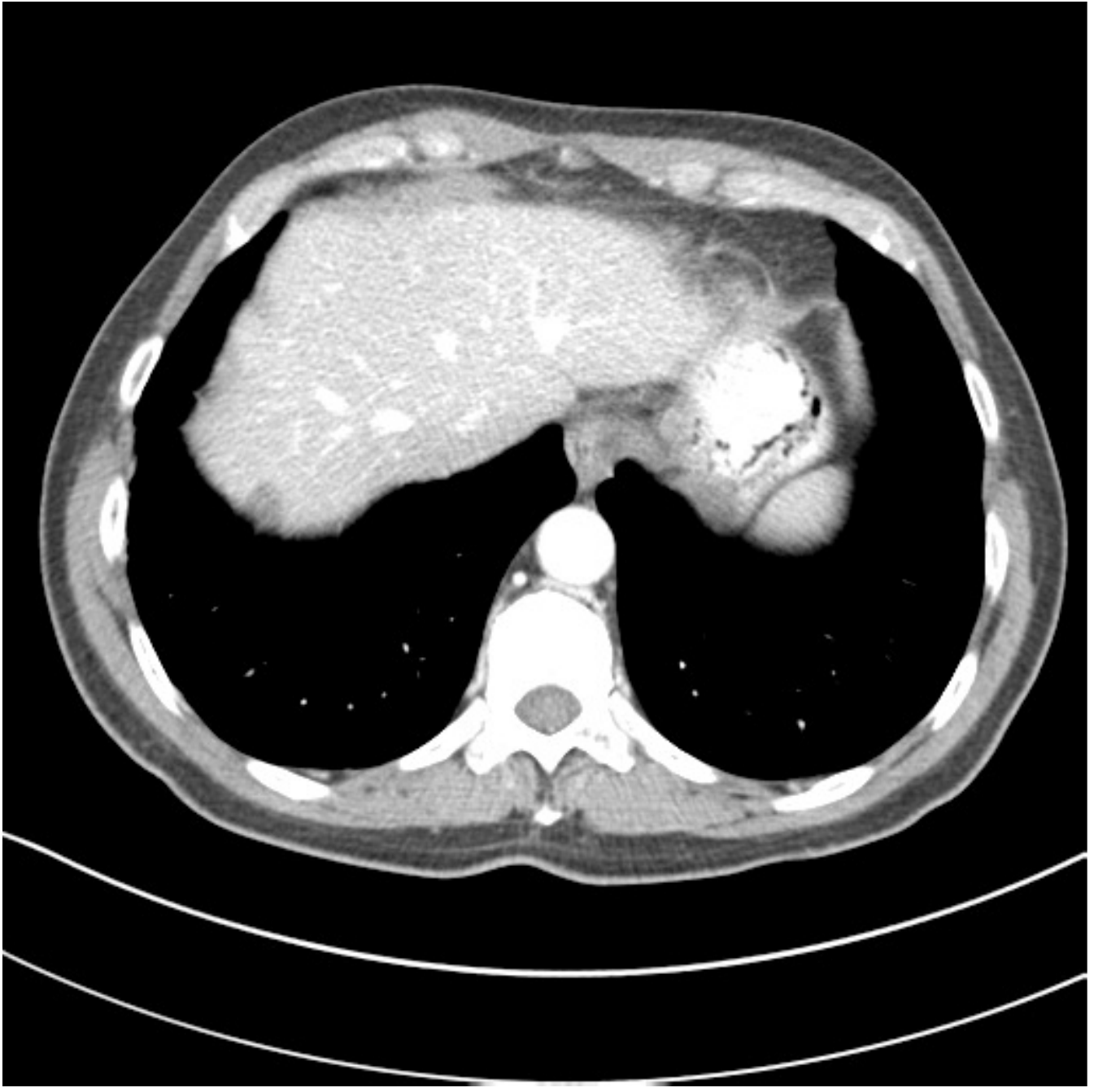} };
			\spy on (-0.6,0.5) in node [right] at (1.1,1.5);
			\node [align = center,white, font=\bf] at (-0.9,1.75) {\small Reference};
	\end{tikzpicture}
	\caption{Reconstructed images of L067 slice20 using FBP, PWLS-ULTRA, FBPConvNet, SUPER-FCN-DataTerm, SUPER-FCN-ULTRA, and the reference image, respectively.}
	\label{fig:Data-term-L067s20}
\end{figure}

\BLUE{Section~\uppercase\expandafter{\romannumeral4}.F~(2)} of our manuscript~\cite{SUPER-as-submit} shows reconstructions of one test sample (slice 150 of patient L192) using FBPConvNet, SUPER-FCN-DataTerm, PWLS-ULTRA, and SUPER-FCN-ULTRA. 
Fig.~\ref{fig:Data-term-L067s20} shows the comparisons for another test slice. 
We observe the similar phenomenon as in the main paper that SUPER-FCN-DataTerm outperforms the standalone FBPConvNet method and PWLS-ULTRA method, while the unsupervised regularizer involved SUPER-FCN-ULTRA method further improves the reconstruction qualities.

\begin{figure}[!htp]
    \centering
	\begin{subfigure}[h]{0.22\textwidth}
		\centering
		\includegraphics[width=1\textwidth]{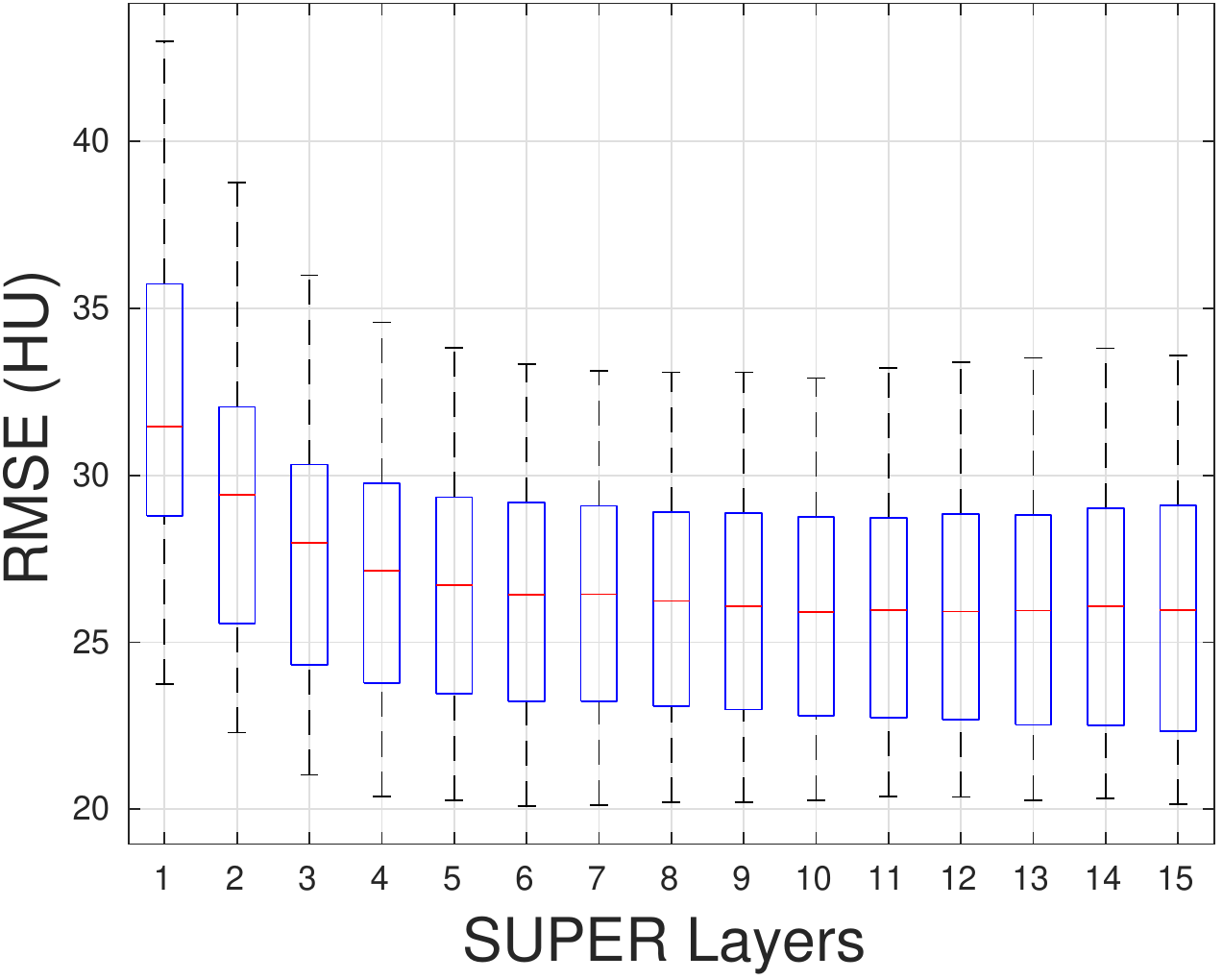}
		\caption{SUPER-FCN-EP}
	\end{subfigure}
	\hfil
	\begin{subfigure}[h]{0.22\textwidth}
		\centering
		\includegraphics[width=1\textwidth]{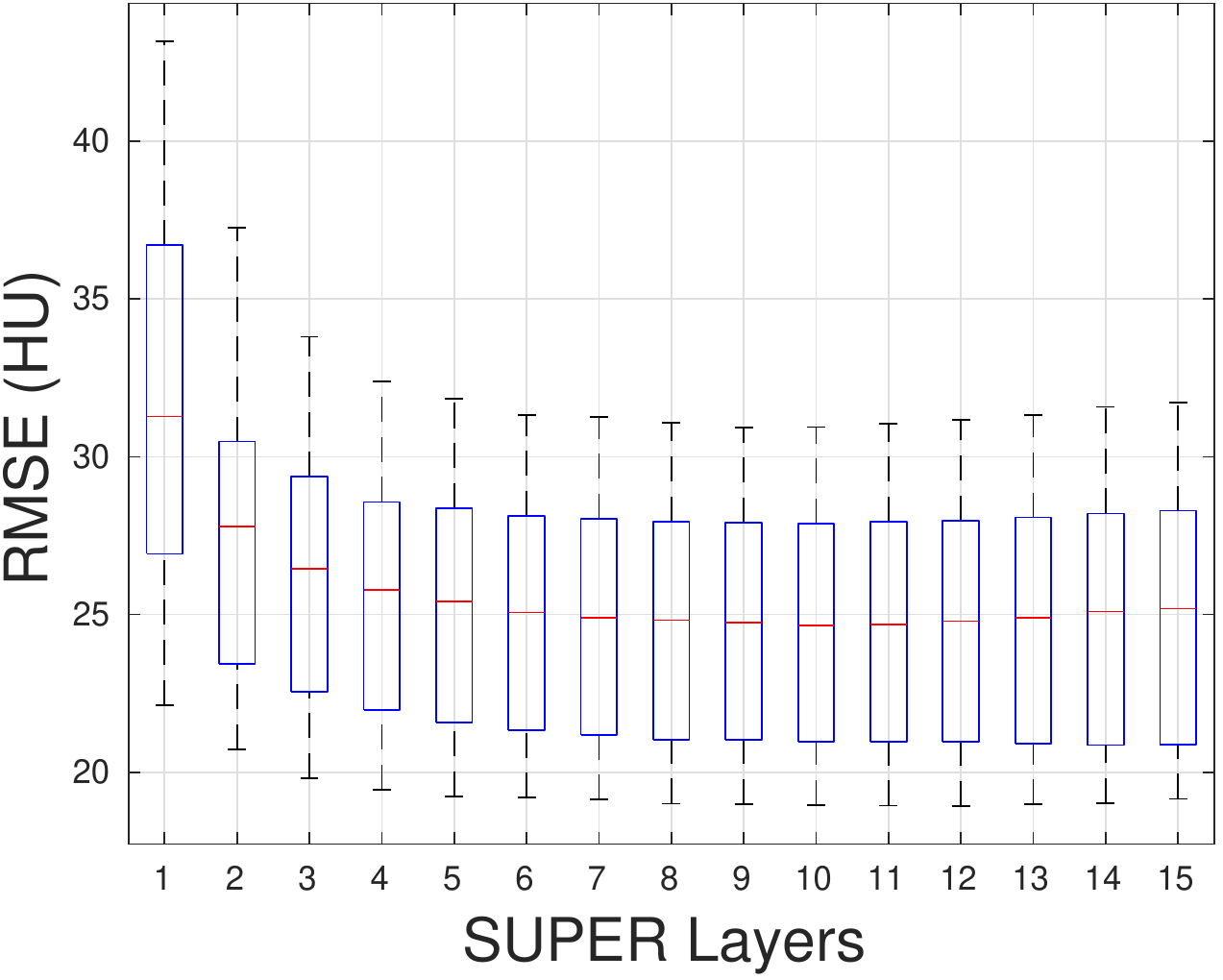}
		\caption{SUPER-FCN-ULTRA}
	\end{subfigure}
    \caption{RMSE spread of 20 test slices over SUPER layers of the SUPER-FCN-EP and SUPER-FCN-ULTRA algorithms.}
    \label{fig:FCN-rmse-layers}
    \vspace{-0.1in}
\end{figure}

 \begin{figure}[!htp]
    \centering
   		\begin{subfigure}[h]{0.22\textwidth}
		\centering
		\includegraphics[width=1\textwidth]{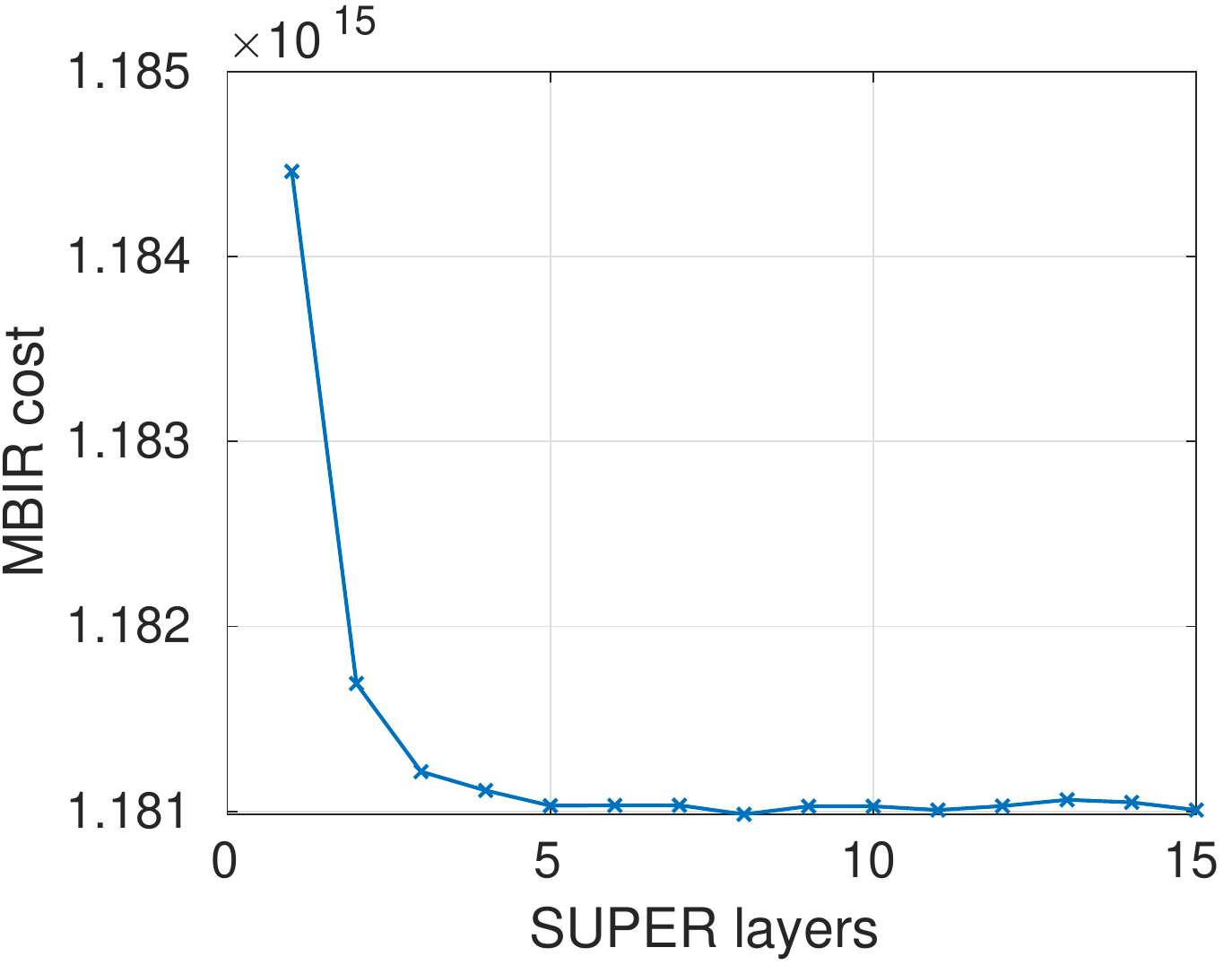}
		\caption{Slice 100 of patient L067}
	\end{subfigure}
	\begin{subfigure}[h]{0.22\textwidth}
		\centering
	\includegraphics[width=1\textwidth]{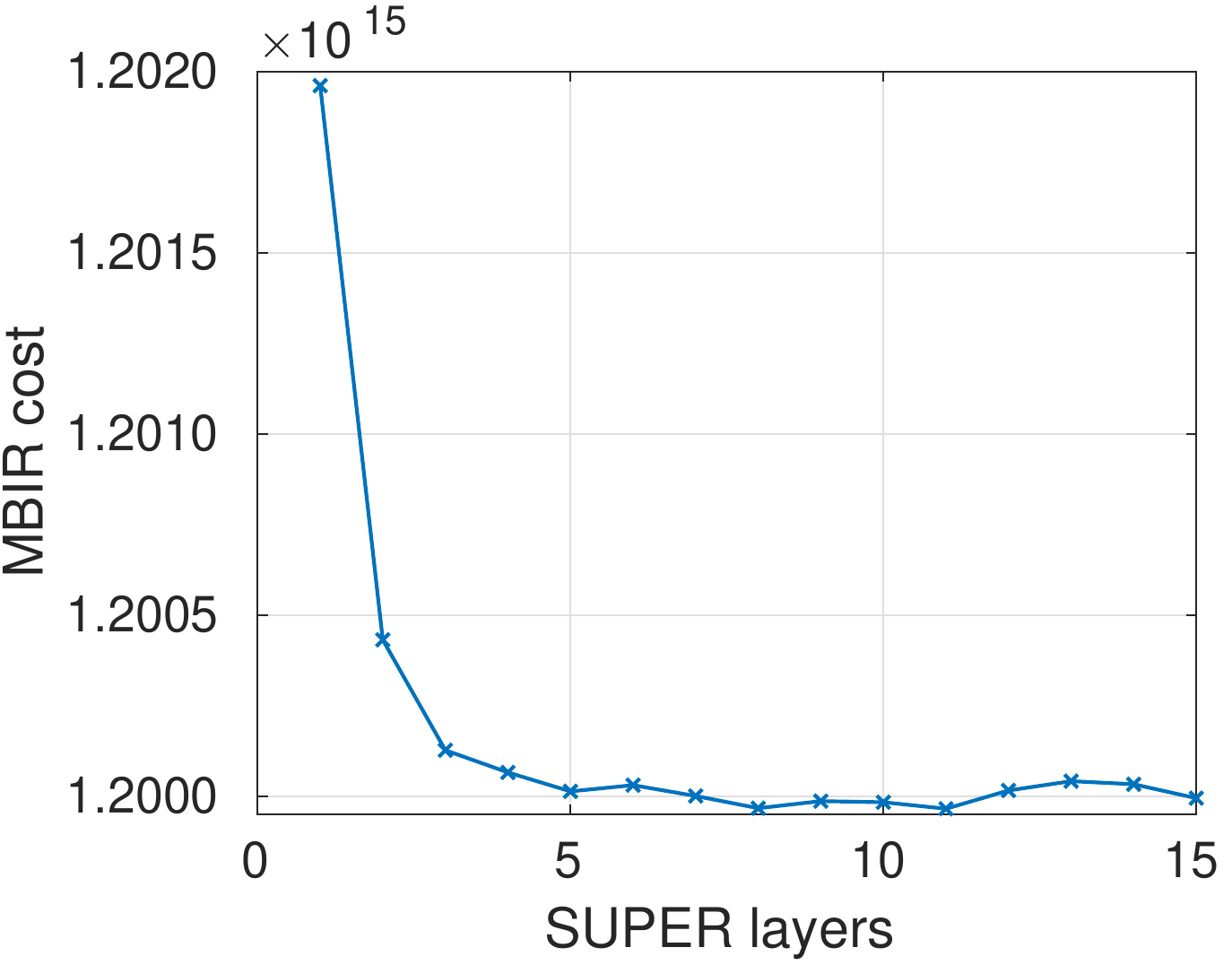}
		\caption{Slice 100 of patient L192}
	\end{subfigure}
   		\begin{subfigure}[h]{0.22\textwidth}
		\centering
		\includegraphics[width=1\textwidth]{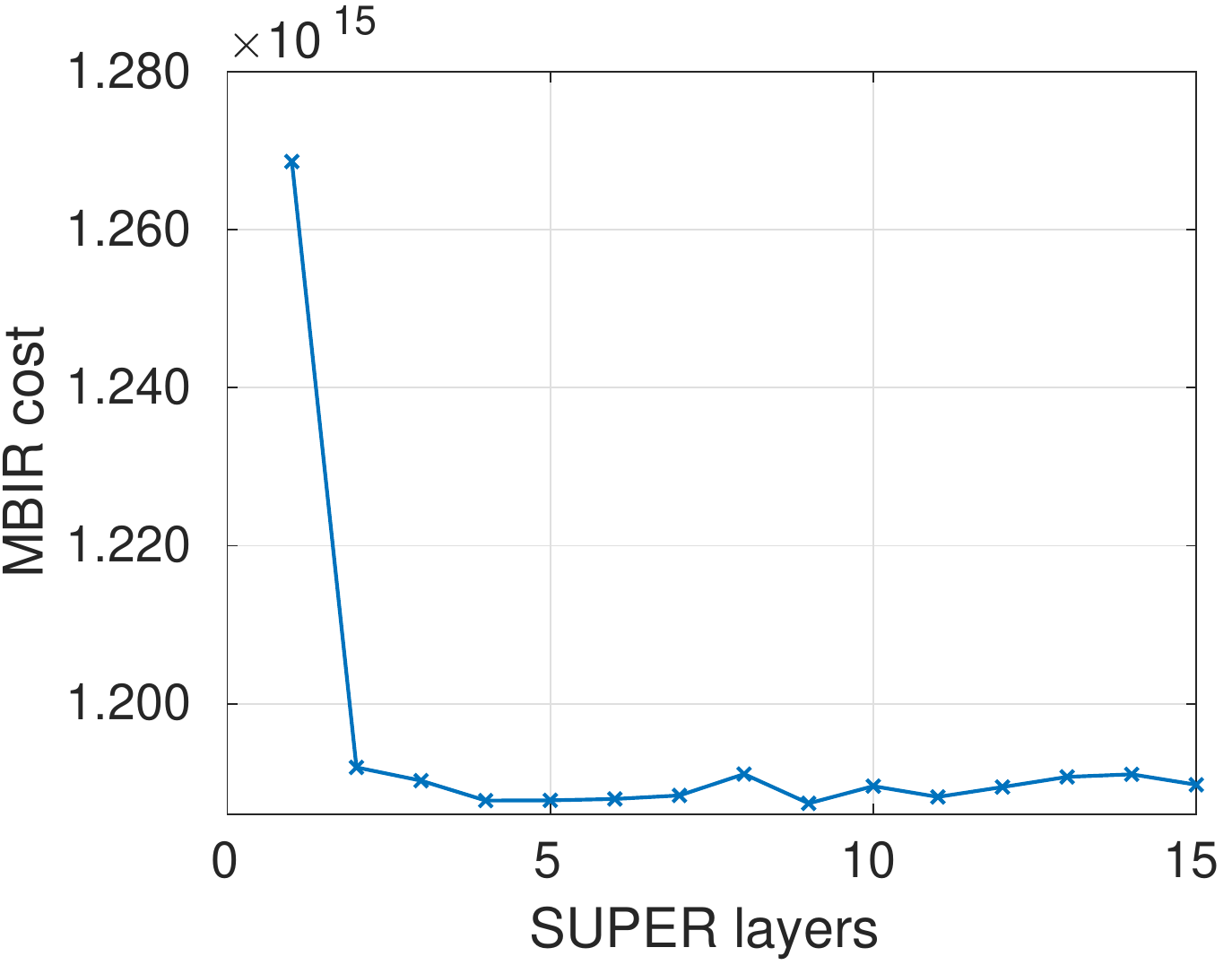}
		\caption{Slice 100 of patient L067}
	\end{subfigure}
   		\begin{subfigure}[h]{0.22\textwidth}
		\centering
		\includegraphics[width=1\textwidth]{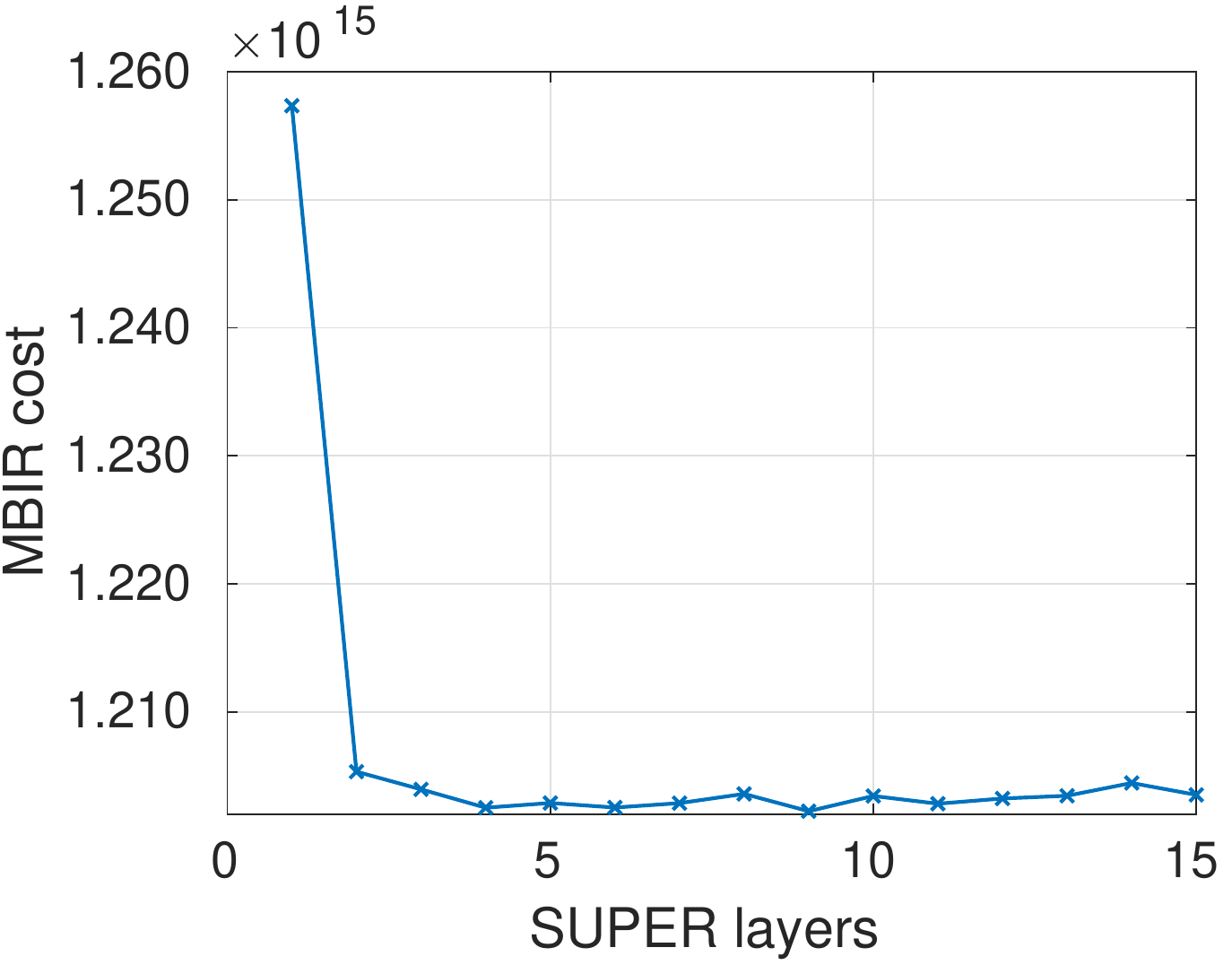}
		\caption{Slice 100 of patient L192}
	\end{subfigure}
    \caption{\BLUE{ULTRA-based reconstruction's cost function (achieved cost in (P0)) plotted over the SUPER layers of SUPER-WRN-ULTRA (the first row) and SUPER-FCN-ULTRA (the second row), when reconstructing two selected test slices.}}
    \label{fig:WRN_cost_converge}
    \vspace{-0.15in}
\end{figure}

\begin{figure}[!htp]
    \centering
   		\begin{subfigure}[h]{0.23\textwidth}
		\centering
		\includegraphics[width=1\textwidth]{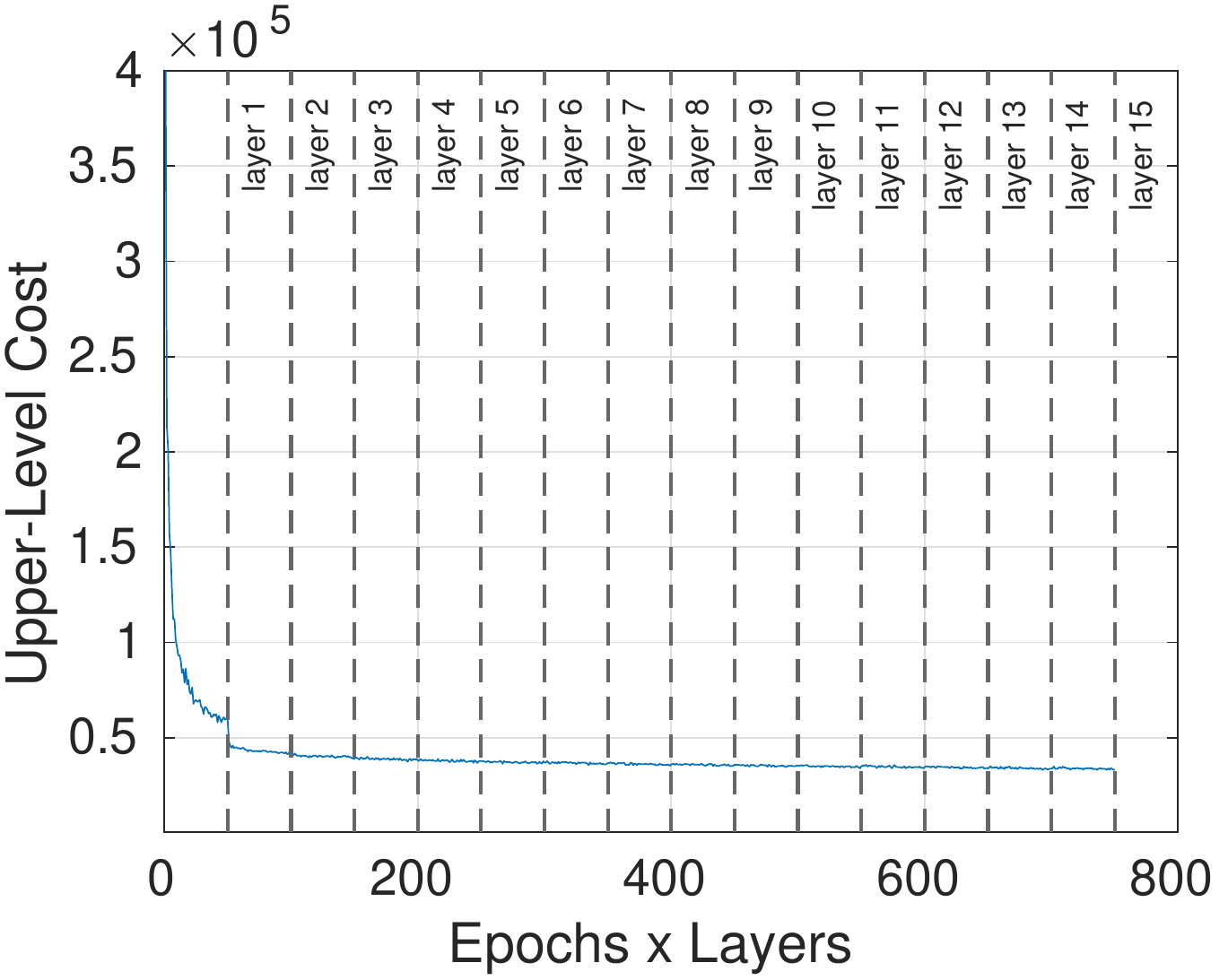}
		\caption{Cost correspond to (P1).}
		\label{fig:wrs_upperloss_xprev}
	\end{subfigure}
	\hfill
	\begin{subfigure}[h]{0.245\textwidth}
		\centering
	\includegraphics[width=1\textwidth]{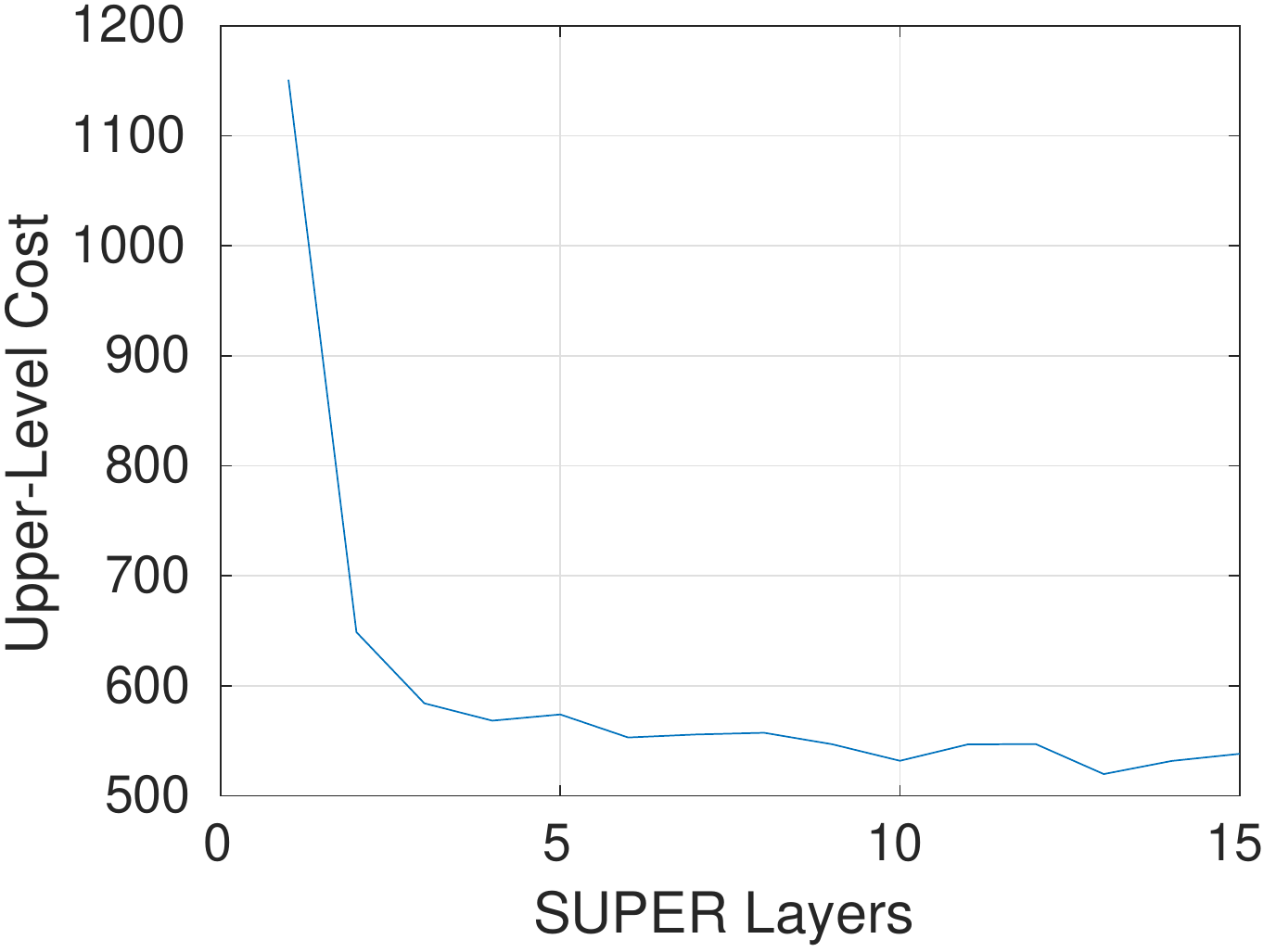}
	\caption{Upper-level cost in Eq. (5).}
	\label{fig:wrs_upperloss_xcur}
	\end{subfigure}
    \caption{\BLUE{Convergence of Training. (a) The cost corresponding to (P1) is plotted over training epochs and over all SUPER layers sequentially. (b) The cost corresponding to the upper-level cost of (5) is plotted over each SUPER layer based on reconstructions and network weights from the same ($l$th) layer, i.e., $\hat{\x}^{(l)}_{\bm{\theta}^{(l)}}(\y)$ and $\bm{\theta}^{(l)}$.}}
    \label{fig:WRN_trainingloss}
    \vspace{-0.15in}
\end{figure}

\subsection{Convergence Behavior of SUPER Reconstruction}
We have shown in Fig.~6 in the manuscript the convergence behavior in terms of RMSE of test slices using WavResNet based SUPER. Here, Fig.~\ref{fig:FCN-rmse-layers} demonstrates a similar RMSE convergence behavior (over SUPER layers) using FBPConvNet based SUPER.
We also plot \BLUE{in Fig.~\ref{fig:WRN_cost_converge} the ULTRA-based iterative module's costs of two test examples at the end (MBIR) iteration in each SUPER layer to indicate the achieved (P0) cost in each SUPER layer.}
In these two test examples, the nonconvex cost function decreases quickly during initial SUPER layers and varies only slightly in later layers for both SUPER-WRN-ULTRA and SUPER-FCN-ULTRA methods. The same behavior happens for other test samples as well. This indicates that our proposed algorithm can achieve \BLUE{empirically} stable results even with nonconvex cost functions.
\BLUE{In Fig.~\ref{fig:WRN_trainingloss}, we plot the training cost (P1) as well as the upper-level cost corresponding to the bilevel problem in Eq.~(5) of~\cite{SUPER-as-submit} for SUPER-WRN-ULTRA. 
The training loss in (P1) is plotted over training epochs and over each SUPER layer sequentially.
It is obvious that the training loss is convergent within $15$ SUPER layers (with 50 epochs in each layer). Fig.~\ref{fig:wrs_upperloss_xcur} roughly reflects the upper-level cost of the bilevel problem (Eq.~(5)), as we plugged (back) into the upper-level cost (as network inputs), the approximate reconstructions from the lower-level cost, using networks trained in the \textit{current} layer.
We observe that the upper-level cost in the bilevel problem decreases dramatically in the first five SUPER layers and then tends to converge with small oscillations within 15 layers, which provided practical stable reconstructions in our experiments.
This indicates potential for the proposed alternating training algorithm as a heuristic for the bilevel problem in Eq.~(5) of~\cite{SUPER-as-submit}.
}

\subsection{\BLUE{Examples with Various Dose Levels and Generalization}}
\begin{figure}[!htp]
	\centering 
	\vspace{-0.05in}
	\begin{subfigure}{1\textwidth}\leftskip10pt
		\scalebox{0.95}{
			\centering 
			\begin{tikzpicture}
			\begin{scope}
			[spy using outlines={rectangle,green,magnification=2.3,size=10mm, connect spies}]
			\node {\includegraphics[width=0.22\textwidth]{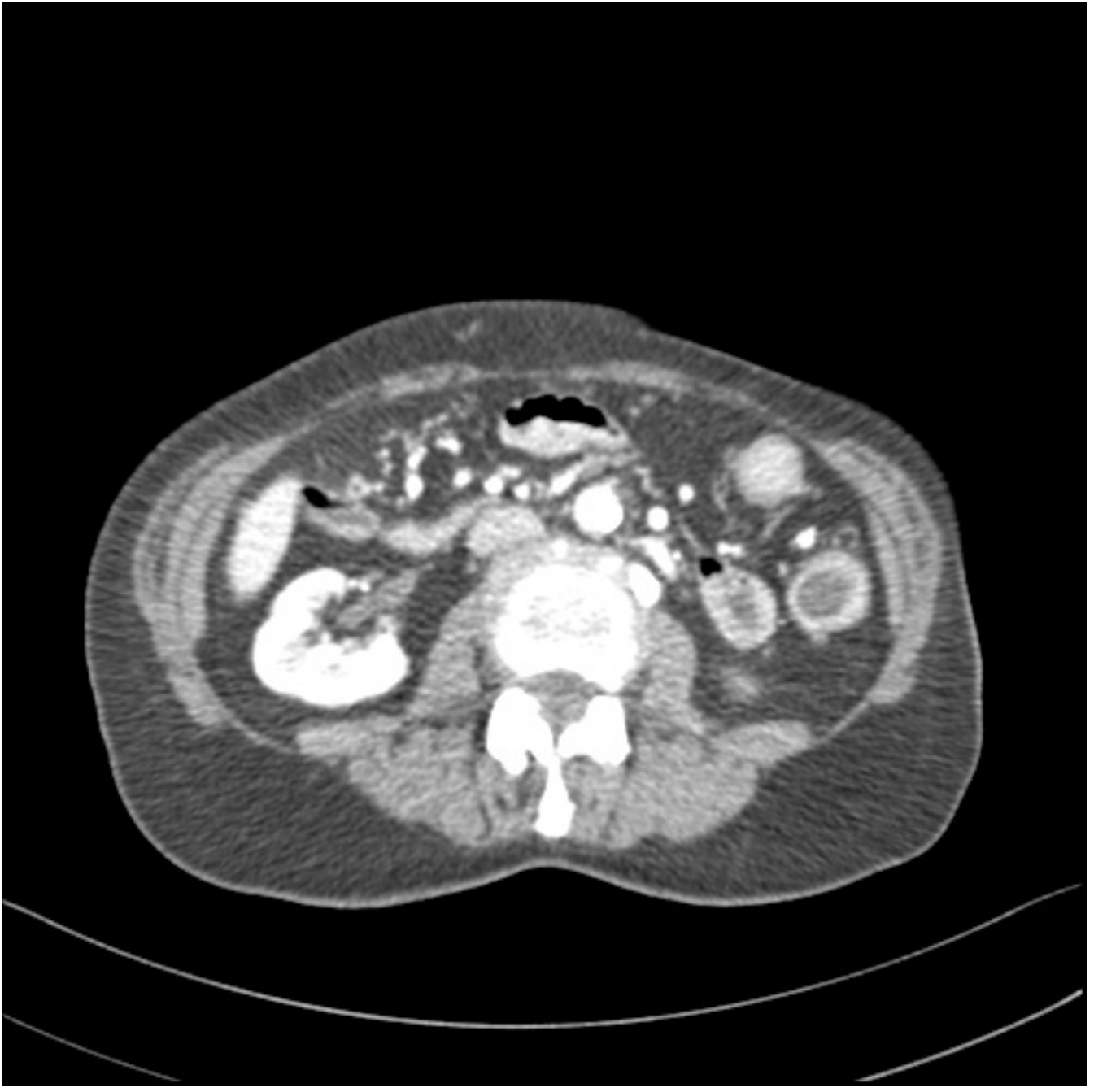} };
			\spy on (0.5,0.1) in node [right] at (0.95,1.4);
			\spy on (-0.25,-1.0) in node [left] at (-1,-1.65);
			\end{scope}
			\draw [->,>=stealth,red,line width=1pt] (1.0,1.0) -- (1.4,1.17);
			\draw [->,>=stealth,red,line width=1pt] (-1.8,-1.7) -- (-1.45,-1.6);
			\node [align = center,white, font=\bf] at (-0.1,1.72) {\small FBP};
			\node [white, font=\footnotesize] at (0.7,-1.7) {RMSE =26.7~HU};
			\end{tikzpicture}
				\hspace{-0.1in}
			\begin{tikzpicture}
			\begin{scope}
			[spy using outlines={rectangle,green,magnification=2.3,size=10mm, connect spies}]
			\node {\includegraphics[width=0.22\textwidth]{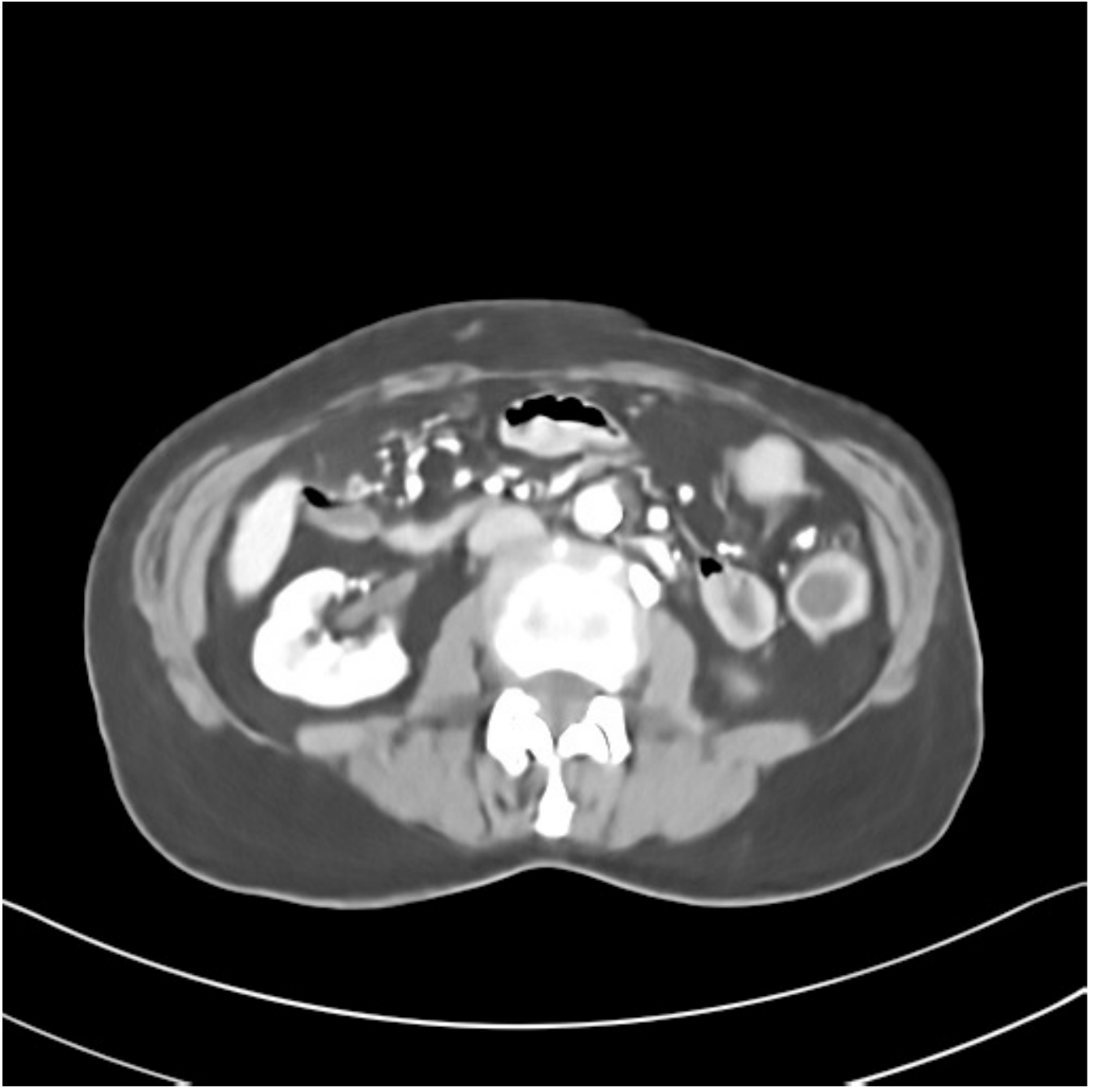} };
			\spy on (0.5,0.1) in node [right] at (0.95,1.4);
			\spy on (-0.25,-1.0) in node [left] at (-1,-1.65);
			\end{scope}
			\draw [->,>=stealth,red,line width=1pt] (1.0,1.0) -- (1.4,1.17);
			\draw [->,>=stealth,red,line width=1pt] (-1.8,-1.7) -- (-1.45,-1.6);
			\node [align = center,white, font=\bf] at (-0.3,1.72) {\small PWLS-ULTRA};
			\node [white, font=\footnotesize] at (0.7,-1.7) {RMSE = 16.4~HU};
			\end{tikzpicture}}
\end{subfigure}
	\vfil \vspace{-0.04in}
	\begin{subfigure}{1\textwidth}\leftskip10pt
		\scalebox{0.95}{
			\centering 
			\begin{tikzpicture}
			\begin{scope}
			[spy using outlines={rectangle,green,magnification=2.3,size=10mm, connect spies}]
			\node {\includegraphics[width=0.22\textwidth]{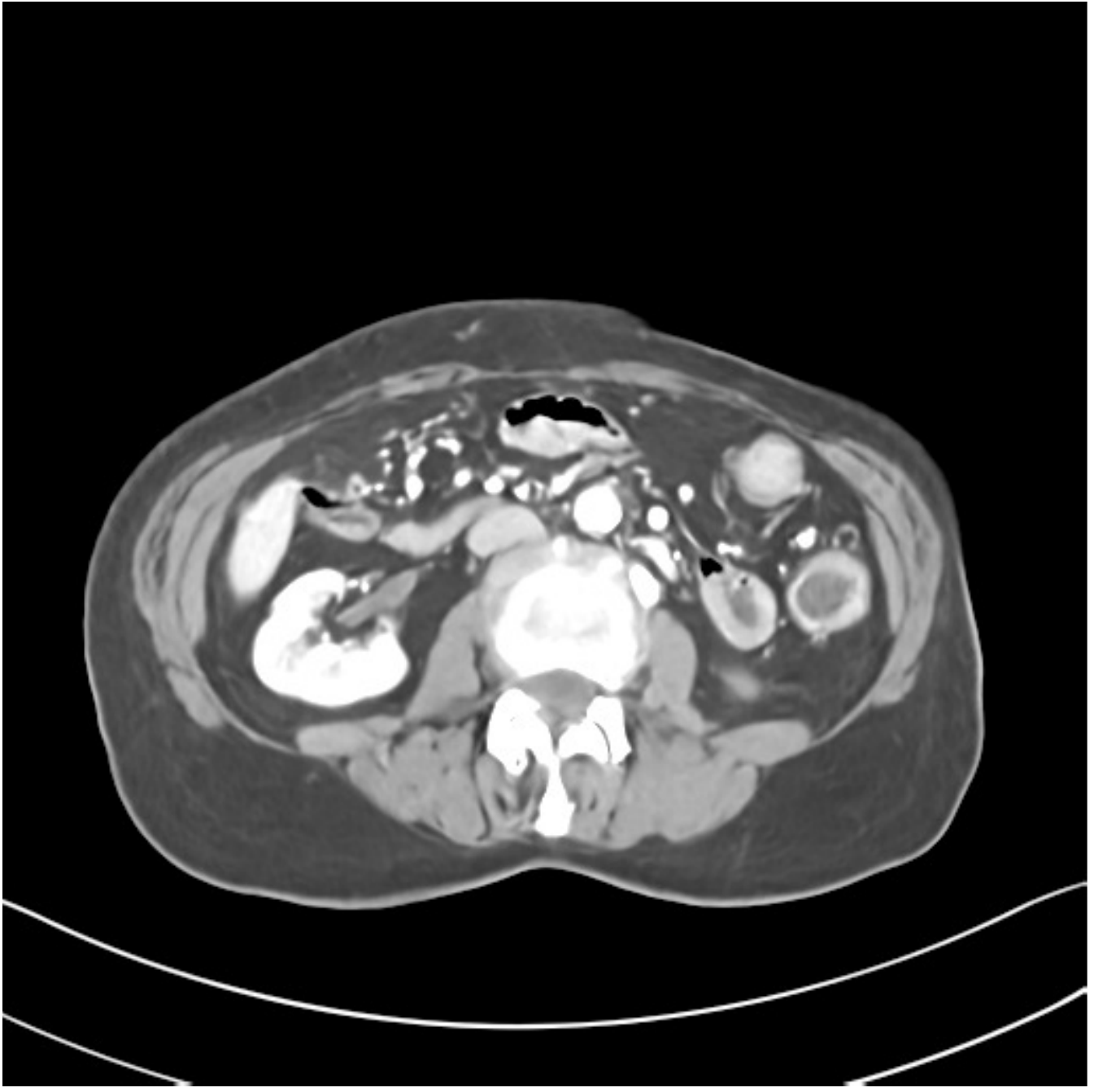} };
	\spy on (0.5,0.1) in node [right] at (0.95,1.4);
			\spy on (-0.25,-1.0) in node [left] at (-1,-1.65);
			\end{scope}
			\draw [->,>=stealth,red,line width=1pt] (1.0,1.0) -- (1.4,1.17);
			\draw [->,>=stealth,red,line width=1pt] (-1.8,-1.7) -- (-1.45,-1.6);
			\node [align = center,white, font=\bf] at (-0.3,1.72){\small FBPConvNet};
			\node [white, font=\footnotesize] at (0.7,-1.7) {RMSE = 12.2~HU};
			\end{tikzpicture}
				\hspace{-0.1in}
			\begin{tikzpicture}
			\begin{scope}
			[spy using outlines={rectangle,green,magnification=2.3,size=10mm, connect spies}]
			\node {\includegraphics[width=0.22\textwidth]{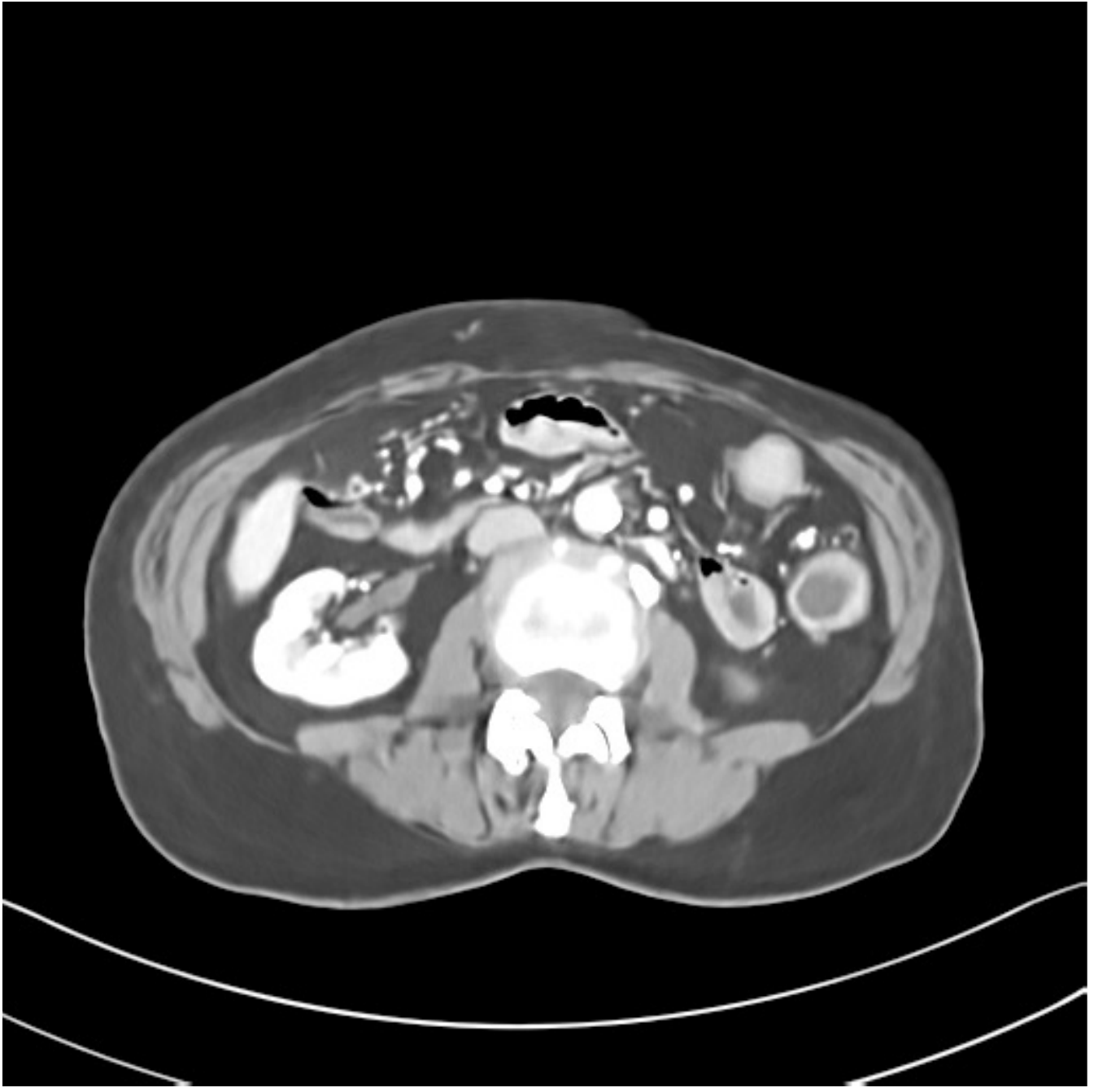} };
			\spy on (0.5,0.1) in node [right] at (0.95,1.4);
			\spy on (-0.25,-1.0) in node [left] at (-1,-1.65);
			\end{scope}
			\draw [->,>=stealth,red,line width=1pt] (1.0,1.0) -- (1.4,1.17);
			\draw [->,>=stealth,red,line width=1pt] (-1.8,-1.7) -- (-1.45,-1.6);
			\node [align = center,white, font=\bf] at (-0.5,1.72) {\small SUPER-FCN-ULTRA};
			\node [yellow, font=\footnotesize] at (0.7,-1.7) {RMSE = 11.1~HU};
			\end{tikzpicture}}
			\end{subfigure}
	\vspace{-0.05in}
	\caption{\BLUE{Reconstructed images of slice 100 of patient L192 under $I_0 = 1\times 10^5$ using different methods.}}
	\label{fig:super-fcn-1e5}
\end{figure}
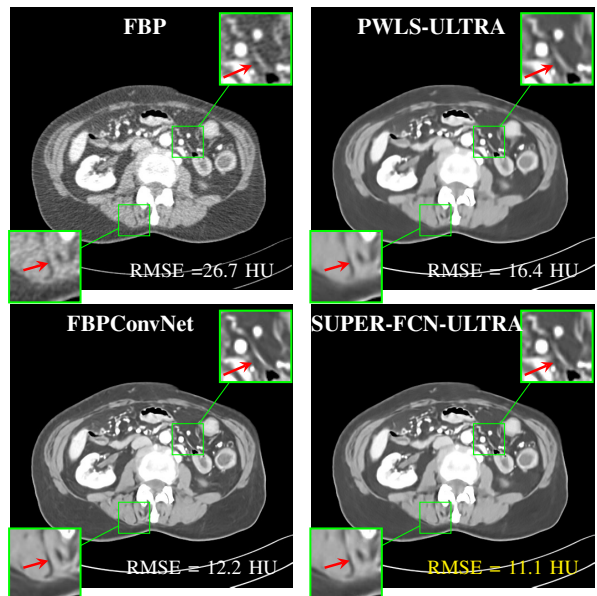

\begin{figure}[!htp]
	\centering 
	\vspace{-0.05in}
	\begin{subfigure}{1\textwidth}\leftskip10pt
		\scalebox{0.95}{
			\centering 
			\begin{tikzpicture}
			\begin{scope}
			[spy using outlines={rectangle,green,magnification=2.3,size=10mm, connect spies}]
			\node {\includegraphics[width=0.22\textwidth]{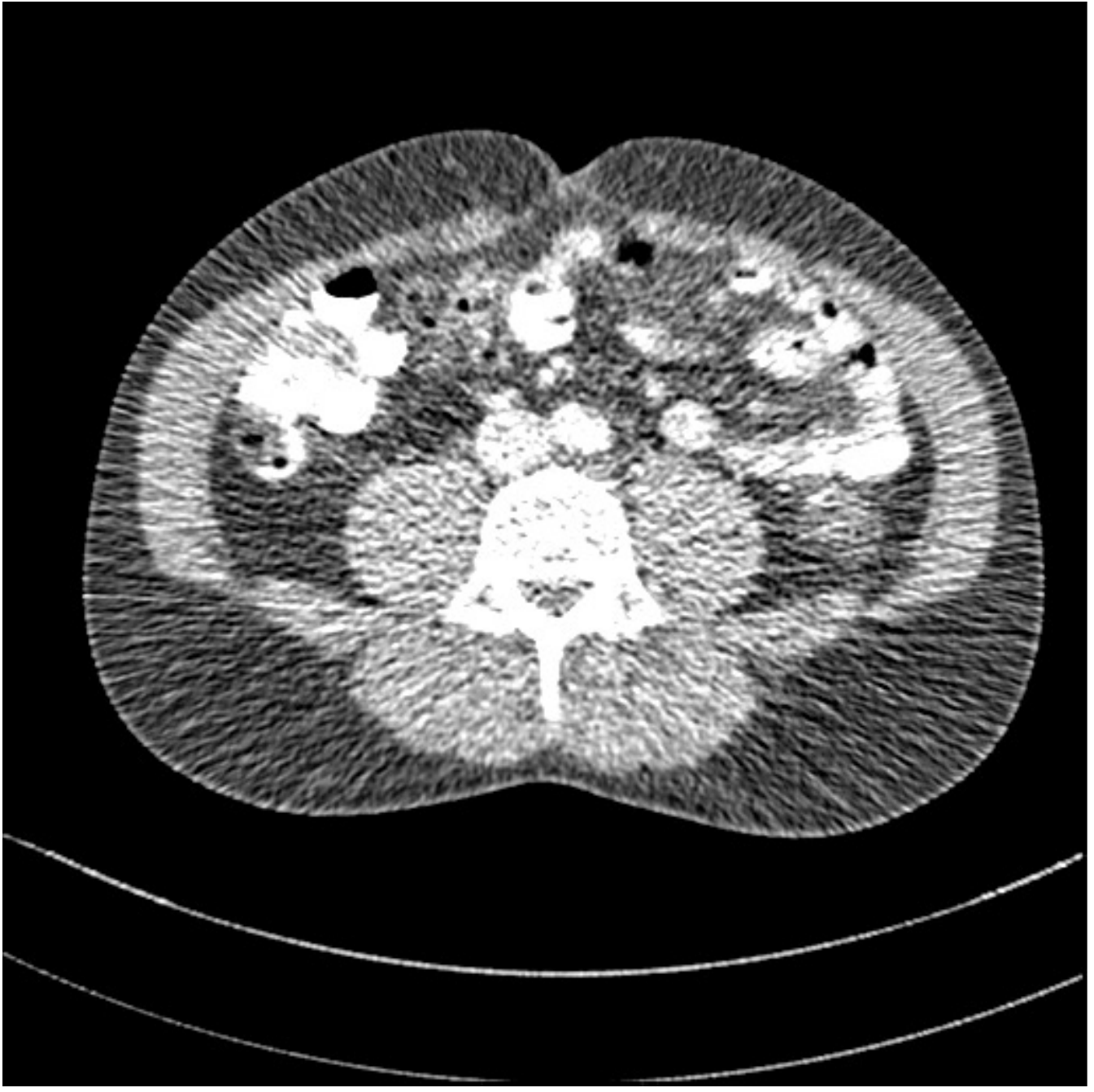} };
	\spy on (0.75,-0.52) in node [right] at (1.1,1.5);
 		\spy on (-0.3,-0.2) in node [left] at (-1,-1.65);
			\end{scope}
			\node [align = center,white, font=\bf] at (-0.1,1.72) {\small FBP};
			\node [white, font=\footnotesize] at (0.7,-1.7) {RMSE =54.8~HU};
			\end{tikzpicture}
			\begin{tikzpicture}
			\begin{scope}
			[spy using outlines={rectangle,green,magnification=2.3,size=10mm, connect spies}]
			\node {\includegraphics[width=0.22\textwidth]{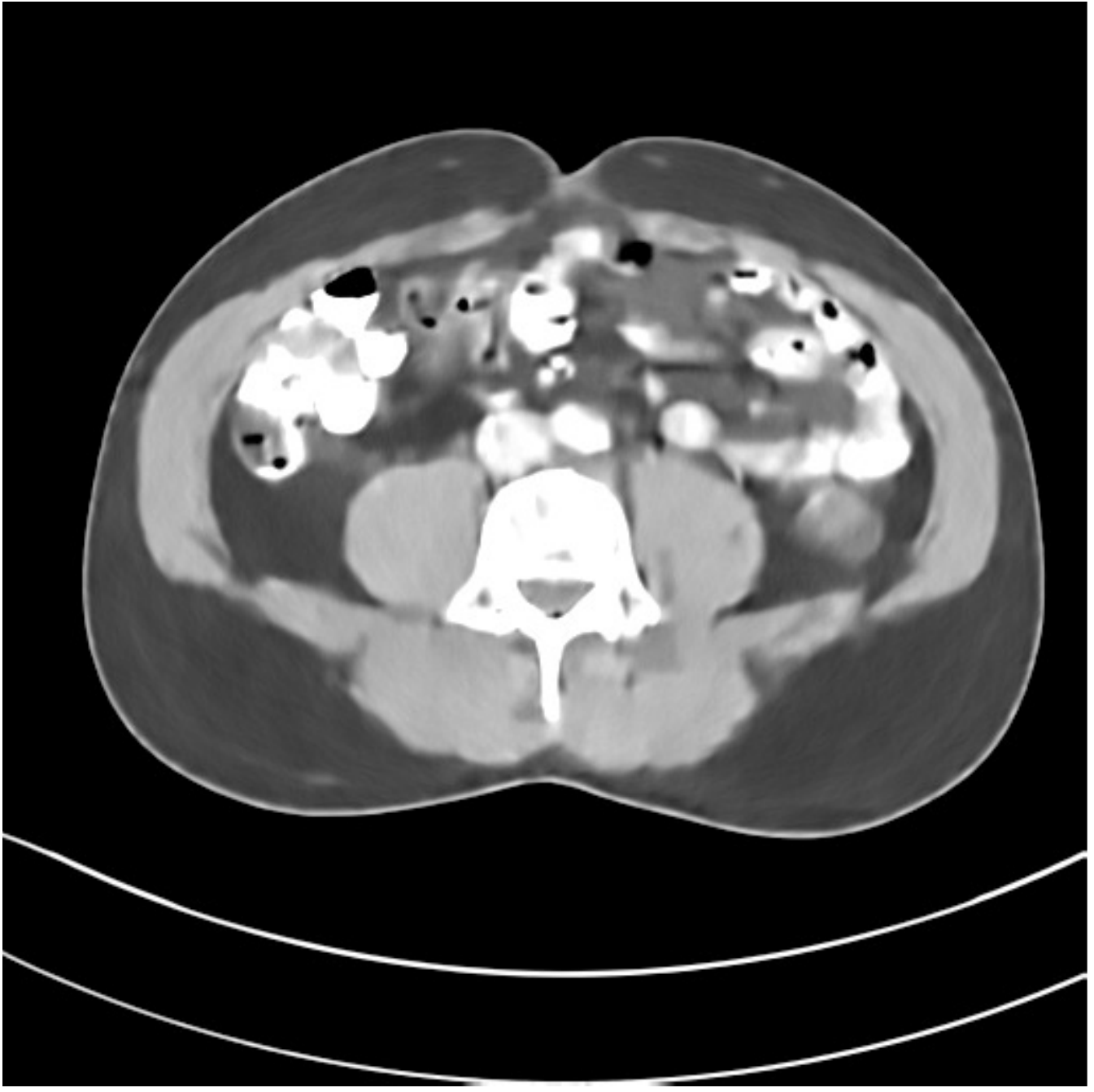} };
		\spy on (0.75,-0.52) in node [right] at (1.1,1.5);
 		\spy on (-0.3,-0.2) in node [left] at (-1,-1.65);
			\end{scope}
		
			\node [align = center,white, font=\bf] at (-0.3,1.72) {\small PWLS-ULTRA};
			\node [white, font=\footnotesize] at (0.7,-1.7) {RMSE = 24.8~HU};
			\end{tikzpicture}}
	\end{subfigure}
	\vfil \vspace{-0.04in}
	\begin{subfigure}{1\textwidth}\leftskip10pt
		\scalebox{0.95}{
			\centering 
			\begin{tikzpicture}
			\begin{scope}
			[spy using outlines={rectangle,green,magnification=2.3,size=10mm, connect spies}]
			\node {\includegraphics[width=0.22\textwidth]{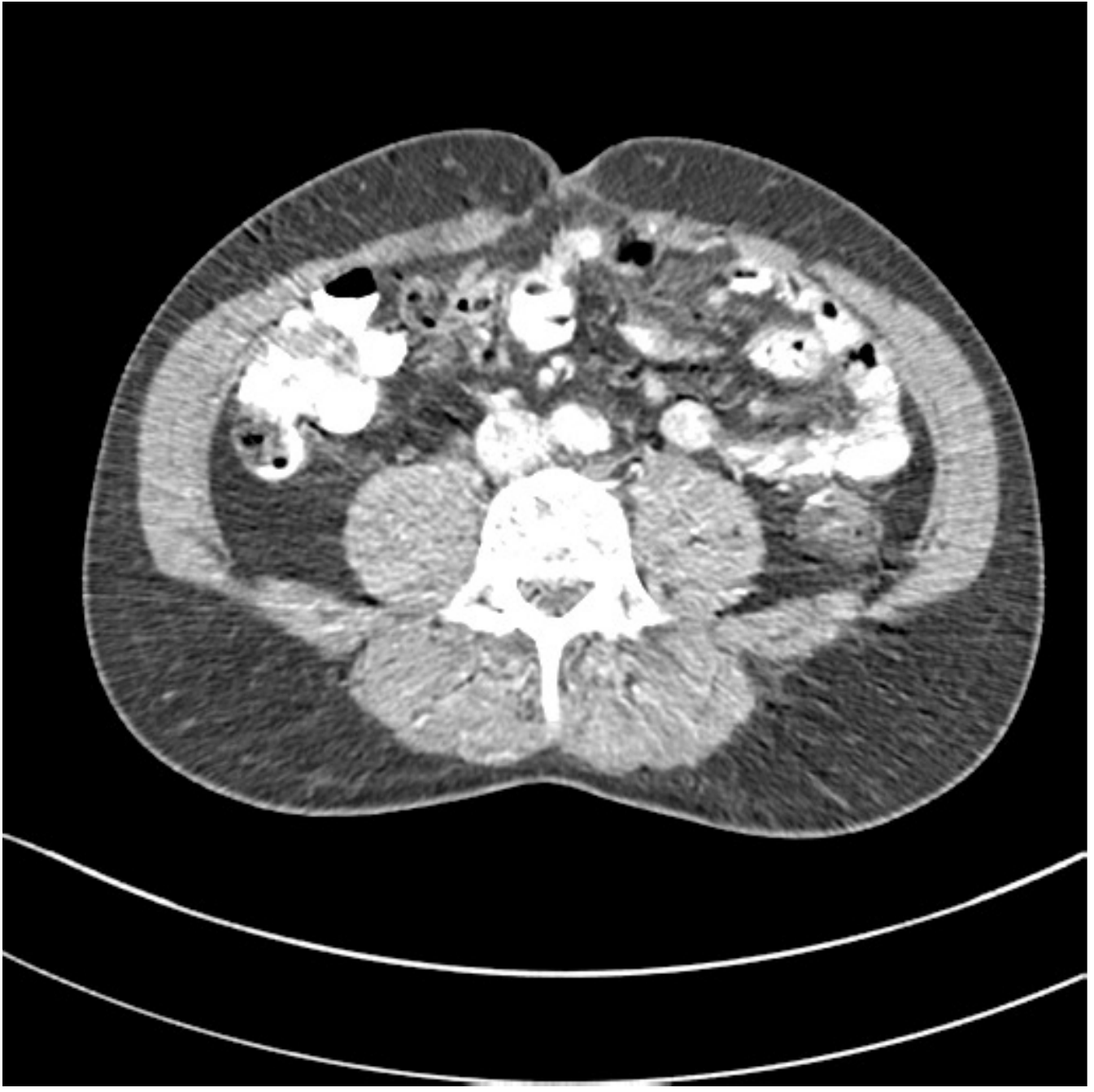} };
\spy on (0.75,-0.52) in node [right] at (1.1,1.5);
 		\spy on (-0.3,-0.2) in node [left] at (-1,-1.65);
			\end{scope}
		
			\node [align = center,white, font=\bf] at (-0.3,1.72){\small FBPConvNet};
			\node [white, font=\footnotesize] at (0.7,-1.7) {RMSE = 29.9~HU};
			\end{tikzpicture}
			\begin{tikzpicture}
			\begin{scope}
			[spy using outlines={rectangle,green,magnification=2.3,size=10mm, connect spies}]
			\node {\includegraphics[width=0.22\textwidth]{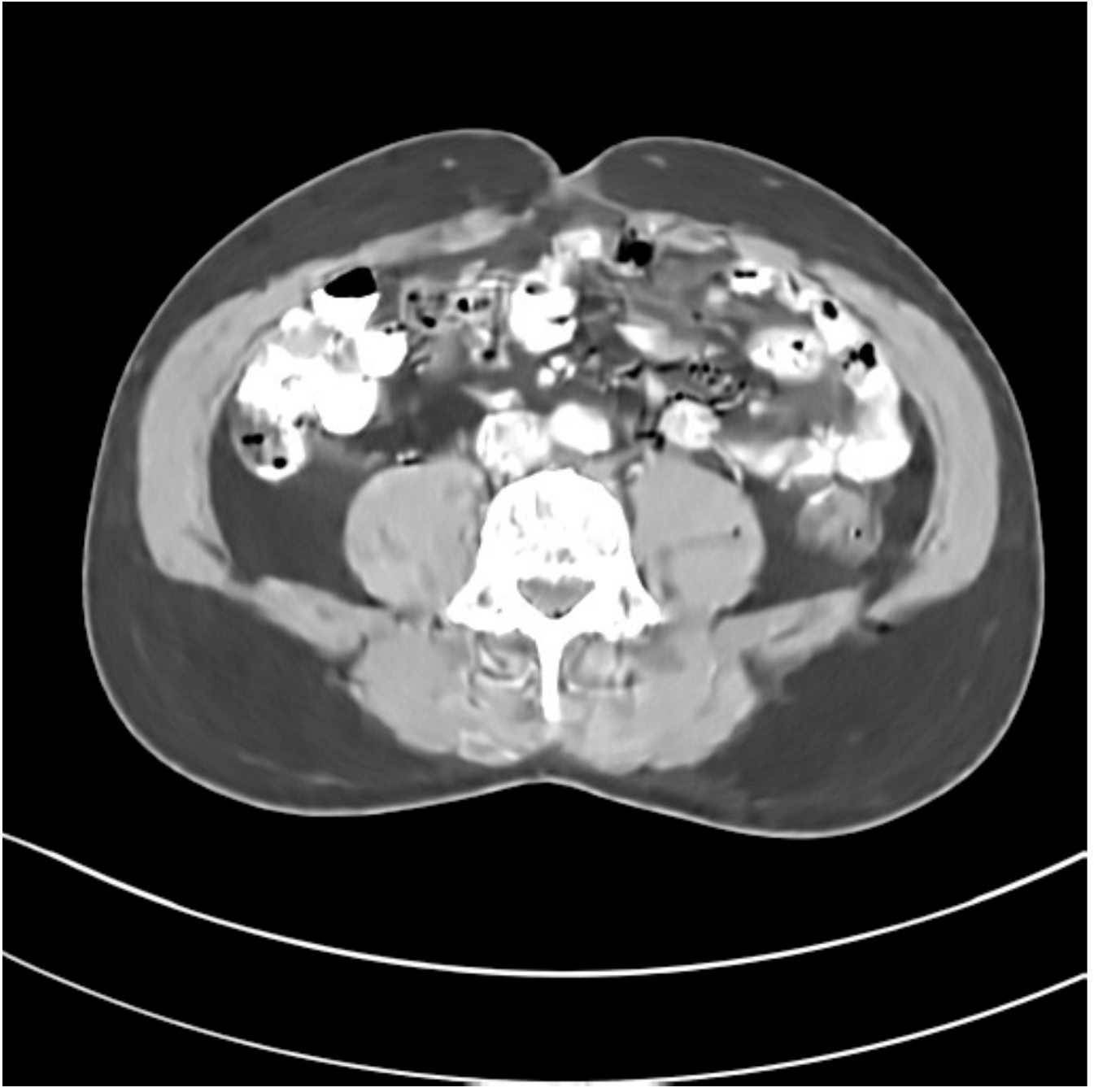} };
		\spy on (0.75,-0.52) in node [right] at (1.1,1.5);
 		\spy on (-0.3,-0.2) in node [left] at (-1,-1.65);
			\end{scope}
		
			\node [align = center,white, font=\bf] at (-0.5,1.72) {\small SUPER-FCN-ULTRA};
			\node [yellow, font=\footnotesize] at (0.7,-1.7) {RMSE = 21.5~HU};
			\end{tikzpicture}}
	\end{subfigure}
	\vspace{-0.05in}
	\caption{\BLUE{Reconstructed images of slice 100 of patient L067 under $I_0 = 2\times 10^4$ using different methods. FBPConvNet and SUPER-FCN-ULTRA used networks trained under $I_0 = 1\times 10^5$.}}
	\label{fig:super-fcn-2e4}
	\vspace{-0.2in}
\end{figure}
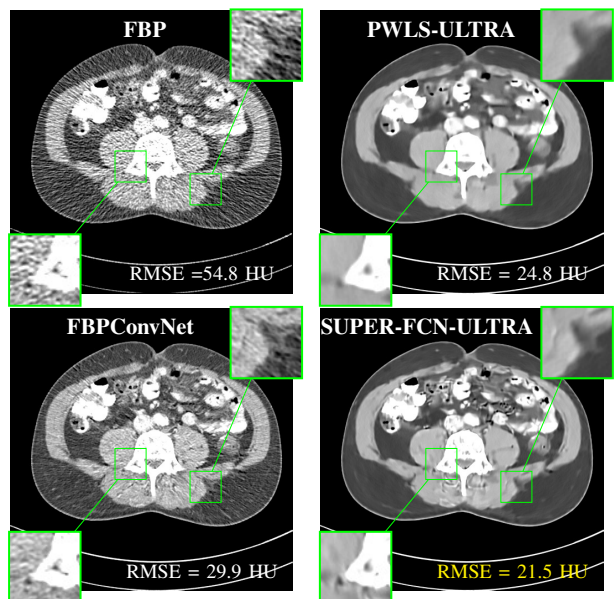

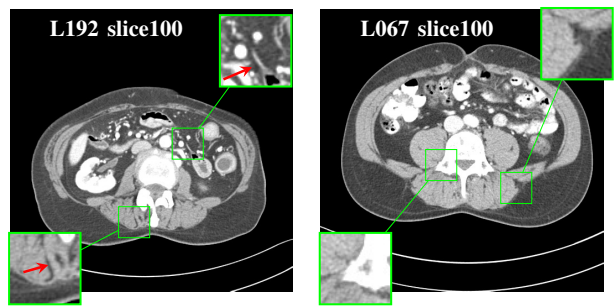
\begin{figure}[!htp]
	\centering 
	\vspace{-0.05in}
	\begin{subfigure}{1\textwidth}\leftskip10pt
		\scalebox{0.95}{
			\centering 
			\begin{tikzpicture}
			\begin{scope}
			[spy using outlines={rectangle,green,magnification=2.3,size=10mm, connect spies}]
			\node {\includegraphics[width=0.22\textwidth]{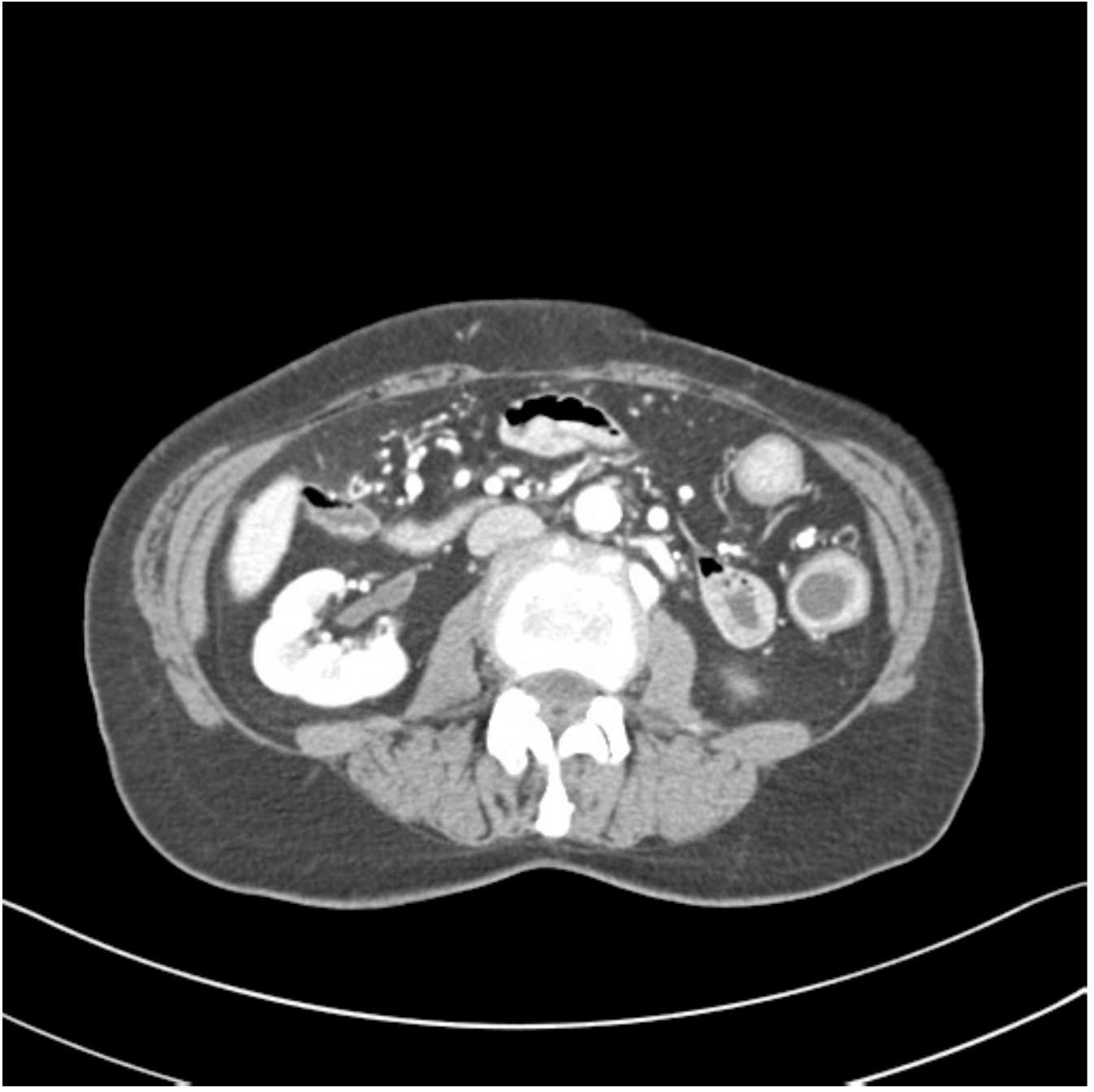} };
			\spy on (0.5,0.1) in node [right] at (0.95,1.4);
			\spy on (-0.25,-1.0) in node [left] at (-1,-1.65);
			\end{scope}
			\draw [->,>=stealth,red,line width=1pt] (1.0,1.0) -- (1.4,1.17);
			\draw [->,>=stealth,red,line width=1pt] (-1.8,-1.7) -- (-1.45,-1.6);
			\node [align = center,white, font=\bf] at (-0.5,1.72) {\small L192 slice100};
			\end{tikzpicture}
			\begin{tikzpicture}
			\begin{scope}
			[spy using outlines={rectangle,green,magnification=2.3,size=10mm, connect spies}]
			\node {\includegraphics[width=0.22\textwidth]{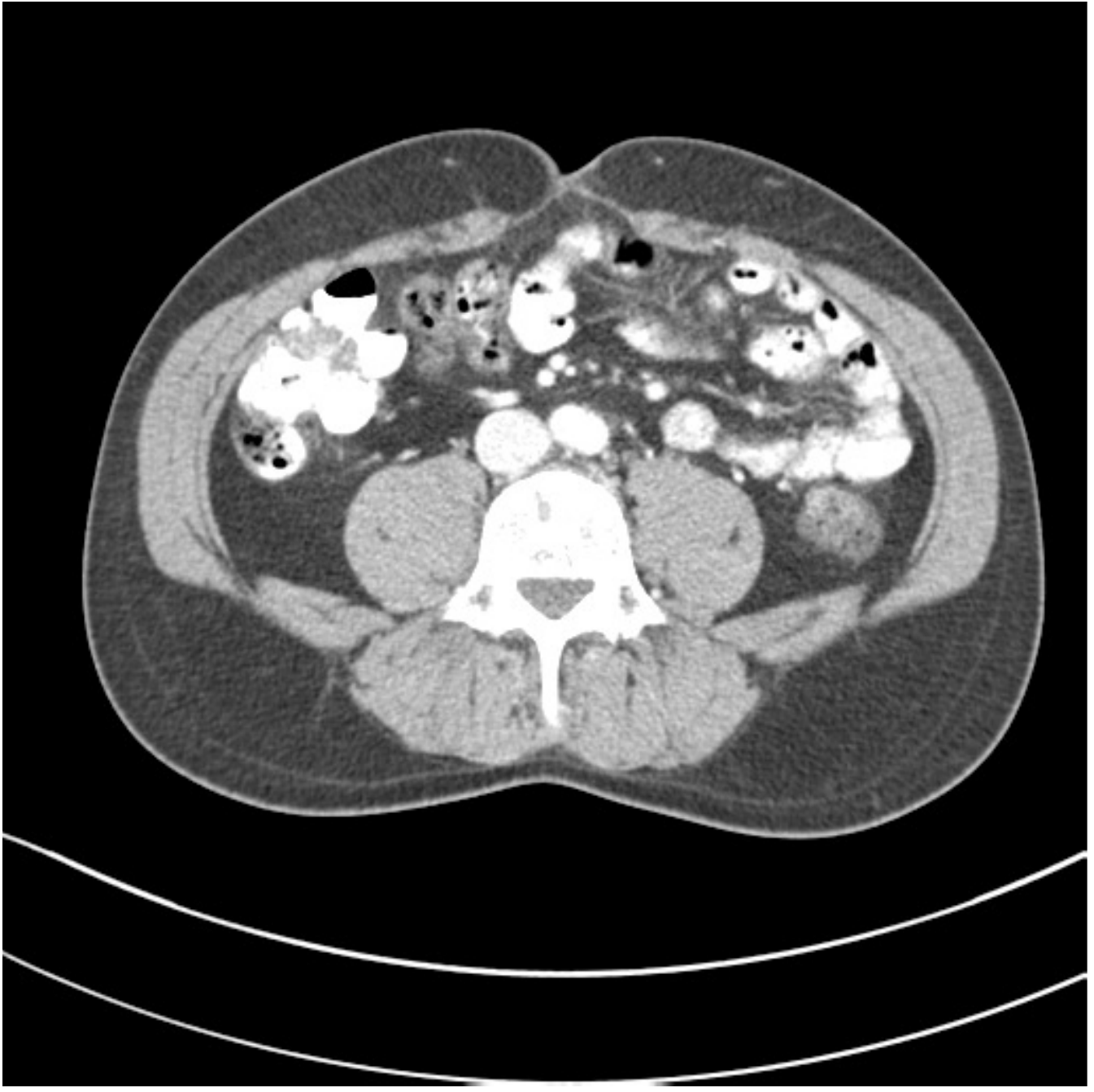} };
		\spy on (0.75,-0.52) in node [right] at (1.1,1.5);
 		\spy on (-0.3,-0.2) in node [left] at (-1,-1.65);
			\end{scope}
			\node [align = center,white, font=\bf] at (-0.5,1.72) {\small L067 slice100};
			\end{tikzpicture}}
			\end{subfigure}
\caption{\BLUE{Reference images corresponding to Fig.~\ref{fig:super-fcn-1e5} and Fig.~\ref{fig:super-fcn-2e4}, respectively.}}
	\label{fig:vary-dose-reference}
\end{figure}

\BLUE{This subsection shows examples of reconstructions with different dose levels (corresponding to Section IV.I in the main paper).
Fig.~\ref{fig:super-fcn-1e5} shows an example test slice reconstructed by different methods under $I_0 = 1\times 10^5$ in both training and testing. In this figure, the SUPER reconstructed image looks cleaner and sharper than the ones reconstructed by the constituent standalone supervised and unsupervised iterative methods.
In Fig.~\ref{fig:super-fcn-2e4}, we show reconstructions for dose $I_0=2\times 10^4$ using the supervised networks trained under the very different dose $I_0=1\times 10^5$. In this example, the image provided by FBPConvNet is much noiser than the PWLS-ULTRA reconstruction, while combining these two models in SUPER, i.e., SUPER-FCN-ULTRA, achieved better results than both of its standalone counterparts in terms of RMSE values and image quality (less noise and sharper edges).
This indicates that it is still possible that SUPER-FCN-ULTRA outperforms (generalizes better than) the recent unsupervised PWLS-ULTRA when significant dose mismatch appears between training and testing.
We show the reference images of these two examples in Fig.~\ref{fig:vary-dose-reference}.
}

\end{document}